\RequirePackage{rotating}
\documentclass{article}
\usepackage{booktabs} 
\usepackage[ruled]{algorithm2e} 

\SetAlFnt{\small}
\SetAlCapFnt{\small}
\SetAlCapNameFnt{\small}
\SetAlCapHSkip{0pt}
\IncMargin{-\parindent}

\usepackage[numbers]{natbib}
\usepackage{graphicx}
\usepackage{adjustbox}
\usepackage{subcaption}
\usepackage{float}
\usepackage{romannum}
\usepackage{subcaption}
\usepackage{amsfonts}
\usepackage{siunitx}
\usepackage{rotating}
\usepackage{xcolor}

\DeclareGraphicsExtensions{.eps}

\newcommand{\rev}[1]{#1}
\newcommand{\revm}[1]{#1}

\begin{document}
\title{Community Detection in Multiplex Networks}
\date{Originally published on ACM Computing Surveys\\version 1, January 2020}

\author{Matteo Magnani\thanks{Infolab, Uppsala University, Uppsala, Sweden, matteo.magnani@it.uu.se} \and
Obaida Hanteer\thanks{IT University of Copenhagen, Copenhagen, Denmark, obha@itu.dk} \and
Roberto Interdonato\thanks{CIRAD, UMR TETIS, Montpellier, France, roberto.interdonato@cirad.fr} \and
Luca Rossi\thanks{IT University of Copenhagen, Copenhagen, Denmark, lucr@itu.dk} \and
Andrea Tagarelli\thanks{University of Calabria, Calabria, Italy, andrea.tagarelli@unical.it}}

\maketitle

\begin{abstract}

A multiplex network models different modes of interaction among same-type entities. 
In this article we provide 
a taxonomy o\rev{f} community detection algorithms in multiplex networks. We characterize the different algorithms based on various properties and we discuss the type of communities detected by each method.  We then provide an extensive \rev{experimental} evaluation of the reviewed methods to answer three main questions\rev{:} to what extent the evaluated methods are able to detect ground\rev{-}truth communities, to what extent different methods produce similar community structures and to what extent the evaluated methods are scalable. \rev{One} goal of this survey is to \rev{help} scholars and practitioners \rev{to choose} the right method\rev{s} for the data and the task at hand\rev{, while also emphasizing when such choice is problematic.}
\end{abstract}

\section{Introduction}
\label{intro}


Multiplex network analysis \rev{has} emerged as a promising approach to investigate complex systems.
A multiplex network is a model used to represent multiple modes of interaction or different \rev{types of} relationships among entities of the same type (e.g. people). Th\rev{is} model ha\rev{s} been used to study a large variety of systems across disciplines\rev{,} ranging from living organisms and human societies to transportation systems and critical infrastructures. For example, a description of the full protein-protein interactom\rev{e}\footnote[1]{\rev{A}n interactom\rev{e} is the totality of protein--protein interactions happen\rev{ing} in a cell\rev{.}} \rev{involves}, for some organisms, up to seven distinct modes of interaction among thousands of protein molecules \citep{DeDomenico2015}. Another example is in air transportation systems when modeling the connections between airports through direct flights; here, the different commercial airlines can be seen as different modes of connection among airports \citep{Cardillo2013}. 

\begin{figure}[!htpb]
	\centering
	\includegraphics[width=0.6\textwidth]{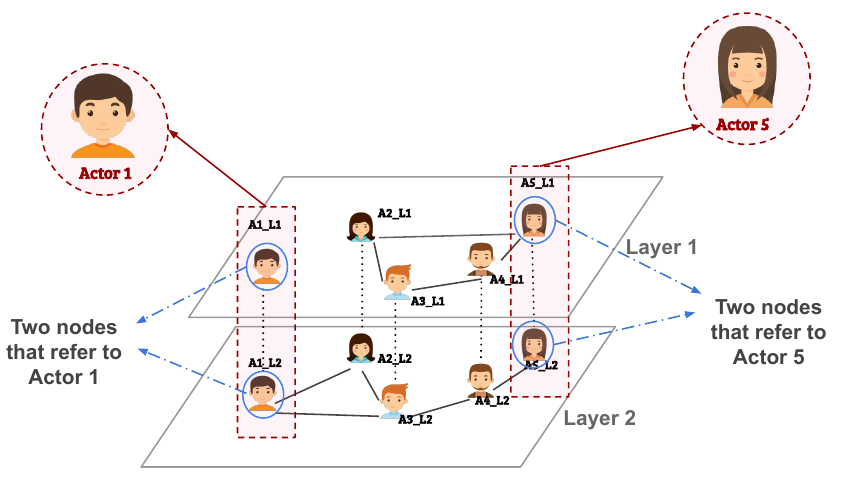}
	\caption{A\rev{n} example of a multiplex network with two types of interaction among five actors. This is represented as five nodes replicated in two layers. The two nodes representing the same actor (e.g. the same person) are linked by a dotted line}
	\label{fig:mpx}
\end{figure}

Figure \ref{fig:mpx} shows a typical layered representation of a multiplex network, where each layer corresponds to a type of interaction and nodes \rev{(also called vertices)} in different layers can be associated to the same actor, e.g. the same person or the same airport. Here, we adopt the term \emph{actor} from the field of social network analysis, where multiplex networks have been first applied, and the term \emph{layer} from recent generalizations of the original multiplex model~\citep{Brodka2011a,Magnani2011b,Kivela2014,DickisonMagnaniRossi2016}. 

A core task in network analysis is to identify and understand communities, also known as clusters or cohesive groups; that is, to explain why groups of entities (actors) belong together based on the explicit ties among them and/or the implicit ties induced by some similarity measures given some attributes of these entities. Since members of a community tend to share common properties, revealing the community structure in a network can provide a better understanding of the overall functioning of the network. 

\rev{Unfortunately, c}ommunity detection methods \rev{for} simple graphs are not sufficient to deal with the complexity of the multiplex 
\rev{model, for three main reasons. First, without allowing the analysis of subsets of the layers some communities may become hidden by edges in irrelevant layers. This is a common problem also in traditional multivariate data analysis, where several preprocessing methods have been developed to remove irrelevant information and algorithms have been extended to explore subsets of the data dimensions, as done by subspace clustering methods. Second, algorithms not explicitly representing the different layers cannot differentiate between different types of multiplex communities, e.g., those present on a single layer and those made of specific combinations of layers. Third, without a concept of layer it is not possible to include the same actor in different communities depending on the layer where the actor is active. In other words, community detection methods for simple graphs cannot conceptually represent (and thus discover) some types of communities that can only be defined on multilayer networks, although this does not 
imply that
non-multilayer methods will always be outperformed by multilayer algorithms.}

\rev{To address the above limitations, 
several community detection algorithms for multiplex networks have been recently proposed, based on different definitions of community and different computational approaches.}
Recent works have  provided a partial overview of \rev{existing algorithms}. \citep{Kim2015} proposed some criteria to compare multi-layered community detection algorithms, but without \rev{any} experimental evaluation. Similarly, \citep{Bothorel2015} highlighted the conceptual differences among different clustering methods over attributed graphs, including edge-labeled graphs that can be used to represent multiplex networks, but only provided a taxonomy of the different algorithms without any experimental analysis. 
\citep{Loe2015} instead performed a pairwise comparison of the different clusterings produced by some existing algorithms. The work by \citep{Huang2020} provides a more general overview on multilayer networks (which include multiplex networks), but without comprehensive experiments.


This article provides a systematic review and experimental comparison of existing methods\rev{, with the aim of} simplify\rev{ing} the choice and the setup of the most appropriate algorithm for the task at hand. \rev{We test the accuracy of the different methods with respect to some given ground truth on both synthetic and real networks and we study their scalability in terms of the size of the network, both vertically (number of layers) and horizontally (number of actors).} At the same time, \rev{we} highlight weaknesses and strengths of specific methods and of the current state-of-the-art as a whole\rev{, showing how even the most sophisticated methods fail to identify some types of communities}.


The focus of this survey is on algorithms explicitly designed to discover communities in multiplex networks through the analysis of the network structure.  Several community detection algorithms have been proposed \rev{to} deal with models \rev{related to but} not \rev{compatible} with the multiplex model\rev{, such as}
Heterogeneous Information Networks \citep{Sun2013,Sun2009,Zhou2013,Sun2009a} and \rev{b}ipartite \rev{n}etworks \citep{Barber2007,Guimera2007}, and are not included in our article. Since we focus on \rev{network structure}, graph clustering on attributed networks \citep{Boden2012a,Ruan, Silva,Xu,Zhou, Qi,Li} \rev{is also not} included in our analysis. For a survey on attributed graph clustering we refer the reader to \citep{Bothorel2015}. 


The rest of this work  is organized as follows.  \rev{Section~\ref{sec:mpx_community} provides some basic definitions used throughout the article.} In Section~\ref{sec:taxonomy} we introduce a taxonomy of existing multiplex community detection methods. \rev{Section~\ref{sec:theory} provides a theoretical comparison of the reviewed algorithms, while} Section~\ref{sec:experiments} presents the experimental settings and the evaluation datasets used in our experiments\rev{. T}he results \rev{of the experimental analysis are given} in Section~\ref{sec:results}. We  summarize our main findings and indicate usage guidelines emerged from our experiments in Section~\ref{sec:discussion}. 

\section{\rev{Multiplex networks and communities}}\label{sec:prel}

\rev{A multiplex network is a special case of a multilayer network.
A \textit{multilayer network} is defined as a tuple $(A, L, V, E)$, where $A$ is a set of actors, $L$ is a set of layers, and $(V, E)$ is a graph on $V \subseteq A \times L$.  Notice that this definition does not require all the actors to be present in all the layers, and allows actors to be present in some layers without having any neighbor on those layers.} 

\rev{In multiplex networks $E$ is restricted to intra-layer edges, that is, an edge $((v_1, l_1), (v_2, l_2))$ is allowed only if $l_1 = l_2$. In the following we use $a$, $l$, $v$, and $e$ to refer to the cardinality of, respectively, $A$, $L$, $V$, and $E$. We use the terms vertex or node to indicate the elements of $V$, that is, actors inside a layer.}

\begin{figure}[!htpb]
	\centering
	\begin{subfigure}[t]{0.32\textwidth}
		\centering
		\includegraphics[width=\textwidth]{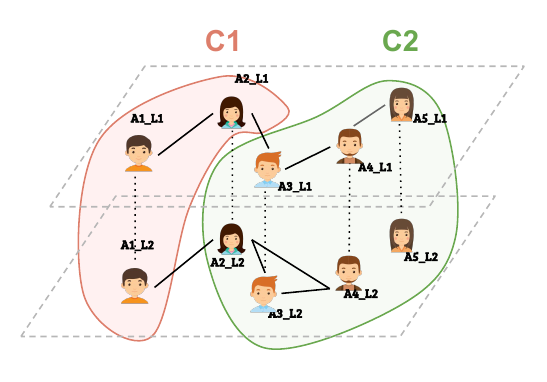} 
		\caption{Total}
		\label{fig:total}
	\end{subfigure}
	\begin{subfigure}[t]{0.32\textwidth}
		\centering
		\includegraphics[width=\textwidth]{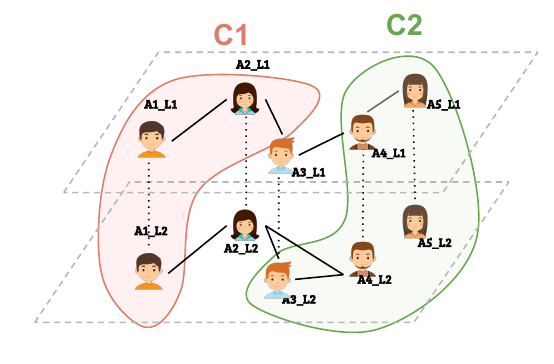} 
		\caption{Partial}
		\label{fig:partial}
	\end{subfigure}
	\begin{subfigure}[t]{0.32\textwidth}
		\centering
		\includegraphics[width=\textwidth]{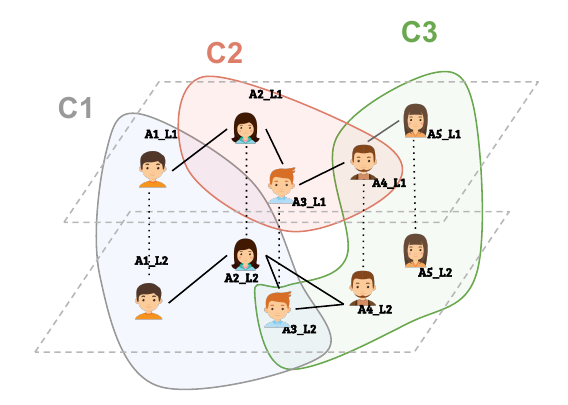} 
		\caption{Node-overlapping}
		\label{fig:node_overlapping}
	\end{subfigure}
	\begin{subfigure}[t]{0.32\textwidth}
		\centering
		\includegraphics[width=\textwidth]{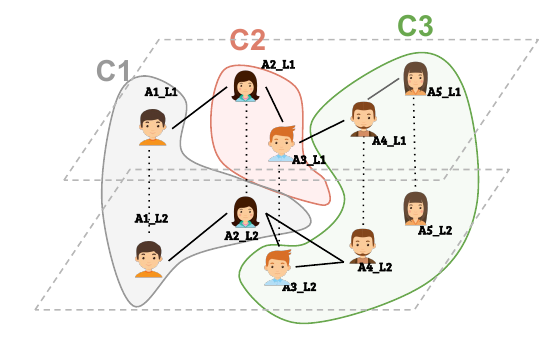} 
		\caption{Node-disjoint}
		\label{fig:node_partitioning}
	\end{subfigure}
	\begin{subfigure}[t]{0.31\textwidth}
		\centering
		\includegraphics[width=\textwidth]{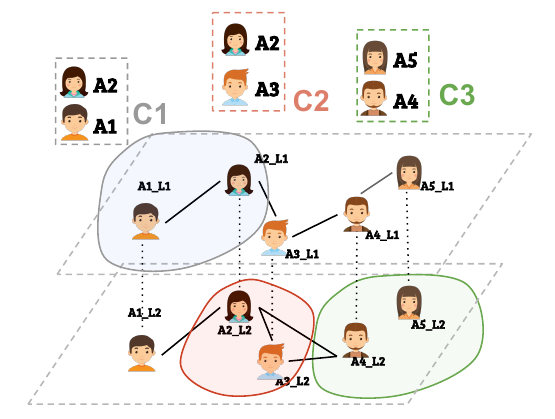} 
		\caption{Actor-overlapping}
		\label{fig:actor_overlapping}
	\end{subfigure}
	\begin{subfigure}[t]{0.31\textwidth}
		\centering
		\includegraphics[width=\textwidth]{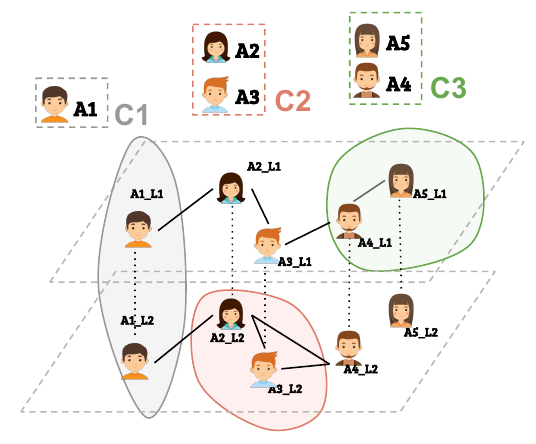} 
		\caption{Actor-disjoint}
		\label{fig:actor_partitioning}
	\end{subfigure}
	\caption{Different types of clustering on a multiplex network\rev{. In (c) the two overlapping nodes are A4\_L1 and A3\_L2. In (e) A2 is the overlapping actor}}
	\label{fig:clustering_types}
\end{figure}

\label{sec:mpx_community}
The \rev{most common} output of a community detection algorithm for multiplex networks is a set of communities  $\mathcal{C}$ = \(\{C_1, C_2, \dots , C_k\} \) such that each community contains  a non-empty subset of $V$. \rev{$\mathcal{C}$ is a representation of the  \emph{community structure} of the network.} \rev{Sometimes the term \emph{cluster} is also used} as a synonym  of community, \rev{although the term community can be interpreted more broadly to also refer to the subgraph induced by its nodes, or even more broadly to indicate the real-world concept it represents, e.g., a group of people with shared norms, values or objectives in a social network. A few community detection methods discover clusters of edges instead of clusters of nodes or actors. Keeping the above considerations in mind, the term} 
\emph{clustering} \rev{is also used to refer to the} set of \rev{all} communities. Figure~\ref{fig:clustering_types} illustrates different possible types of clusterings on a multiplex network.

\begin{figure}[!htpb]
	\centering
	\begin{subfigure}[t]{0.36\textwidth}
		\centering
		\includegraphics[width=\textwidth]{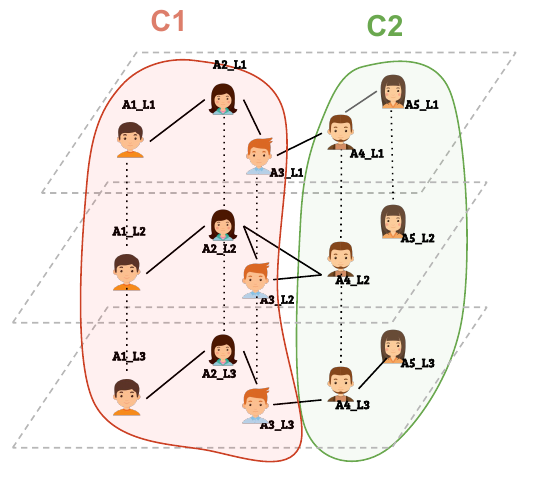} 
		\caption{Pillar communities}
		\label{fig:pillar}
	\end{subfigure}
	\begin{subfigure}[t]{0.36\textwidth}
		\centering
		\includegraphics[width=\textwidth]{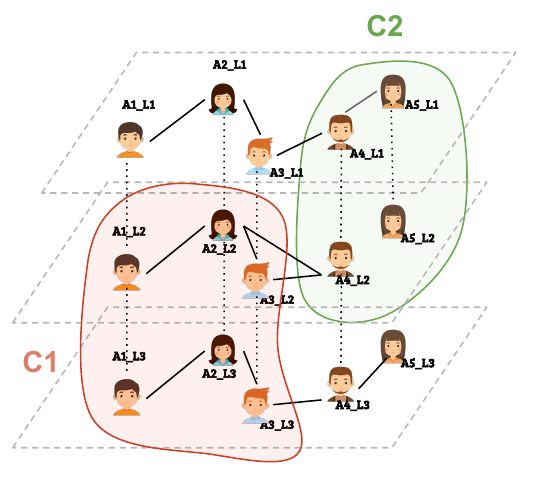} 
		\caption{Semi-pillar communities}
		\label{fig:spillar}
	\end{subfigure}
	\caption{Pillar and semi-pillar multiplex community structures}
	\label{fig:pillar_vs_spillar}
\end{figure}


A  clustering $\mathcal{C}$ is \textit{total} if every node in $V$ belongs to at least one community, and it is \textit{partial} otherwise.
We also call a clustering \textit{node-overlapping} if there is at least a node that belongs to more than one cluster, otherwise the clustering is called \textit{node-disjoint}. Analogously, if there is at least an actor belonging to more than one cluster we call the clustering \textit{actor-overlapping},  otherwise it is called \textit{actor-disjoint}. Notice that a node-overlapping clustering is also actor-overlapping, while an actor-overlapping clustering may or may not be node-overlapping. 

\rev{Finally, a} multiplex community is called \rev{\textit{semi-pillar} on layers $L' \subset L$ if for each actor $a \in A$ in the community all nodes in $\{ (a,l) \in V : l \in L' \}$ are included in the community. When $L' = L$ we talk of a \textit{pillar} community (Figure \ref{fig:pillar_vs_spillar}). Please notice that two pillar communities are either disjoint or both actor- and node-overlapping.}

\section{A taxonomy of the reviewed algorithms}
\label{sec:taxonomy}

\begin{figure}[!t]
	\includegraphics[width=\textwidth]{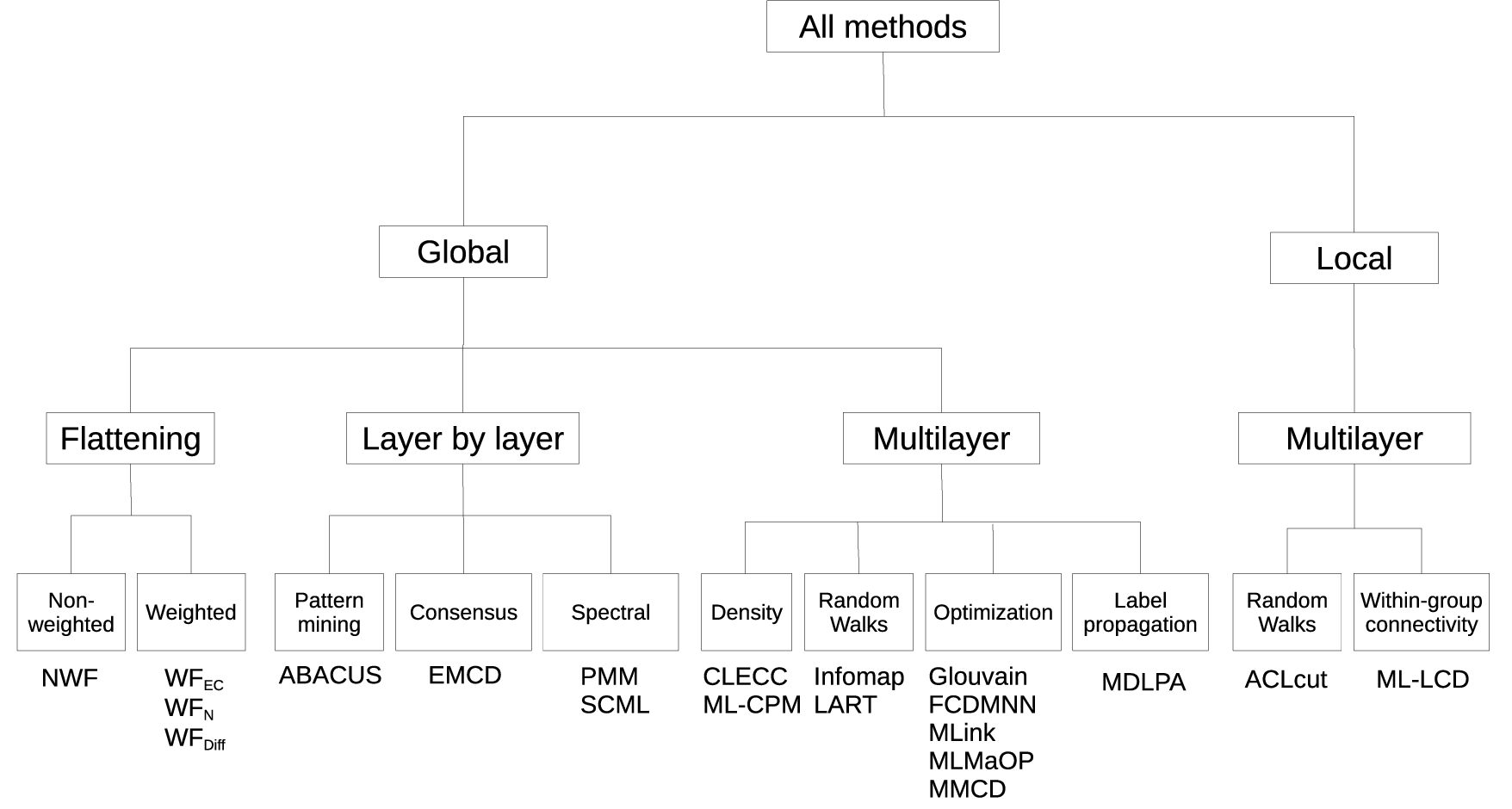}
	\caption{A taxonomy of multiplex community detection algorithms}
	\label{fig:taxonomy}

\captionof{table}{Multiplex community detection algorithms covered in this survey}
        \begin{center}
	    \begin{tabular}{ lll }
		\toprule
		Algorithm  & Notation & Reference  \\
		\toprule
		Non-Weighted Flattening & \textsf{NWF} & \citep{Berlingerio2011c}  \\
		Weighted Flattening (Edge Count) &   \textsf{WF\_EC} &\citep{Berlingerio2011c} \\
		Weighted Flattening (Neighbourhood)&   \textsf{WF\_N} & \citep{Berlingerio2011c}  \\
		Weighted Flattening (Differential) &   \textsf{WF\_Diff} & \citep{Kim2016} \\
		\hline
		Frequent pattern mining-based community discovery & \textsf{ABACUS} &  \citep{Berlingerio2013} \\
		Ensemble-based Multi-layer Community Detection &  \textsf{EMCD} &\citep{Tagarelli2017}\\
		Principal Modularity Maximization & \textsf{PMM} &   \citep{Tang2009UncoverningGV,Tang2012}\\
		Subspace  Analysis  on  Grassmann  Manifolds & \textsf{SCML} &   \citep{Dong2014}\\
		\hline
		\revm{Cross-Layer Edge Clustering Coefficient (based on)} & \textsf{\revm{CLECC}} &   \revm{\citep{Brodka2013}}\\
		Multi Layer Clique Percolation Method & \textsf{ML-CPM} &   \citep{Afsarmanesh2018}\\
		Locally Adaptive Random Transitions & \textsf{LART} &\citep{Kuncheva2015}\\
		Modular Flows on Multilayer Networks & \textsf{Infomap} & \citep{DeDomenico2015_infomap,Edler2017}   \\
		Generalized Louvain & \textsf{GLouvain} &  \citep{Mucha2010science,Jutla}\\
		Fast algorithm for comm.~detection based on multiplex net.~modularity & \textsf{FCDMNN} &  \citep{Zhai2018}\\
		Multilink community detection & \textsf{MLink} &  \citep{Mondragon2018}\\
		\revm{Multi-Layer Many-objective OPtimization algorithm} & \textsf{\revm{MLMaOP}} &  \revm{\citep{Pizzuti2017}}\\
		Multilevel memetic algorithm for composite community detection & \textsf{MNCD} &  \citep{Ma2018}\\
		Multi Dimen\rev{s}ional Label Propagation & \textsf{\revm{MDLPA}} & \citep{Boutemine2017}\\
		\hline
		Andersen-Chung-Lang cut & \textsf{ACLcut} &\citep{JEUB2017} \\
		Multilayer local community detection &\textsf{ML-LCD}&\citep{Interdonato2017}\\
        \bottomrule
    \end{tabular}
    \end{center}
    \label{table:algorithms}
\end{figure}

In this section we provide a taxonomy of multiplex community detection methods \rev{with three levels of classification. The top-level distinction is between \emph{global} or \emph{local} methods, respectively discovering all communities in the input network or generating a single community around one or more seed nodes. The results of these two types of algorithms are not directly comparable without arbitrary choices in the selection of seed nodes, so we treat them in separate sections in our experimental evaluation. The second level regards the way in which the algorithms handle the presence of multiple layers: reducing them to a single layer (flattening), processing each layer independently  \revm{(e.g., performing single-layer community detection)} to then merge the results \revm{of the processing}, or considering all the layers at the same time. The last level of the taxonomy groups the algorithms based on more specific approaches, such as optimizing an objective function, considering the behavior of a random walker or identifying dense subgraphs.} Figure \ref{fig:taxonomy} and Table \ref{table:algorithms} show an overview of the related methods. \rev{Please notice that Section~\ref{sec:theory}, describing some theoretical properties of the algorithms such as whether they are deterministic or not, can also be used to differentiate between different types of algorithms.}

\subsection{\rev{Global methods}}


Global methods are designed to discover all possible communities in a network, thus requiring knowledge \revm{of} the whole network structure. \rev{As it happens for many multiplex data analysis methods \cite{DickisonMagnaniRossi2016}, global community detection algorithms can also be grouped into three typical main classes, described in the following.}



\subsubsection{\rev{Flattening}}

\begin{figure}
	\includegraphics[width=\textwidth]{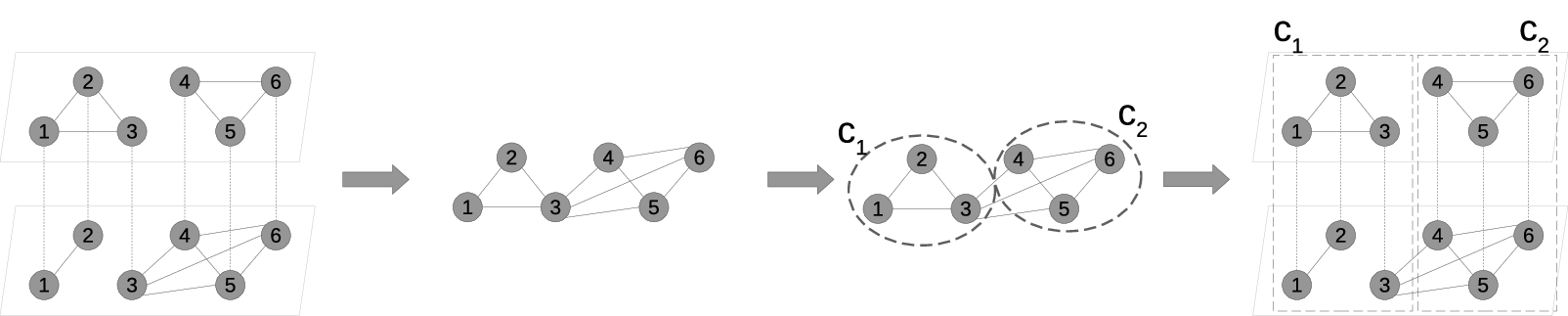}
	\caption{\rev{The general process used by flattening methods: a single-layer network is first constructed merging edges from the different layers, then a traditional community detection algorithm is applied to the flattened network, and its result can be used to induce communities on the original network}}
	\label{fig:com-flat}
\end{figure}

The first approach consists \rev{in} simplifying the multiplex network into a graph by merging its layers, using a so-called \textit{flattening} algorithm, then applying a traditional community detection algorithm. \rev{This process is illustrated in Figure~\ref{fig:com-flat}.}

 \rev{The algorithms belonging to this class are defined by the flattening method and by the single-layer community detection algorithm applied to the flattened network. The simplest flattening method consists in creating an unweighted graph where two nodes are adjacent if their corresponding actors are adjacent on any of the input layers \rev{\citep{Berlingerio2011c}}. The advantage of this approach is that the resulting graph is easier to handle, because there are more clustering algorithms for simple graphs than for weighted graphs and weights often imply an additional level of complexity, e.g., deciding a threshold above which weighted edges should be considered. A potential disadvantage is that an unweighted flattening is more susceptible to noise.}

\rev{Weighted flattenings reflect} some structural properties of the original multiplex network in the form of weights assigned to the output edges \rev{\citep{Berlingerio2011c,Kim2016}}. \rev{In theory these methods are less susceptible to noise, but the resulting communities may be biased towards edges appearing on several layers, and the results can be more difficult to interpret because of the weights.}

\rev{In general, the algorithms in this class are only able to identify pillar communities, and some communities may emerge because of edges spread on different layers that would not constitute a community on any individual layer, because of the flattening process.}

\subsubsection{\rev{Layer by layer}}

\begin{figure}
	\includegraphics[width=\textwidth]{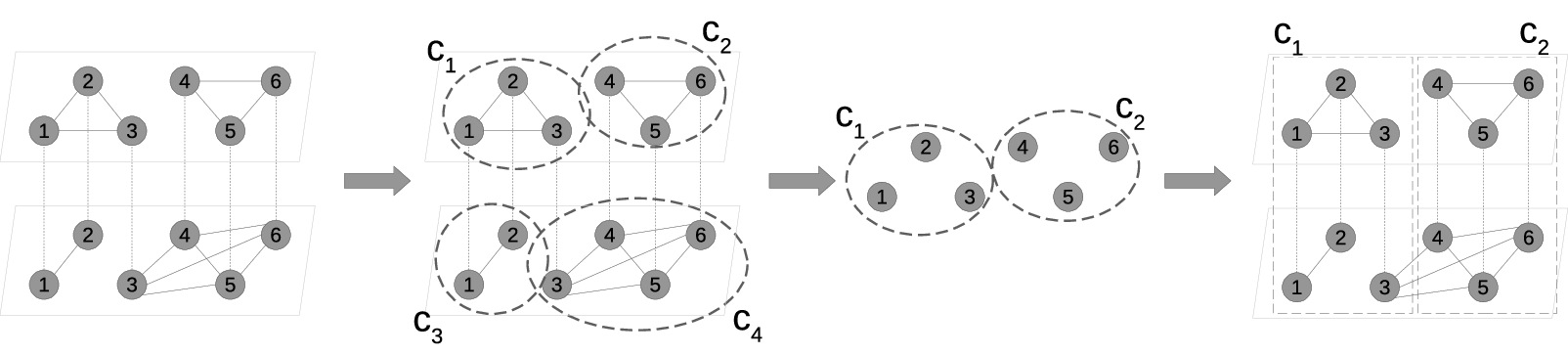}
	\caption{\rev{The general process used by layer-by-layer methods: communities are identified in each layer, the information obtained from each layer is used to cluster the actors, and this clustering can be used to induce communities on the original network}}
	\label{fig:com-lbl}
\end{figure}

\rev{While the methods in the previous class merge the layers and then apply traditional community detection algorithms, layer-by-layer methods first \revm{process each layer (e.g., applying traditional community detection algorithms)}, then merge the results \revm{of the processing}. This is illustrated in Figure~\ref{fig:com-lbl}.}

\rev{As a consequence of the layer-by-layer community detection step, these methods include actors in the same community only when they are part of the same community in at least one layer. Also, due to the merging of layer-specific communities, these methods can in principle only identify pillar communities.  
}

\rev{We have identified three types of layer-by-layer approaches in the literature. The \emph{pattern mining} approach exploits association rule mining methods, which are among the main data mining tasks used to find objects that frequently co-occur together in different transactions. (A typical example of transaction is the basket of products bought together by a customer at a supermarket.) ABACUS considers each single-layer community as a transaction, so that the final communities contain actors that are part of the same community in at least a minimum number of layers \rev{\citep{Berlingerio2013}}.}

\rev{The second way to merge the result of single-layer community detection methods is based on a notion of \textit{consensus}: given a set (or ensemble) of community structure solutions from the individual layers, the goal  is to find a single, meaningful solution that is representative of the input ensemble, by optimizing an  objective function  that is designed to aggregate information from the individual solutions in the ensemble. While early approaches such as the one in~\citep{ConClus2012} are limited to use a clustering ensemble method as a black-box tool for combining multiple  clustering  solutions from a single-layer network,   
the first well-principled formulation of the ensemble-based multilayer community detection (EMCD) problem, provided in~\citep{Tagarelli2017}, does not limit aggregation at node membership level, but rather it accounts   for intra-community and inter-community connectivity. 
 The consensus solution discovered by EMCD is the one with maximum multilayer modularity from a search space of candidates delimited by topological upper-bound and lower-bound solutions, respectively, of the input multilayer network. 
}

\rev{Finally, some methods in the literature process the layer-specific adjacency matrices, or derived matrices, 
 and extend spectral-clustering for simple graphs by  exploiting the relationship between the eigenvectors and eigenvalues in the constructed matrices and the presence of clusters in the corresponding graphs.   
As an example, the principal modularity maximization (PMM) method~\cite{Tang2009UncoverningGV} extracts structural features  from the various layers, then concatenates the  features and performs PCA to select the top eigenvectors. Using these eigenvectors, a low-dimensional embedding is computed to capture the principal patterns across the layers, finally a simple $k$-means is applied to assign nodes to communities.   
 Further details on this class of approaches can be found in~\citep{TangLiu2010}. }

\subsubsection{\rev{Multilayer}}

\begin{figure}
	\centering
	\includegraphics[width=.5\textwidth]{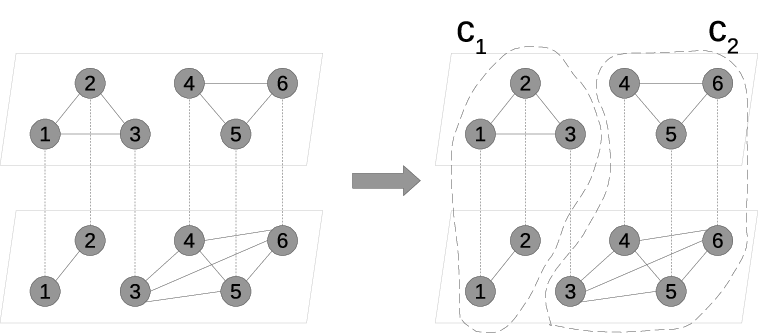}
	\caption{\rev{\revm{M}ultilayer methods \revm{discover} communities directly on the multiplex data}}
	\label{fig:com-ml}
\end{figure}

\rev{The third class of algorithms operates directly on the multiplex network model, as shown in Figure~\ref{fig:com-ml}. As an example, a method belonging to this class based on a random walker would allow the walker to switch from one layer to the other.}

\rev{Various approaches originally developed for simple graphs have been extended to the multilayer case. \emph{Density-based methods} first identify dense regions of the network, then include adjacent regions in the same community. A popular method for simple graphs is clique percolation, where dense regions correspond to cliques and adjacency consists in having common nodes. The multilayer clique percolation method \revm{(ML-CPM)} extends this process by looking for cliques spanning multiple layers, and redefining adjacency so that both common nodes and common layers are required \rev{\citep{Afsarmanesh2018}}. \revm{CLECC uses a different but related approach, identifying sparse locations of the network having a low cross-layer clustering coefficient \citep{Brodka2013}. The higher the proportion of common neighbors across all layers (or any number of layers provided as input), the higher the cross-layer clustering coefficient.}

\rev{Methods based on \emph{random walks} consider that an entity randomly following the edges in a network would tend to get trapped inside communities, because of the higher edge density between nodes inside the same community, less frequently moving from one community to the other. LART \rev{\citep{Kuncheva2015}} and Infomap \rev{\citep{DeDomenico2015_infomap}} are both based on this consideration, with Infomap using a shortest information coding approach to identify the corresponding communities.}

\rev{Several of the reviewed algorithms in the multilayer class use an objective function that, given an assignment of the nodes to communities, returns a higher value when there are more edges inside communities and less edges across communities. Once the objective function has been defined, then different optimization methods can be used to identify a community assignment corresponding to a high value of the function. Generalized Louvain} \revm{(GLouvain)} \rev{\citep{Mucha2010science,Jutla}}, the best-known method in this class, uses an extended version of modularity, and has been analyzed in more detail in \cite{hanteer_unspoken_2020}.} \revm{While GLouvain has become the most popular modularity-based method for multiplex networks, it is worth mentioning the alternative approach used by \cite{Pizzuti2017}, aimed at obtaining a high modularity in each individual layer instead of a global extended definition of modularity.} \rev{This class also includes a method returning a different type of communities with respect to the ones generated by the other algorithms, where edges are grouped instead of actors and nodes \cite{Mondragon2018}.}

\rev{Finally, the multilayer class includes an algorithm based on \emph{label propagation} \rev{\citep{Boutemine2017}}. A traditional label propagation method would start assigning a different label to each node, then having each node replace its label with one that is frequent among its neighbors, until some stopping condition is satisfied. The multilayer version of this approach follows the same idea, weighting the contribution of each neighbor based on their similarity with the node on the different layers. For example, two nodes being adjacent on all layers and having the same neighbors on all layers would have a higher probability of getting the same label.}

\subsection{\rev{Local methods}}

\rev{L}ocal methods (also known as \textit{node-centric}) are \textit{query-dependent}, i.e., they are designed to discover the community \rev{around} a set of \rev{input} query nodes. \rev{Please notice that the term \emph{local} has also been used with other meanings in the literature, for methods finding global community structures using only neighborhood information when processing vertices in the graph.}
\rev{At the time of writing, we recognize the availability of two methods able to discover multiplex local communities: \textsf{ML-LCD}~\citep{Interdonato2017} and \textsf{ACLcut}~\citep{JEUB2017}.
\textsf{ML-LCD} searches for the local community associated to a seed actor without having a complete knowledge of the network graph, through an incremental exploration of the neighborhood of the query actor, according to   the optimization of a criterion function based on  the internal and external connectivity of the local community. 
\textsf{ACLcut} exploits the solution of a personalized PageRank approximated for an input seed-set (i.e., a set of query actors) in order to find the local communities, using a sweep cut method 
to sample local communities based on the lowest conductance values.
Both methods operate directly on the multiplex network model, so that the \textit{Local} branch of our hierarchy only includes the \textit{Multilayer} class. Nevertheless (even if, to the best of our knowledge, there are no such examples in literature) it is in theory possible to easily design multiplex local community detection methods that operate through flattening or layer-by-layer schemes, by exploiting existing single-layer local community detection methods such as \textsf{LCD}~\citep{ChenZG09} and \textsf{Lemon}~\citep{LiHBH15}.}

\subsection{\rev{Selection of algorithms}}


\rev{In the following sections we will provide a detailed comparative analysis of a large subset of the algorithms in our taxonomy. We include at least one representative method for each leaf in the taxonomy. In those cases where different well-known methods inside the same leaf show significant differences, either theoretically or experimentally, we have also included them, as detailed in the following.}

\rev{We only focus on a selection of the flattening methods, with one representative for each class (unweighted and weighted), because of the small variation between the different approaches and because the features and performance of these algorithms are determined more by the single-layer approach used to implement them than by the way in which weights are assigned. While the main interest of this article is on multilayer-specific methods, we still considered it important to test some flattening methods in detail, because as we will see in our comparative analysis these simpler approaches can still produce good and sometimes better results than more sophisticated methods.}

\rev{We include all the methods from the layer-by-layer class (ABACUS, EMCD, PMM and SCML), because they are representative of  different ways to merge the results of the single-layer algorithms. PMM has been first published in conference proceedings \citep{Tang2009UncoverningGV} and then abstracted and extended in a journal article \citep{Tang2012}. We use the conference version, because the code for the journal version is not available.}

\rev{From the multilayer class we include \revm{at least one representative method for each sub-class. Among the modularity-opimization methods we have selected} GLouvain \revm{because it} is the \revm{best-known} optimization algorithm \revm{as witnessed by its large number of citations}. MLink \revm{has only been} included in the scalability analysis \revm{because it} produces link communities that are not directly comparable with the ones produced by other methods.}

\rev{We also include all the local methods (ACLcut and ML-LCD), because they use significantly different approaches.}

\section{\rev{Theoretical analysis}}\label{sec:theory}

\rev{In this section we present some theoretical properties of the reviewed algorithms. We describe the types of community structures that can be returned by each algorithm, we indicate some features of the algorithms themselves such as whether they are deterministic, and we discuss parameter setting and computational complexity.}

\rev{These properties should be considered in combination with the results of our experimental evaluation. For example, the fact that \emph{in theory} an algorithm is able to produce some types of multiplex communities does not imply that these types of communities will be found in practice. 
Nonetheless, knowing that some algorithms are not able to return some types of communities or that their execution time grows exponentially with respect to the number of layers can be useful to choose which algorithms to use in specific situations.}

\subsection{\rev{Types of community structures}}

In Section~\ref{sec:prel} we have described different properties of multiplex community structures.
\rev{Table~\ref{table:com_type} indicates which ones are associated to each reviewed algorithm. In particular,
\begin{itemize}
    \item[\textbf{(NPC)}] if the algorithm can generate \revm{N}on-\revm{P}illar \revm{C}ommunities;
    \item[\textbf{(AO)}] if it can generate \revm{A}ctor-\revm{O}verlapping community structures;
    \item[\textbf{(NO)}] if it can generate \revm{N}ode-\revm{O}verlapping community structures;
    \item[\textbf{(Pa)}] if it can generate \revm{Pa}rtial community structures.
\end{itemize}
}
\rev{An algorithm not satisfying these properties (i.e., those with an `$\times$' in the table) would, respectively, only be able to produce pillars, only partition the actors and nodes, and force all nodes to belong to at least one community. Notice that this can be perfectly fine in some cases, so satisfying or not the properties above does not mean that the algorithm is worse or better. These properties should only be used as an indication about the appropriateness of the algorithm for specific scenarios.}



\begin{table}[!htpb]
	\caption{Types of clustering \rev{produced} by the reviewed methods \rev{and algorithmic properties. The second column recalls the class of the algorithm (G-Flat: global flattening, G-LBL: global layer by layer, G-ML: global multilayer, L-ML: local multilayer). Columns NP\revm{C} (Non-Pillar), AO (Actor-Overlapping), NO (Node-Overlapping) and Pa (Partial) indicate if the algorithm can (\checkmark) or cannot ($\times$) produce that type of community structure. Columns LR (Layer Relevance), Det (Deterministic), AK (Automated selection of the number of communities) and SS (Subgraph Structure) refer to the functioning of the algorithm.} (*) indicates that the answer  depends on the single-layer clustering algorithm used by the method. \rev{(-) indicates that the property is not relevant for the algorithm.} \revm{The last column indicates the time complexity of the method if studied in the original paper or easily derivable from the algorithm: $a$: number of actors, $e$: number of edges, $l$: number of layers, $c$: number of communities, ARM: cost to compute closed association rules, $|C|$: size of local community, $d$: maximum node degree, $\delta$: complexity of the single-layer community detection algorithm used as a sub-procedure, $\Phi$: parameter depending on the used subprocedure (see Section~\ref{sec:comp.comp}).}}
     \begin{minipage}{\columnwidth}
        \begin{center}
	    \begin{tabular}{ll | llll | llll | l}
		\toprule
		Algorithm & \rev{Category} & NP\revm{C} & AO & NO & Pa & LR & Det & AK & SS & \revm{Compl} \\
		\toprule 
		NWF & \rev{G-Flat} & \rev{$\times$} & * & * & * & $\times$ & * & * & * & \revm{$\mathcal{O}(e + \delta)$} \\
		WF$_{EC}$ & \rev{G-Flat} & \rev{$\times$} & * & * & * & $\times$ & * & * & * & \revm{$\mathcal{O}(e + \delta)$} \\
		\hline
		ABACUS & \rev{G-LBL} & \rev{$\times$} & \rev{\checkmark} & \rev{\checkmark} & \rev{\checkmark} & $\times$ & * & \checkmark & \checkmark & \revm{$\mathcal{O}(l\delta)$ + ARM}  \\
		EMCD & \rev{G-LBL} & \rev{$\times$} & \rev{$\times$} & \rev{$\times$} & \rev{$\times$} & \checkmark & * & \checkmark & \checkmark & \revm{$\mathcal{O}(l\delta) + \mathcal{O}(i(e+lc))$} \\
		PMM & \rev{G-LBL} & \rev{$\times$} & \rev{$\times$} & \rev{$\times$} & \rev{$\times$} & $\times$ & $\times$ & \rev{$\times$} & \rev{$\times$} & -- \\
		SCML & \rev{G-LBL} & \rev{$\times$} & \rev{$\times$} & \rev{$\times$} & \rev{$\times$} & $\times$ & \rev{$\times$} & $\times$ & \rev{$\times$} & -- \\
		\hline
		ML-CPM & \rev{G-ML} & \rev{\checkmark} & \rev{\checkmark} & \rev{\checkmark} & \rev{\checkmark} &$\times$ & \checkmark & \checkmark &  \checkmark & \revm{$\geq \mathcal{O}(l3^\frac{a}{3})$} \\
		Infomap & \rev{G-ML} & \rev{\checkmark} & \rev{\checkmark} & \rev{\checkmark} & \rev{\checkmark} & \rev{\checkmark}  & \rev{$\times$} & \checkmark & \checkmark  & -- \\
		LART & \rev{G-ML} & \rev{\checkmark} & \rev{\checkmark} & \rev{$\times$} & \rev{$\times$} & \rev{\checkmark} & \rev{$\times$} & \checkmark & \checkmark & -- \\
		GLouvain & \rev{G-ML} & \rev{\checkmark} & \rev{\checkmark} & \rev{$\times$} & \rev{$\times$} & \checkmark & $\times$ & \checkmark & \checkmark & -- \\
		MDLPA & \rev{G-ML} & \rev{\checkmark} & \rev{\checkmark} & \rev{$\times$} & \rev{$\times$} & \rev{\checkmark} & $\times$ & \checkmark & \checkmark & \revm{$\mathcal{O}(ea + e2^l + ei)$} \\
		\hline 
		ML-LCD & \rev{L-ML} & \rev{$\times$} & \rev{-} & \rev{-} & \rev{-} & \rev{\checkmark} & \rev{\checkmark} & -  & \rev{\checkmark} & \revm{$\mathcal{O}(|C|^2 d \Phi)$} \\
		ACLcut & \rev{L-ML} &  \rev{\checkmark} & \rev{-} & \rev{-} & \rev{-}  & \rev{$\times$} & \rev{$\times$} & - &  \rev{\checkmark} & --  \\
        \bottomrule 
    \end{tabular}
    \end{center}
\end{minipage}
	\label{table:com_type}
\end{table}

\revm{The following are some considerations summarized in Table~\ref{table:com_type}:}
\rev{\begin{itemize}
    \item For all flattening methods, the type of the resulting community structure (Overlapping/Disjoint and Total/Partial) depends on the single-layer algorithm used after flattening. The choice of the single-layer algorithm can then be made depending on the wanted result.
    \item All flattening methods produce pillar communities, because the actors on different layers are reduced to a single node in the flattened graph.
    \item All multilayer methods can produce non-pillar communities in theory, although our experimental evaluation shows that pillar communities are often returned by some of these methods.
    \item Pillar actor-overlapping communities are always node-overlapping, by definition.
    \item Non-pillar actor-overlapping communities may be or not node-overlapping.
\end{itemize}}


\subsection{\rev{Algorithmic properties}}

In their survey work, \citep{Kim2015} discussed a  classification framework  based on a set of desired properties for multilayer community detection methods. These properties are: multiple layer applicability,  consideration of each layer's importance, flexible layer participation (i.e., every community can have a different coverage of the layers' structure), no-layer-locality assumption (e.g., independence from initialization steps biased by a particular layer),  independence from the order of layers, algorithm insensitivity,  and overlapping layers (e.g., two or more communities can share substructures over different layers).


We observe that the first of the properties \rev{listed above} (multiple layer applicability) \rev{is satisfied by} all methods we reviewed, therefore we do not elaborate on this further. By contrast, the second property  (consideration of each layer's importance) \rev{is also included in our list and further elaborated, as detailed below (Layer Relevance)}. \rev{We collapse the} properties about independence from \rev{the order in which} node\rev{s and} layer\rev{s are examined into a single property, also including stochastic behaviors such as in the case of random walkers (Determinism). As we focus on multiplex networks, we do not treat the case where layers are ordered}. The insensitivity property (i.e., independence or robustness against main tunable input parameters) is instead \rev{replaced by a more specific property on} whether the number of communities is automatically derived (Auto-detection), and \rev{a more general discussion about how to set additional parameters.} \rev{The last property we consider (Subgraph Structure) was not discussed in previous surveys.}

In light of the above considerations we define \rev{the following} properties\rev{, indicated in Table~\ref{table:com_type}}.

\begin{itemize}
	\item[\rev{(\textbf{LR})}] \textbf{Layer relevance.} Some methods take into consideration each layer's importance\rev{, also called relevance in some of the reviewed works,}
	in order to control their contribution to the computation of the multiplex community structure. 
	\rev{Layer relevance is either} learned based on the layer characteristics, or \rev{it can be an input of the algorithm} based on a-priori knowledge (e.g., user preferences). 
	\item[\rev{(\textbf{Det})}] \textbf{Determinism.} This refers to whether a method has a deterministic behavior, \rev{e.g.,} its output is independent from the order of examination of the nodes and/or layers.
	\item[\rev{(\textbf{AK})}] \textbf{Auto-detection of the number of communities.}
	Some methods expect the number of communities to be decided ahead of time while other methods can automatically define the number of communities.
	\item[\rev{(\textbf{SS})}]  \textbf{Subgraph structure.}  
\rev{The primary product of all the reviewed methods are the cluster memberships of nodes. However, some methods also tell us something about the multilayer subgraph structures underlying each community, that is, we can get more information about which edges contributed to the discovery of each community.}
\end{itemize}

\rev{
Different algorithms tune layer relevance (LR) in different ways. The only algorithm allowing to specify weights as input parameters is GLouvain, through the parameter omega \revm{(}$\omega$\revm{)} that gives more or less importance to the fact that the same actor is included in the same community in different layers. However, these weights are assigned to pairs of actors in different layers, not to individual layers, and in practice $\omega$ is set to a single value for the whole network.
In EMCD, the importance of the various layers may be considered by differently setting the resolution parameter in the multilayer modularity. Both LART and \revm{MDLPA} use a concept of layer relevance (that is, how important a layer is for a node or a pair of nodes) to weight the probability of the random walker to switch layer or of a label to be propagated.
ML-LCD is designed to explicitly incorporate layer relevance  weighting schemes in the local community functions. 
}

\rev{Non-determinism is the result of different features in different algorithms: using heuristics to optimize an objective function (such as GLouvain), using non-deterministic clustering algorithms as sub-procedures (as PMM and SCML), using stochastic choices (as LART) or} \revm{the  iterative computations performed by MDLPA and ACLcut, depending on the order in which nodes are processed.}

\revm{The automated selection of the number of communities is a practically important property especially for networks. Traditional clustering algorithms requiring the number of clusters as input, such as k-means, can be run multiple times to optimize $k$ using some measures of clustering quality, but this procedure has not been explored for the algorithms studied in this survey.}

\rev{With regard to the last property, all the methods returning non-pillar communities provide information about which layers define each community. For example, in ML-CPM communities are combinations of adjacent cliques, so all the edges in these cliques can be considered part of the community. As another example, \revm{MDLPA} computes a score for each pair of nodes indicating how likely a label should be propagated from one to the other, leading to a common community. However, also methods not returning information about layers as their primary output could be used to indicate which layers and edges determine each community. EMCD only accounts for those edges from different layers that contribute to maximize the multilayer modularity of the consensus community structure solution. In ABACUS, even if the output of the algorithm is about actors, for each pair of actors included in the same community we could look at which layers determined that assignment.}




\subsection{\rev{Parameter setting}}

\rev{Apart from the number of communities to discover, which is required by some algorithms as input, the reviewed methods have a variety of additional input parameters to set. While explaining the meaning of each parameter goes beyond the aims of this survey, it \revm{is} useful to characterize the methods with respect to how difficult and/or important it is to properly set their parameters.}

\rev{Some methods can be executed parameter-free. This is the case for all flattening methods, except if their single-layer clustering algorithm needs some, and for \revm{MDLPA} and Infomap, although Infomap provides additional options that the interested reader can check on the information-rich website provided by the authors.\footnote{https://www.mapequation.org}}

\rev{ABACUS and ML-CPM require to specify minimum values for the number of layers and actors to be included in a community, which makes them able to identify partial community structures. These parameters affect the result by making it more and more difficult to accept some groups of nodes as a community, and while setting the correct values may require multiple trials, in our opinion the meaning of these parameters is easy to grasp.}

\rev{EMCD requires to specify the co-association threshold,  $\theta$, that may have a  strong impact on the resulting consensus  communities. The original paper presenting this algorithm indicates optimal ranges of values on some networks and suggests that similar values can be used for similar networks.}

\rev{PMM requires to specify the number of structural features, which can be any number between 1 and $\#a-2$. Also in this case different settings can lead to quite different results, and this parameter has a less intuitive meaning if compared with those required by other methods. Similarly, SCML requires a regularization parameter lambda. In addition, both methods require to specify the number of expected communities, as mentioned in the previous section, and the number of times the k-means algorithm used as a sub-procedure should be repeated. In general, different executions of k-means can lead to different results.}

\rev{GLouvain requires only two parameters: $\omega$, weighting inter-layer contributions, and $\gamma$, the so-called resolution parameter. Regarding $\gamma$, we refer the reader to the literature about its usage and shortcomings in the single-layer version of modularity. $\omega$, which in theory can be set individually for each \revm{actor and} pair of \revm{layers} but is more practically set to a single value, has an apparently intuitive meaning: a low value would give priority to intra-layer communities, a higher value would tend to discover communities spanning multiple layers. We refer the reader to \cite{hanteer_unspoken_2020} for a deeper discussion about what can and cannot be identified with different settings of $\omega$.}

\rev{LART requires four parameters: $t$, $\epsilon$, $\gamma$, and \textit{linkage}. While the interpretation of some of these parameters is intuitive, in particular the type of hierarchical clustering to be performed inside the algorithm (\textit{linkage}) and the number of steps to be taken by the random walker ($t$), it is in general difficult to predict what impact each setting would have on the final result, which makes these parameters more difficult to be set if compared with other methods.}

\rev{Regarding the local methods, they naturally take the set of query nodes as an input parameter. \textsf{ML-LCD} has no additional parameters, except for the ones controlling layer weights in the   \textsf{ML-LCD$_{(lwsim)}$} formulation. However, in absence of exogenous information about the importance of each layer, uniform weights can be used without loss of generality. Concerning \textsf{ACLcut}, the main parameters are the ones controlling the random walk generating the input transition tensor. Two alternative models can be used, which differ in how they navigate the multiplex network: a classic random walk, controlled by an uniform interlayer edge weight $\omega$, and a relaxed random walk, controlled by a layer-jumping probability $r$. These parameters are shown to have a major impact in the characteristics of resulting local communities, thus it is not clear how to set them in general cases. 
\textsf{ACLcut} also includes an underlying \textsf{APPR} (Approximated Personalized PageRank) procedure, whose resolution is controlled by two additional parameters: the teleportation parameter
$\gamma$ and the truncation parameter $\epsilon$.
A default value of $0.95$ can be used for $\gamma$, while arbitrary small values can be used for $\epsilon$ (e.g., inversely proportional to the number of nodes in the network).}

\subsection{\rev{Some notes on computational complexity}} \label{sec:comp.comp}

\rev{In most cases, a detailed study of the computational complexity of community detection algorithms is not provided in the original references. This can be explained by the fact that many well-known algorithms have not been developed by computer scientists nor published in computer science venues. However, we also notice that worst-case complexity would often be not particularly informative: execution time typically strongly depends on data and parameter setting, making an experimental analysis more useful in characterizing the methods. 
At the same time, some considerations can be useful to either predict or understand the behaviour of some algorithms in specific situations.}

\rev{For flattening methods, time complexity depends on the flattening step and on the subsequent single-layer community detection step. Basic types of flattening are in $O(e)$, in which case the complexity of the algorithm corresponds to the one of the community detection step. \revm{It is interesting to notice that higher layer similarity for example in terms of edge Jaccard \citep{Brodka2018} would lead to a lower number of edges, possibly resulting in a lower execution time of the single-layer community detection algorithm.}}

\rev{As for layer-by-layer methods, the complexity also depends on the community detection algorithm applied to each layer, but the step where the communities from the different layers are merged can be significantly more expensive than a flattening. ABACUS uses association rule mining, which can in theory generate an exponential number of rules. The actual execution time is however dependent on the input thresholds: the minimum number of layers where actors must be assigned to the same community to be included in the final result (corresponding to the support count measure in association rule mining) and the minimum number of actors in a community to be counted (limiting the transaction size in the association rule mining algorithm). 
\rev{EMCD  linearly scales  with the number of multilayer edges and with the number of consensus communities.} 
\revm{While the paper introducing PMM does not provide a complexity analysis, the algorithm requires two expensive steps: the extraction of $f$ eigenvalues from each layer and a} singular value decomposition on data of size $a \times fl$; therefore, its complexity depends on the number of actors, the number of layers (that is, the data), and on the number of features (which is an input parameter).}

\rev{ML-CPM requires the computation of maximal cliques, that is NP-Hard even on a single layer. This implies that dense regions of the input networks across $m$ or more layers consisting of a few tens of nodes may lead to impractically slow computations. Maximal clique detection can however be very fast in practice for sparser networks with small communities. GLouvain uses a heuristic to optimize an extended modularity objective function, as modularity optimization is already NP-Hard on single networks. In general, label propagation algorithms have a complexity of $O(e i)$, where $i$ is the number of iterations which is often small. However, \revm{MDLPA} also contains a subroutine iterating over all subsets of the layers, to compute pairwise weights to be used when labels are propagated. This makes its complexity exponential in the number of layers $l$.}

\rev{Computational complexity of \textsf{ML-LCD} is proportional to the size of the generated  community, thus the overall upper bound is  $\mathcal{O}(|C|^2 \times d \times \Phi)$, where $|C|$ is the size of the local community, $d$ is the maximum degree of a node in the network and $\Phi$ is the cost of optimizing the $LC$ function. Possible values of $\Phi$ depend on the three alternative formulations and are 
$\mathcal{O}(l d)$ for \textsf{ML-LCD$_{(lwsim)}$}, $\mathcal{O}(l d^2 \log d)$ for \textsf{ML-LCD$_{(wlsim)}$} and $\mathcal{O}(|C| d^2 \log d l^2)$ for \textsf{ML-LCD$_{(clsim)}$}. \revm{The c}omplexity of \textsf{ACLcut} has not been studied in the original paper.}

\section{Experimental evaluation}
\label{sec:experiments}

We devised an experimental evaluation to pursue two main goals in comparing the various methods: one relating to the  quality of the produced communities, the other to efficiency aspects. More specifically, our experiments were carried out to  answer the following research questions: 
\begin{itemize}
	\item[\rev{\textbf{Q1}}] To what extent 
	are the evaluated methods able to detect ground truth communities?
	\item[\rev{\textbf{Q2}}] To what extent do the evaluated methods produce similar community structures? 
	\item[\textbf{Q3}]  To what extent are the evaluated methods scalable? 
\end{itemize}


Two main stages of evaluation were devised: one for global methods (Sect.~\ref{ssec:results_global}),  whose  output is a set of communities, and one for local methods (Sect.~\ref{ssec:results_local}), whose  output is a single community centered around a node (or set of nodes). Due to their structural differences, these two tracks had to be evaluated separately and by means of different criteria. \revm{The reason why we have not tested the algorithms on single-layer networks is that multilayer methods are generalizations of single-layer algorithms, so their results would be exactly the same as those already reported in single-layer studies.} 


\subsection{\rev{Data}} 

\rev{To evaluate the communities discovered by the tested methods, we use} a selection of real datasets widely used in the literature, \rev{representing different application areas and with different characteristics: AUCS (short for Aarhus University Computer Science) \cite{Rossi2015}, a hybrid online/offline network with five types of relationships between employees of a university department; DKPol (short for Dansk Politik) \cite{DBLP:conf/asunam/Hanteer0DM18}, a network with three types of online relations between Danish Members of the Parliament on Twitter, Airports (short for Air Transportation Multiplex) \cite{Cardillo2013}, with flight connections between European airports, and Rattus \cite{DeDomenico2015}, about genetic interactions. AUCS and DKPol also come with some possible community structures, referred to as ground truth in the following: respectively, the research groups at the department, and affiliation to political parties.} \rev{The ground truth for AUCS (research groups) and \rev{DKPol} (parties) is approximately pillar partitioning, as indicated in Table~\ref{table:gt_stats}.}

\begin{table}
	\centering
	\caption{Statistics about the community structures in our networks with ground truth. We denote with \textbf{\#c} the number of communities, with \textbf{sc1} the size of the largest community \rev{(number of nodes)}, with \textbf{sc2/sc1} the ratio between the size of the \revm{second} largest community  \revm{and} the largest, with \textbf{\%n} the percentage of nodes assigned to at least one community, with \textbf{\%p} the percentage of pillars, with \textbf{\%ao} the percentage of actors in more than one community, with \%no the percentage of nodes in more than one community and with \textbf{\%s} the percentage of singleton communities}
	\begin{tabular}{l 
	    rrrrrrrr}
		\toprule
		method & {$\#c$} & {$sc1$} & {$sc2/sc1$} & {$\%n$} & {$\%p$} & {$\%ao$} & {$\%no$} & {$\%s$} \\
		\toprule
        AUCS & 8.00 & 70.00 & 0.57 & 0.90 & 0.90 & 0.00 & 0.00 & 0.12 \\ 
DKPol & 10.00 & 210.00 & 0.87 & 1.00 & 1.00 & 0.00 & 0.00 & 0.00 \\ 
PEP & 10.00 & 30.00 & 1.00 & 1.00 & 1.00 & 0.00 & 0.00 & 0.00 \\
PNP & 10.00 & 90.00 & 0.66 & 1.00 & 1.00 & 0.00 & 0.00 & 0.00 \\
PEO & 10.00 & 39.00 & 1.00 & 1.00 & 0.73 & 0.27 & 0.27 & 0.00 \\
PNO & 10.00 & 99.00 & 0.69 & 1.00 & 0.74 & 0.26 & 0.26 & 0.00 \\
SEP & 20.00 & 20.00 & 1.00 & 1.00 & 0.00 & 1.00 & 0.00 & 0.00 \\
SNP & 20.00 & 60.00 & 0.66 & 1.00 & 0.00 & 1.00 & 0.00 & 0.00 \\
SEO & 20.00 & 26.00 & 1.00 & 1.00 & 0.00 & 1.00 & 0.18 & 0.00 \\
SNO & 20.00 & 66.00 & 0.69 & 1.00 & 0.00 & 1.00 & 0.17 & 0.00 \\
HIE & 18.00 & 40.00 & 0.75 & 1.00 & 0.00 & 1.00 & 0.00 & 0.00 \\
MIX & 6.00 & 30.00 & 0.66 & 0.36 & 0.10 & 0.15 & 0.01 & 0.00 \\

		\bottomrule
	\end{tabular}
\label{table:gt_stats}
\end{table}

\begin{figure}
	\centering
	\begin{subfigure}[t]{0.24\textwidth}
		\centering
		\includegraphics[width=\textwidth]{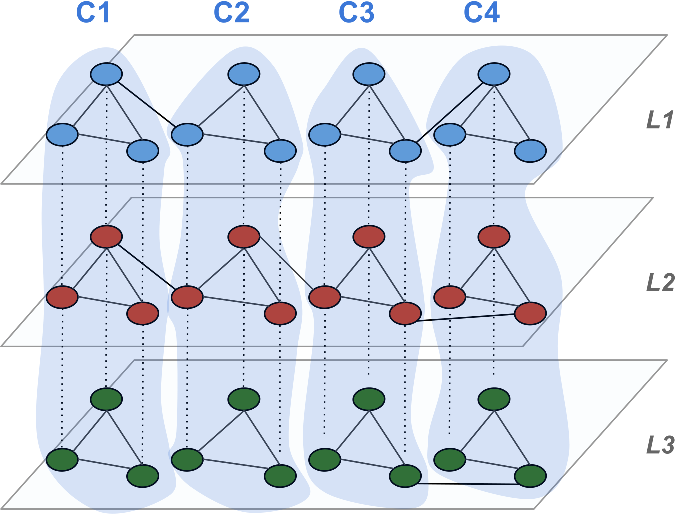} 
		\caption{Pillar Equal Partitioning (PEP)}
		\label{fig:CCSS:pep}
	\end{subfigure}
	\begin{subfigure}[t]{0.24\textwidth}
		\centering
		\includegraphics[width=\textwidth]{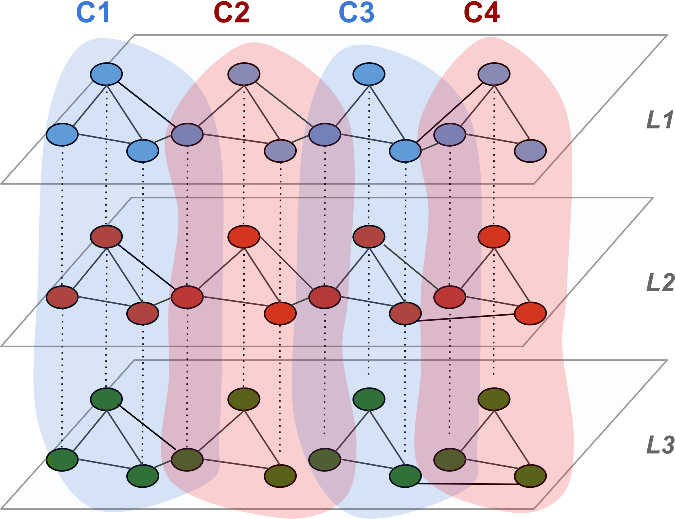} 
		\caption{Pillar Equal Overlapping (PEO)}
		\label{fig:CCSS:peo}
	\end{subfigure}
	\begin{subfigure}[t]{0.24\textwidth}
		\centering
		\includegraphics[width=\textwidth]{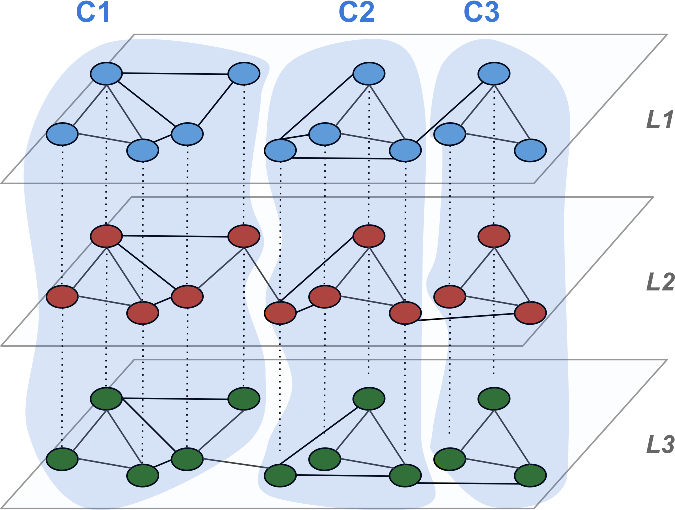} 
		\caption{Pillar Non-Equal Partitioning (PNP)}
		\label{fig:CCSS:pnp}
	\end{subfigure}
	\begin{subfigure}[t]{0.24\textwidth}
		\centering
		\includegraphics[width=\textwidth]{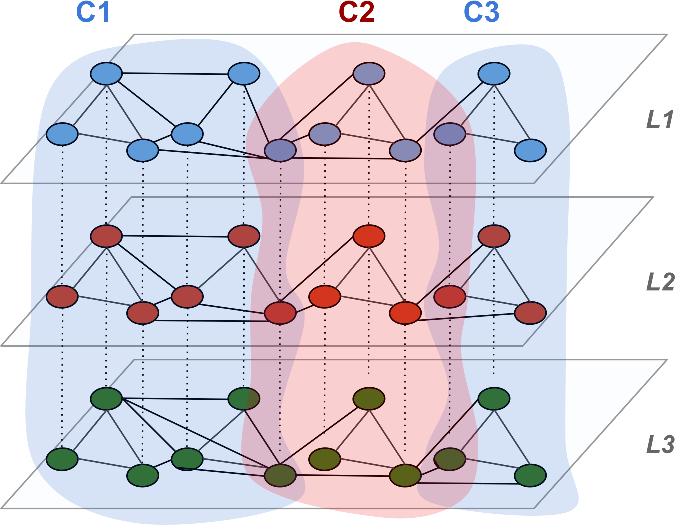} 
		\caption{Pillar Non-Equal Overlapping (PNO)}
		\label{fig:CCSS:pno}
	\end{subfigure}
	\begin{subfigure}[t]{0.24\textwidth}
		\centering
		\includegraphics[width=\textwidth]{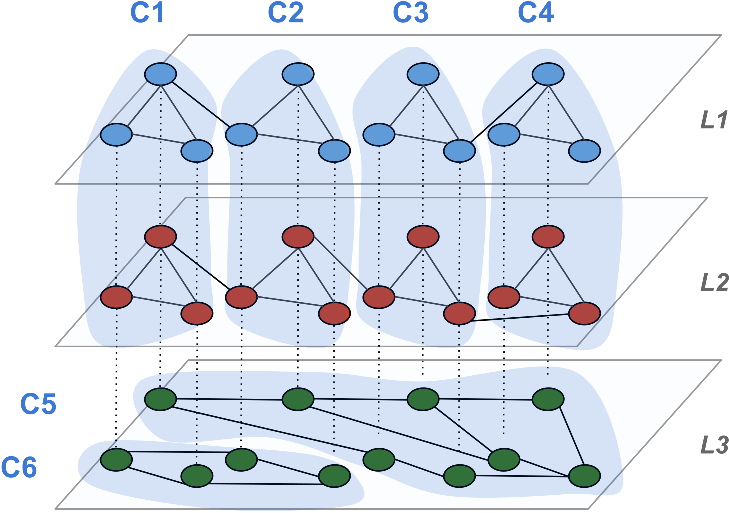} 
		\caption{Semi-pillar Equal Partitioning (SEP)}
		\label{fig:CCSS:sep}
	\end{subfigure}
	\begin{subfigure}[t]{0.24\textwidth}
		\centering
		\includegraphics[width=\textwidth]{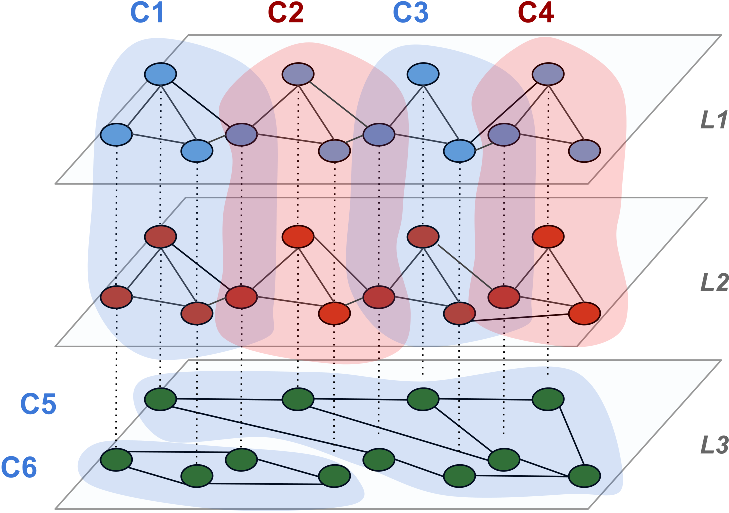} 
		\caption{Semi-pillar Equal Overlapping (SEO)}
		\label{fig:CCSS:seo}
	\end{subfigure}
	\begin{subfigure}[t]{0.24\textwidth}
		\centering
		\includegraphics[width=\textwidth]{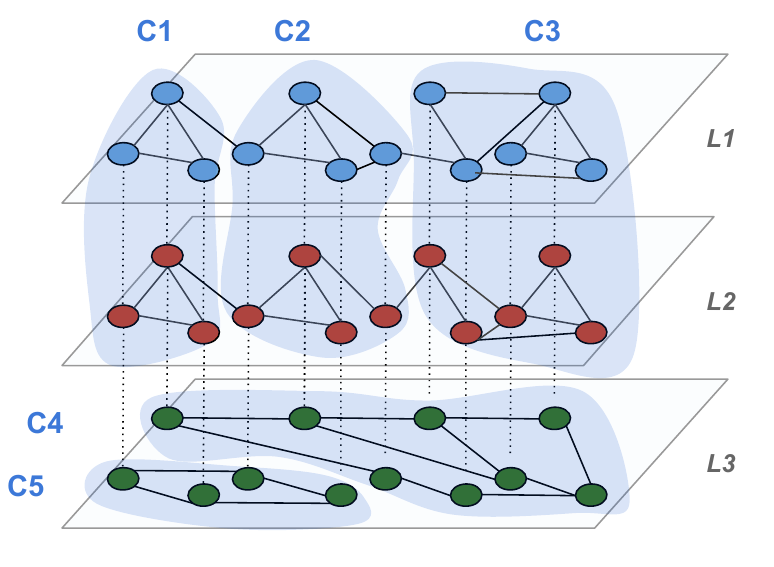} 
		\caption{Semi-pillar Non-equal Part.~(SNP)}
		\label{fig:CCSS:snp}
	\end{subfigure}
	\begin{subfigure}[t]{0.24\textwidth}
		\centering
		\includegraphics[width=\textwidth]{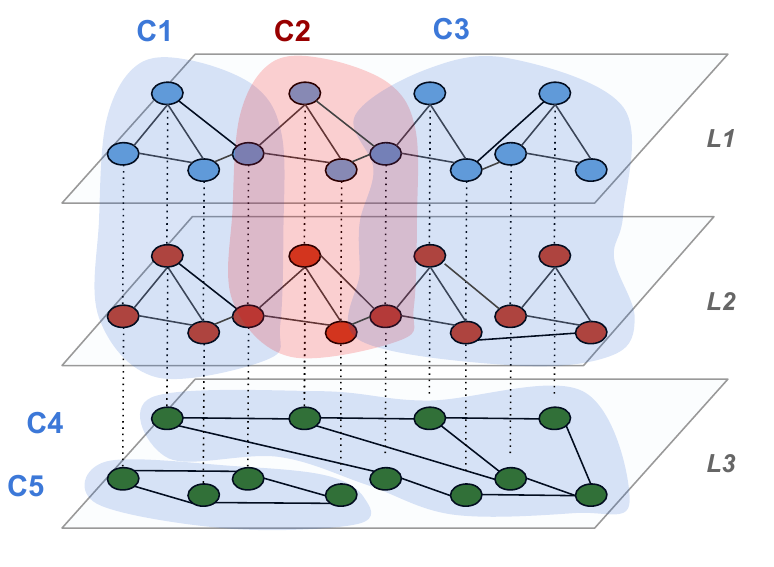} 
		\caption{Semi-pillar Non-equal Over.~(SNO)}
		\label{fig:CCSS:sno}
	\end{subfigure}
	\begin{subfigure}[t]{0.26\textwidth}
		\centering
		\includegraphics[width=\textwidth]{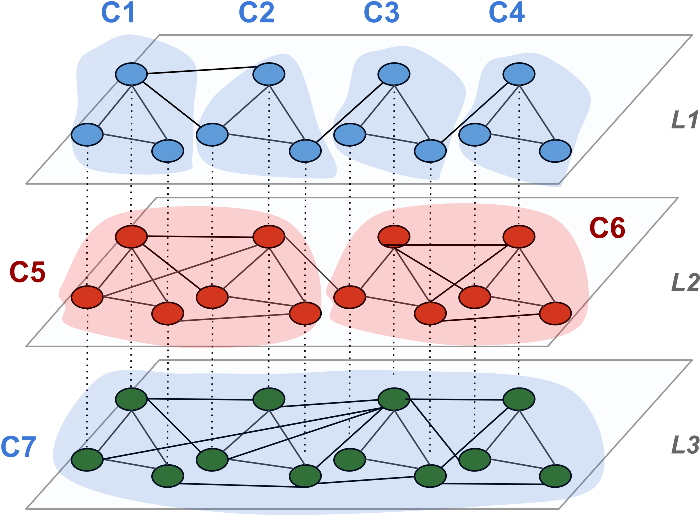} 
		\caption{Hierarchical (HIE)}
		\label{fig:CCSS:HIE}
	\end{subfigure}
	\begin{subfigure}[t]{0.26\textwidth}
		\centering
		\includegraphics[width=\textwidth]{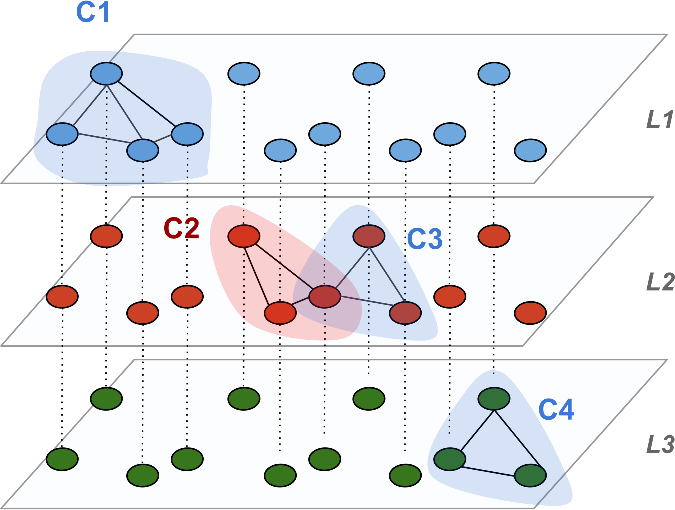} 
		\caption{Mixed (MIX)}
		\label{fig:CCSS:nnm}
	\end{subfigure}
	\caption{An illustration of \rev{the types of} synthetic multiplex networks generated for different possible multiplex community structures\rev{. Equal/Non-Equal refers to the number of nodes (size) in the communities}}
	\label{fig:CCSS}
\end{figure}

\rev{\revm{We have} also generated synthetic datasets forcing specific types of community structures, illustrated in Figure~\ref{fig:CCSS}. \revm{This has two motivations: first, ground truth should be used carefully in cluster analysis, with no single accepted definition of what the correct result should be. So-called ground truths should only be used as part of a broader evaluation, as well known in the field of clustering and also pointed out about community detection \citep{Peel2017}. In addition, the ground truth in the real datasets has a quite simple structure, mostly containing pillar non-overlapping communities. Therefore the synthetic networks are used to check whether the tested algorithms are able to identify specific types of structures.} \revm{We used small datasets to be able to compare all methods including those not scaling well. One should however consider that smaller probabilistically generated networks have a larger structural variability, and when testing scalable methods larger networks can be used to reduce variance in the results. Here we focus on the comparison between methods, which are all tested on the same data.} The code used to generate these networks is available at: https://bitbucket.org/uuinfolab/20csur.}

More in detail, we generated \rev{10} different multiplex networks with 8 different built-in community structures. To keep the focus on the community structure, each of the \rev{10} multiplex networks is comprised of 3 layers, 100 actors, and 300 nodes (100 per layer). After forcing a specific community structure on each multiplex, the edges were generated with a probability $P_{in} = 0.5$ to be internal (within a community) and a probability $P_{ext} = 0.01$ to be external (among communities). The following is a brief description of each multiplex network: 

\begin{itemize}
	\item \textbf{Pillar Equal Partitioning (PEP)}: The community structure in this multiplex is a set of pillar non-overlapping communities that are approximately equal in size. (Figure~\ref{fig:CCSS:pep}).
	
	\item \textbf{Pillar Equal Overlapping (PEO)}:  Similar to PEP in terms of the size of the communities and the pillar structure. The communities in PEO are however overlapping    (Figure~\ref{fig:CCSS:peo}).
	
	\item \textbf{Pillar Non-Equal Partitioning (PNP)}: The community structure in this multiplex is a set of pillar non-overlapping communities. As to the size distribution of the communities, there are few big pillar   communities and many small pillar   communities (Figure~\ref{fig:CCSS:pnp}).
	
	\item \textbf{Pillar Non-Equal Overlapping (PNO)}: 
	Similar to PNO in terms of the community size distribution and the pillar structure. The communities in PNO are however overlapping  (Figure~\ref{fig:CCSS:pno}).
	
	\item \textbf{Semi-pillar Equal Partitioning (SEP)}: The community structure in this multiplex is a set of semi-pillar non-overlapping communities that are approximately equal in size and a set of single-layer communities (Figure~\ref{fig:CCSS:sep}).
	
	\item \textbf{Semi-pillar Equal Overlapping (SEO)}: Similar to SEP except that the semi-pillar communities are overlapping (Figure~\ref{fig:CCSS:seo}). 
	
	\item \rev{\textbf{Semi-pillar Non-Equal Partitioning (SNP)}: The community structure in this multiplex is a set of semi-pillar non-overlapping communities. As to the size distribution of the communities, there are few big pillar   communities and many small pillar   communities (Figure~\ref{fig:CCSS:snp}).}
	
	\item \rev{\textbf{Semi-pillar Non-Equal Overlapping (SNO)}: 
	Similar to SNP in terms of the community size distribution and the pillar structure. The communities in SNO are however overlapping  (Figure~\ref{fig:CCSS:sno}).}
	
	\item \textbf{Hierarchical (\rev{HIE})}: The community structure in this multiplex reflects some hierarchy among communities on the actor level. Some big node-level communities (like $C_7$ in Figure~\ref{fig:CCSS:HIE}) on a layer $L_3$ are constituted of smaller communities on layer $L_2$.
	
	\item \textbf{Mixed (\rev{MIX})}: The community structure in this multiplex is a \rev{small} set of single-layer communities some of which are overlapping (Figure~\ref{fig:CCSS:nnm}).
\end{itemize}

Table \ref{table:gt_stats} provides information about the communities in these multiplex networks and Figure~\ref{fig:CCSS} illustrates the different types of multiplex community structures.

General information about the\rev{se} networks including the mean and standard deviation over the layers for density, degree, average path length and clustering coefficients are reported in Table \ref{table:mpx_info}. More information about the datasets  used in the experiments \rev{is} provided \rev{in} the supplementary online material.

\rev{Then, we generated networks with varying numbers of actors (100 to 10000) and layers (1 to 20) to perform scalability tests. These networks have the same structure indicated as PEP (Pillar Equal Partitioning) in Figure~\ref{fig:CCSS}, because this is the only type of community structure that most of the methods can correctly recover, as we shall see in the results of our experiments.} \revm{While the number of layers and actors varies, the probabilities of node adjacency inside and across communities are set in the same way as for the PEP network in Table \ref{table:mpx_info}.}

Finally, we generate networks using the PEP model and varying the probability of adjacency for nodes in different communities, to study the impact of noise as reported in Section~\ref{appendix_noise}.

\sisetup{separate-uncertainty}

\begin{table}[!htbp]
	\centering
	\caption{ \textbf{Summary of structural characteristics of the evaluation   networks}:  number of layers (\textbf{l}),   number of actors (\textbf{a}),   number of edges (\textbf{e}), and mean/std over the layers of density (\textbf{den}),  \revm{average}  degree (\textbf{a\_deg}),  average path length (\textbf{a\_p\_len}), and    clustering coefficient (\textbf{ccoef})}
	\label{table:mpx_info}
	\begin{tabular}{lrrr
	    S[table-format=1.2(3)]
	    S[table-format=2.2(3)]
	    S[table-format=1.2(3)]
	    S[table-format=1.2(3)]}
		\toprule
		\textbf{Network} & \textbf{l} & \textbf{a} & \textbf{e} & {\textbf{den}} & {\textbf{a\_deg}} & {\textbf{a\_p\_len}} & {\textbf{ccoef}} \\ 
		\toprule
        AUCS & 5 & 61 & 620 & 0.12 \pm 0.07 & 5.21 \pm 2.46 & 2.43 \pm 0.73 & 0.43 \pm 0.1\\
DKPol & 3 & 490 & 20226 & 0.07 \pm 0.08 & 28.85 \pm 44.24 & 3.43 \pm 1.32 & 0.24 \pm 0.26\\
Airports & 37 & 417 & 3588 & 0.06 \pm 0.02 & 3.13 \pm 1.45 & 2.25 \pm 0.34 & 0.07 \pm 0.08\\
Rattus & 6 & 2640 & 3956 & 0.05 \pm 0.07 & 1.62 \pm 0.62 & 2.75 \pm 2.22 & 0.03 \pm 0.08\\

		\bottomrule
	\end{tabular}
	\subcaption{Real datasets}
	\bigskip
	\begin{tabular}{lrrr
	    S[table-format=1.2(3)]
	     S[table-format=2.2(3)]
	    S[table-format=1.2(3)]
	    S[table-format=1.2(3)]
	     }
		\toprule
		\textbf{Network} & \textbf{l} & \textbf{a} & \textbf{e} & {\textbf{den}} & {\textbf{a\_deg}} & {\textbf{a\_p\_len}} & {\textbf{ccoef}} \\ 
		\toprule
        PEP & 3 & 100 & 943 & 0.05 \pm 0.00 & 5.32 \pm 0.32 & 3.39 \pm 0.09 & 0.31 \pm 0.05\\
PNP & 3 & 100 & 1584 & 0.1 \pm 0.01 & 9.51 \pm 0.52 & 2.77 \pm 0.04 & 0.41 \pm 0.02\\
PEO & 3 & 100 & 1487 & 0.09 \pm 0.00 & 8.78 \pm 0.33 & 2.51 \pm 0.02 & 0.28 \pm 0.03\\
PNO & 3 & 100 & 2079 & 0.13 \pm 0.00 & 12.71 \pm 0.44 & 2.29 \pm 0.03 & 0.37 \pm 0.01\\
SEP & 3 & 100 & 966 & 0.06 \pm 0.00 & 5.45 \pm 0.14 & 3.36 \pm 0.06 & 0.34 \pm 0.02\\
SNP & 3 & 100 & 1360 & 0.08 \pm 0.02 & 7.96 \pm 2.32 & 3.01 \pm 0.39 & 0.38 \pm 0.03\\
SEO & 3 & 100 & 1314 & 0.08 \pm 0.02 & 7.63 \pm 2.03 & 2.8 \pm 0.49 & 0.28 \pm 0.01\\
SNO & 3 & 100 & 1762 & 0.11 \pm 0.05 & 10.65 \pm 4.54 & 2.63 \pm 0.63 & 0.37 \pm 0.02\\
HIE & 3 & 100 & 1820 & 0.11 \pm 0.06 & 11.05 \pm 5.64 & 2.76 \pm 0.58 & 0.41 \pm 0.05\\
MIX & 3 & 100 & 388 & 0.02 \pm 0.01 & 2.21 \pm 0.78 & 2.78 \pm 0.29 & 0.42 \pm 0.05\\

		\bottomrule
	\end{tabular}
	\subcaption{Synthetic datasets with a controlled community structure}
\end{table}

\subsection{\rev{Detailed setting for each method}} 

\rev{For all methods based on a single-layer algorithm, we use Louvain. Using the same algorithm makes the comparison fairer; however we must point out how this deviates from some original publications. We also tested the methods using the single-layer algorithm mentioned in the original references (e.g., label propagation). We think that the relevance of these methods for this paper lies in the way they deal with the multilayer structure rather than the specific algorithm that is used on the single-layer network. Within this perspective, using Louvain provides more stable, more accurate and more comparable results in general.}

With respect to parameter setting, in general we used the default values proposed by the original works. In some specific cases, where \rev{different parameter settings are expected to be used to identify different types of community structures} (i.e., GLouvain, ML-CPM, ABACUS, ACLcut, ML-LCD, and Infomap),  we \rev{tested multiple settings as detailed in the following.} 

\begin{itemize}

	\item \rev{For \textsf{ABACUS},} two main parameters have effect on filtering out possible multiplex communities \rev{when single-layer communities are merged into the final result}, namely, the minimum number of actors in a community (\rev{$k$}) and the minimum number of single-layer communities in which the actors must have been grouped together (\rev{$m$}). We use this algorithm with two settings, \textsf{ABACUS$_{31}$} with (\rev{$k$}=3, \rev{$m$}=1) and  \textsf{ABACUS$_{42}$} with (\rev{$k$}=4, \rev{$m$}=2) which filters out the communities that are not expanded over multiple layers.
	
	\item \rev{\textsf{PMM} takes three parameters: the number of communities to return, the number of structural features, and the number of times k-means should be executed as a subroutine, that we set to 5. The number of communities has been set to the number of known communities in the data where that is known, and to an arbitrary number (10) for Airports and Rattus. The fact that we used knowledge about the expected result to setup the algorithm should be considered when the different methods are compared. We did not find heuristics to set the number of structural features (Ell), so we used two settings: low and constant (Ell = 10), and high and dependent on the number of actors (Ell = $a/2$); these are among the settings returning good results for AUCS and PEP, for which a ground truth compatible with the results that PMM can return exists. However, please notice that the results may vary very significantly by varying this parameter, and we set it based on knowledge of the expected result. This should also be considered when looking at the experimental results.}
		
	\item \rev{\textsf{SCML} takes two parameters: the number of communities, for which the same settings and reflections for PMM apply, and lambda, set to the default value .5.}
						
	\item \rev{\textsf{EMCD} takes one parameter, theta, for which different settings can lead to significantly different results. The original reference contains an evaluation of appropriate ranges of theta for datasets with different statistics.  We based our settings on these considerations: .03 for Airports and Rattus, .01 for DKPol, .2 for AUCS, .1 for the synthetic networks.}
    
	\item \textsf{ML-CPM}: two main parameters \rev{can influence} the results and the \rev{execution time} of the algorithm, namely, the minimum number of actors that form a multilayer clique (\rev{$k$}), and the minimum number of layers to be considered when counting the multilayer cliques (\rev{$m$}). To be more inclusive, we defined two settings for these parameters, \textsf{ML-CPM$_{31}$} with (\rev{$k$}=3, \rev{$m$}=1) which allows single-layer communities but could be computationally very expensive with large networks, and \textsf{ML-CPM$_{42}$} with (\rev{$k$}=4, \rev{$m$}=2) which is less expensive computationally, but forces the communities to be expanded over at least two layers. 
	
	\item \rev{\textsf{LART} has been executed with default parameter settings: $t = 9$ (number of steps for random walker to take), eps = 1  (for binary matrices this will mean adding a self-loop to each node on each layer),
    gamma = 1 (recommended by the authors), and linkage = average (determining the type of hierarchical clustering performed in the algorithm).}
	
	\item \textsf{Infomap} can be used to find both overlapping and non-overlapping communities. Consequently, we  included it twice in our experiments, i.e., forcing a non-overlapping community discovery \rev{(}\textsf{Infomap$_{no}$}\rev{)}, and accepting overlapping communities \rev{(}\textsf{Infomap$_{o}$}\rev{)}.
	
	\item \rev{For} \textsf{GLouvain} we defined  two settings, \textsf{GLouvain$_{h}$} to denote high weight assigned to the inter-layer edges ($\omega = 1$),   and  \textsf{GLouvain$_{l}$} to refer to \rev{a} low value for the inter-layer edge  weight  ($\omega = 0.1$). The motivation is that high values for $\omega$ favor the identification of pillar communities and may prevent the identification of  actor-overlapping communities that the algorithm can retrieve with a low $\omega$.

	\item \rev{\textsf{MLink} takes two input parameters leading to different types of results. As we have not analyzed the resulting communities, for which we refer to the original reference, we use the default values used in the original implementation for scalability analysis.}
    
	\item \rev{\textsf{\revm{MDLPA}} has no input parameters.}

	\item \rev{For} \textsf{ACLcut}\rev{,} two settings were used. One with a classical random walker \textsf{ACLcut$_{c}$}, and another with a relaxed random walker \textsf{ACLcut$_{r}$}.

	\item \rev{For} \textsf{ML-LCD} \rev{we used three}  settings \rev{corresponding to different ways} to optimize the $LC$ function   during the selection of nodes to join a local community, namely,  \textsf{ML-LCD$_{(lwsim)}$},  for the layer-weighted similarity based $LC$, \textsf{ML-LCD$_{(wlsim)}$}  for the within-layer similarity based $LC$, and  \textsf{ML-LCD$_{(clsim)}$}  for the cross-layer similarity based $LC$. 
\end{itemize}

\subsection{\rev{Software}} 


\rev{The following experiments have been performed using a combination of original code (LART in \revm{P}ython2.7, EMCD in \revm{J}ava, PMM, SCML, and M\revm{L}ink in \revm{MATLAB}, Infomap in C++) and the implementations of the other algorithms available in the multinet library (NWF, WF$_{EC}$, ABACUS, \revm{ML-}CPM, GLouvain, \revm{MDLPA}, all written in C++ and also available for R and \revm{P}ython). We also use the multinet library for basic functions to read networks, communities, to compute the Omega index, etc. Infomap was also run from inside multinet, but the code is the one from the authors with minor adaptations to make it compatible with the requirements of the CRAN repository. The implementation of ABACUS uses code from https://borgelt.net/eclat.html for the association rule mining subroutine. All the algorithms are available at https://bitbucket.org/uuinfolab/20csur, except ACLCut which has not been ported to the latest version of the multinet library. The \revm{MATLAB} code in this repository is run using \revm{O}ctave. All the \revm{MATLAB} code could be executed in \revm{O}ctave, except the internal edge clustering subroutine used by M\revm{L}ink. As we did not compare the results of Mlink with other algorithms, we skipped that part of the execution, which does not affect our conclusions about its scalability.}

\subsection{\rev{Assessment criteria}} 

In order to measure pairwise similarity between two global community structures, we use the Omega index which is a well known measure  \citep{Collins1988} that can be applied to situations where both, one, or neither of the clusterings being compared is overlapping \citep{murray2012omega}. It does so by averaging the number of agreements on both clusterings and then adjusting that by the expected \rev{number of} agreement\rev{s between the two} clusterings in case they were generated at random. An agreement is when \revm{two nodes are clustered together} \rev{in the same number of clusters} ($j$) in both clusterings. The values of $j$ start from 0, meaning that if \revm{two nodes are never clustered together in both clusterings}, this still counts as an agreement.

Given two clusterings $\mathcal{C}_1$, $\mathcal{C}_2$, the similarity between them using Omega index is given by
	\begin{equation}
	    \text{Omega ($\mathcal{C}_1$,$\mathcal{C}_2$)} = \frac{\text{Observed ($\mathcal{C}_1$,$\mathcal{C}_2$)} - \text{Expected ($\mathcal{C}_1$,$\mathcal{C}_2$)}}{1-\text{Expected ($\mathcal{C}_1$,$\mathcal{C}_2$)}}
	\end{equation}
	\begin{equation}
	    \text{Observed ($\mathcal{C}_1$,$\mathcal{C}_2$)} = \frac{1}{N} \sum_{j=0}^{l} A_j
	\end{equation}
	\begin{equation}
	    \text{Expected ($\mathcal{C}_1$,$\mathcal{C}_2$)} = \frac{1}{N^2} \sum_{j=0}^{l} N_{(j,1)}N_{(j,2)}
	\end{equation}
	 Where Observed ($\mathcal{C}_1$,$\mathcal{C}_2$) refers to the observed agreement represented by the average number of agreements between  $\mathcal{C}_1$ and $\mathcal{C}_2$, $l$ is the maximum number of times a pair appears together in both $\mathcal{C}_1$ and  $\mathcal{C}_2$ at the same time, N is the total number of possible pairs, $A_j$ is the number of pairs that are grouped together $j$ times in both clusterings, and $N_{(j,1)}$, $N_{(j,2)}$ \rev{indicate} the number\rev{s} of pairs that have been grouped together $j$ times in $\mathcal{C}_1$, $\mathcal{C}_2$ respectively. Theoretically, values \rev{of the Omega index} are in the range [-1,1]. However, in practice, Omega index returns 1 for two identical clusterings, and values close to 0 when one of the two input clusterings is a totally random reordering of the other one.

\rev{To clarify the formulas above, we provide two examples. First, to understand the meaning of each part of the formulas, consider two equal overlapping clusterings of four elements 1, 2, 3, and 4: $\mathcal{C}_1 = \{\{1,2,3\}, \{2,3,4\}\}$ and $\mathcal{C}_2 = \{\{1,2,3\}, \{2,3,4\}\}$. In this case the number of possible pairs $N$ is 6 ($\{1,2\}, \{1,3\}, \{1,4\}\dots$). $A_0=1$, because only the pair $\{1,4\}$ does not appear inside a same cluster in both clusterings. $A_1=4$, corresponding to pairs $\{1,2\}, \{1,3\}, \{2,4\}$, and $\{3,4\}$, all appearing together once in each clustering. Only the pair $\{2,3\}$ is assigned to two different clusters in each clustering, therefore $A_2=1$. The other values to compute the omega index are $N_{(0,1)}=1, N_{(0,2)}=1, N_{(1,1)}=4, N_{(1,2)}=4, N_{(2,1)}=1, N_{(2,2)}=1$. As a result, we  have: 
$\text{Observed ($\mathcal{C}_1$,$\mathcal{C}_2$)} = \frac{1}{6} (1+4+1)
$ and $\text{Expected ($\mathcal{C}_1$,$\mathcal{C}_2$)} = \frac{1}{36} (1\cdot1+4\cdot4+1\cdot1)$.
The corresponding Omega index is 1, as expected because the two clusterings are identical. Now consider the two clusterings $\mathcal{C}_1 = \{\{1,2\}, \{3,4\}\}$ and $\mathcal{C}_2 = \{\{1,2\}, \{3\}, \{4\}\}$. We now have 
$\text{Observed ($\mathcal{C}_1$,$\mathcal{C}_2$)} = \frac{1}{6} (4+1)$ and $\text{Expected ($\mathcal{C}_1$,$\mathcal{C}_2$)} = \frac{1}{36} (4\cdot5 + 2\cdot1)$ with Omega index 0.57.}


The reason why we choose the Omega index is that it is, by  definition, a valid measure when one, both or none of the two clusterings is \rev{overlapping} as we discuss in \cite{hanteer2019_sim}. In addition, Omega index is an adjusted similarity measure that accounts for the by-chance agreements that might still exist between any two random clusterings over the same node-set. 



For measuring similarity between two local communities $s_1$, $s_2$, we use \rev{the} Jaccard coefficient:  
\begin{equation}
JC = \frac{N(s_1,s_2)}{N(s_1)+N(s_2)-N(s_1,s_2)} 
\end{equation}
where $N(s_1)$ refers to the number of actors in solution $s_1$ and $N(s_1,s_2)$ refers to the number of common actors between two solutions $s_1$, $s_2$. The values of \rev{the} Jaccard coefficient lie in the range [0,1] where 1 means perfect similarity and 0 means perfect dissimilarity.

In order to measure the accuracy of the solutions obtained by global methods with respect to a ground truth (Section \ref{sssec:results_glob_acc}), we resort again to \rev{the} Omega index. 
The accuracy of local community detection methods (Section \ref{sssec:local_acc_anyalisis}) has been evaluated by comparing pairwise similarities (using \rev{the} Jaccard index) between a given actor (i.e., seed node) and the ground truth community it belongs to. The average Jaccard index over all actors is then used as the final accuracy score.    

	


\section{Results}
\label{sec:results}

In this section we present the experimental results of our comparative evaluation. Results of the comparative evaluation of global methods are reported in Section~\ref{ssec:results_global}, while results related to the evaluation of local methods are reported in Section~\ref{ssec:results_local}.




\subsection{Global Methods}
\label{ssec:results_global}

In this section we report the experimental results of the comparative evaluation of global multiplex community detection methods. The section is structured as follows: Section~\ref{sssec:results_glob_stats} reports on the main properties of the community structures detected by the evaluated methods in different datasets. Section~\ref{sssec:results_glob_acc} presents the results of the accuracy analysis\rev{.} Section~\ref{sssec:results_glob_pw} discusses the results of the pairwise comparison between different methods. 
Section~\ref{sssec:results_glob_scal} focuses on scalability.

\subsubsection{Basic descriptive statistics}
\label{sssec:results_glob_stats}

As the first step of our comparative analysis, we analyzed the  structural properties of the different community structures identified by the evaluated methods. Table\revm{s}~\ref{table:global_real_stats:1} \revm{and \ref{table:global_real_stats:2}} present the statistics concerning the community structures obtained on the smallest (AUCS) and largest (Airports) of the real-world multiplex networks taken into account. We also present the general statistics about the communities detected on \rev{DKPol} and Rattus datasets in Table\rev{s \ref{table:global_real_stats:dkpol} and \ref{table:global_real_stats:rattus}}. Statistics of each method occupy one row in each a table (multiple settings of the input parameters for some methods are represented as separated entries for the same method). Since some of the methods are non-deterministic, we  \revm{executed} each method 10 times, and provide \rev{mean and standard deviation}. 
It can be observed how \textsf{LART} generates a number of communities which is higher than that of most other methods on all real networks. However a large percentage of these communities appear to be singletons, indicating that this algorithm mostly fails in aggregating nodes into communities.
Other algorithms which appear to generate a relatively high number of communities regardless of the network structure are \rev{\textsf{Infomap$_o$}} and \textsf{ABACUS}, both variants. Interestingly, \rev{both retrieve a large number of communities without retrieving any singleton, showing a different behavior from \textsf{LART}. \rev{The discovery of many communities by \textsf{Infomap$_o$} and \textsf{ABACUS} is associated to a high percentage of node overlapping.}
As regards to the size of the largest community, higher values correspond to \textsf{PMM$_l$} and \textsf{Infomap$_o$}. On the other end, \textsf{ABACUS} (both variants) and \textsf{\revm{ML-}CPM$_{42}$} assign a small number of nodes to the largest communities, in both the AUCS and the Airports networks. \rev{This can be explained by the strong requirements that \textsf{ABACUS} and (even more) \textsf{ML-CPM} have to cluster nodes together.}
Concerning $sc2/sc1$, we can observe how the values tend to be all relatively high for the smallest (\rev{AUCS}) and largest (Airports) network\rev{s}, indicating that in these cases the largest communities for each identified community structure have comparable sizes. \rev{An algorithm grouping most of the nodes together, and thus not able to structure them into separate communities, would have a very low value for $sc2/sc1$.}}

\rev{The values found in columns $\%n$, $\%p$, $\%ao$ and $\%no$ can be explained as follows:} 
\begin{itemize}
\item With regards to the percentage $\%n$ of nodes assigned to at least one community, as we discussed in Section \ref{sec:mpx_community}, certain methods\footnote{\textsf{NWF}, \textsf{WF$_{EC}$}, \textsf{GLouvain} (both variants), \textsf{LART}, \textsf{Infomap} (both variants)} are forced to provide a community assignment for each node: in these cases the value of $\%n$ will \rev{always} be  $1$. 

\item Regarding the percentage $\%p$ of pillars, both flattening methods always return pillar communities (since the information about layers is lost during the flattening process). \rev{\textsf{Infomap} and \textsf{GLouvain} can detect non-pillar clusters in theory. Data show how \textsf{Infomap} can return non-pillars both in the overlapping and in the non-overlapping version, while only \textsf{GLouvain$_{l}$} returns non-pillar communities.} 

\item The percentage of overlapping actors ($\%ao$) and nodes ($\%no$) mainly depends on the properties of the specific methods whether they allow overlapping (on the node level or the actor level) or not.  

\item \revm{T}he percentage of singleton communities $\%s$ appears to be extremely high in the case of \textsf{LART} and \textsf{EMCD} and high in the case of \textsf{PMM$_l$}. It should be noted that, with the exception of \textsf{Infomap}, that returns a small fraction of singletons in the Airports network, the methods that return singletons in the AUCS network return a larger percentage of singletons in the Airports network suggesting that the behaviour is not induced by the network but amplified by its complexity.
\end{itemize}

\begin{table}
	\centering
	\caption{Statistics about the community structures obtained on the \rev{AUCS} network (results averaged over 10 runs). We denote with \textbf{\#c} the number of communities, with \textbf{sc1} the size of the largest community \rev{(number of nodes)}, with \textbf{sc2/sc1} the ratio between the size of the \revm{second} largest community \rev{and} the largest, with \textbf{\%n} the percentage of nodes assigned to at least one community, with \textbf{\%p} the percentage of pillars, with \textbf{\%ao} the percentage of actors in more than one community, with \%no the percentage of nodes in more than one community and with \textbf{\%s} the percentage of singleton communities}
	\resizebox{\textwidth}{!}{%
		\begin{tabular}{l 
	    S[table-format=1.2(2)]
	    S[table-format=3.2(1)]
	    S[table-format=3.2(1)]
	    S[table-format=1.2(2)]
	    S[table-format=1.2(2)]
	    S[table-format=1.2(2)]
	    S[table-format=1.2(2)]
	    S[table-format=1.2(2)]}
		\toprule
		method & {$\#c$} & {$sc1$} & {$sc2/sc1$} & {$\%n$} & {$\%p$} & {$\%ao$} & {$\%no$} & {$\%s$} \\
		\toprule
        NWF & 5.00 & 75.00 & 0.92 \pm 0.01 & 1.00 & 1.00 & 0.00 & 0.00 & 0.00 \\ 
WF$_{EC}$ & 5.00 & 75.00 & 0.92 \pm 0.01 & 1.00 & 1.00 & 0.00 & 0.00 & 0.00 \\ 
ABACUS$_{31}$ & 46.50 \pm 3.44 & 29.90 \pm 2.54 & 0.96 \pm 0.02 & 0.70 & 0.00 & 0.96 & 0.61 & 0.00 \\
ABACUS$_{42}$ & 25.50 \pm 3.58 & 29.20 \pm 2.52 & 0.96 \pm 0.03 & 0.59 \pm 0.01 & 0.00 & 0.67 \pm 0.01 & 0.39 \pm 0.01 & 0.00 \\
EMCD & 11.00 & 70.00 & 0.92 & 1.00 & 1.00 & 0.00 & 0.00 & 0.45 \\ 
PMM$_l$ & 8.00 & 103.00 \pm 21.00 & 0.49 \pm 0.14 & 1.00 & 1.00 & 0.00 & 0.00 & 0.05 \pm 0.08 \\
PMM$_h$ & 8.00 & 79.50 \pm 16.94 & 0.75 \pm 0.17 & 1.00 & 1.00 & 0.00 & 0.00 & 0.02 \pm 0.05 \\
SCML & 8.00 & 66.00 \pm 3.00 & 0.94 \pm 0.11 & 1.00 & 1.00 & 0.00 & 0.00 & 0.00 \\ 
ML-CPM$_{31}$ & 40.00 & 59.00 & 0.61 & 0.61 & 0.00 & 0.93 & 0.46 & 0.00 \\
ML-CPM$_{42}$ & 11.00 & 18.00 & 0.88 & 0.27 & 0.00 & 0.34 & 0.11 & 0.00 \\
LART & 48.60 \pm 0.66 & 51.00 \pm 2.00 & 0.58 \pm 0.02 & 1.00 & 0.98 & 0.01 & 0.00 & 0.91 \\
Infomap$_{no}$ & 5.09 \pm 0.30 & 86.00 \pm 8.30 & 0.80 \pm 0.06 & 1.00 & 1.00 & 0.00 & 0.00 & 0.00 \\
Infomap$_o$ & 20.60 \pm 0.80 & 157.00 \pm 44.00 & 0.49 \pm 0.12 & 1.00 & 0.30 \pm 0.02 & 0.69 \pm 0.02 & 0.69 \pm 0.02 & 0.00 \\
GLouvain$_l$ & 7.50 \pm 0.67 & 80.20 \pm 7.34 & 0.85 \pm 0.08 & 1.00 & 0.44 \pm 0.08 & 0.55 \pm 0.08 & 0.00 & 0.00 \\
GLouvain$_h$ & 5.00 & 76.50 \pm 4.50 & 0.85 \pm 0.04 & 1.00 & 1.00 & 0.00 & 0.00 & 0.00 \\
MDLPA & 6.70 \pm 0.78 & 87.50 \pm 18.74 & 0.65 \pm 0.17 & 1.00 & 1.00 & 0.00 & 0.00 & 0.00 \\ 

		\bottomrule
	\end{tabular}}
	\subcaption*{AUCS\rev{. l = 5, a =  61, e =  620}}
\label{table:global_real_stats:1}
\end{table}

\begin{table}
	\centering
	\caption{Statistics about the community structures obtained on the Airports network (results averaged over 10 runs). We denote with \textbf{\#c} the number of communities, with \textbf{sc1} the size of the largest community \rev{(number of nodes)}, with \textbf{sc2/sc1} the ratio between the size of the \revm{second} largest community \rev{and} the largest, with \textbf{\%n} the percentage of nodes assigned to at least one community, with \textbf{\%p} the percentage of pillars, with \textbf{\%ao} the percentage of actors in more than one community, with \%no the percentage of nodes in more than one community and with \textbf{\%s} the percentage of singleton communities}
	\resizebox{\textwidth}{!}{%
	\begin{tabular}{l 
	    S[table-format=4.2(3)]
	    S[table-format=5.2(1)]
	    S[table-format=3.2(1)]
	    S[table-format=1.2]
	    S[table-format=1.2(3)]
	    S[table-format=1.2(3)]
	    S[table-format=1.2(3)]
	    S[table-format=1.2(3)]}
		\toprule
		method & {$\#c$} & {$sc1$} & {$sc2/sc1$} & {$\%n$} & {$\%p$} & {$\%ao$} & {$\%no$} & {$\%s$} \\
		\toprule
        NWF & 6.80 \pm 0.60 & 4229.10 \pm 653.56 & 0.77 \pm 0.16 & 1.00 & 1.00 & 0.00 & 0.00 & 0.00 \\ 
WF$_{EC}$ & 6.70 \pm 0.45 & 4417.80 \pm 532.90 & 0.73 \pm 0.11 & 1.00 & 1.00 & 0.00 & 0.00 & 0.00 \\
ABACUS$_{31}$ & 5320.30 \pm 89.42 & 84.00 & 0.95 & 0.12 & 0.00 & 0.63 & 0.10 & 0.00 \\
ABACUS$_{42}$ & 4086.40 \pm 69.71 & 84.00 & 0.95 & 0.09 & 0.00 & 0.32 \pm 0.01 & 0.07 & 0.00 \\
EMCD & 314.00 & 2035.00 & 0.30 & 1.00 & 1.00 & 0.00 & 0.00 & 0.97 \\ 
PMM$_l$ & 10.00 & 14289.40 \pm 246.98 & 0.03 \pm 0.01 & 1.00 & 1.00 & 0.00 & 0.00 & 0.48 \pm 0.15 \\
PMM$_h$ & 10.00 & 2171.90 \pm 153.49 & 0.86 \pm 0.07 & 1.00 & 1.00 & 0.00 & 0.00 & 0.00 \\
SCML & 10.00 & 4336.39 \pm 1128.68 & 0.46 \pm 0.21 & 1.00 & 1.00 & 0.00 & 0.00 & 0.00 \\ 
ML-CPM$_{31}$ & 62.00 & 93.00 & 0.72 & 0.04 & 0.00 & 0.35 & 0.00 & 0.00 \\
ML-CPM$_{42}$ & 3.00 & 8.00 & 1.00 & 0.00 & 0.00 & 0.00 & 0.00 & 0.00 \\
Infomap$_{no}$ & 7.70 \pm 1.55 & 10330.40 \pm 4920.32 & 0.17 \pm 0.22 & 0.80 \pm 0.35 & 0.79 \pm 0.35 & 0.00 & 0.00 & 0.08 \pm 0.10 \\
Infomap$_o$ & 34.10 \pm 2.02 & 6034.70 \pm 993.70 & 0.83 \pm 0.11 & 1.00 & 0.36 & 0.63 & 0.63 & 0.00 \\
GLouvain$_l$ & 11.20 \pm 0.60 & 6161.80 \pm 382.63 & 0.35 \pm 0.06 & 1.00 & 0.50 & 0.49 & 0.00 & 0.00 \\
GLouvain$_h$ & 9.50 \pm 1.11 & 5372.40 \pm 333.32 & 0.57 \pm 0.13 & 1.00 & 1.00 & 0.00 & 0.00 & 0.00 \\ 

		\bottomrule
	\end{tabular}}
	\subcaption*{Airports\rev{. l = 37, a =  417, e =  3588}}
\label{table:global_real_stats:2}
\end{table}

\begin{table}
	\centering
	\caption{\rev{Statistics about the community structures obtained on the \rev{DKPol} network (results averaged over 10 runs). We denote with \textbf{\#c} the number of communities, with \textbf{sc1} the size of the largest community (number of nodes), with \textbf{sc2/sc1} the ratio between the size of the  \revm{second} largest community  \revm{and} the largest, with \textbf{\%n} the percentage of nodes assigned to at least one community, with \textbf{\%p} the percentage of pillars, with \textbf{\%ao} the percentage of actors in more than one community, with \%no the percentage of nodes in more than one community and with \textbf{\%s} the percentage of singleton communities. ML-CPM$_{31}$ is not present because it takes too long to produce a result, for all executions.}}
	\resizebox{\textwidth}{!}{%
	\begin{tabular}{l 
	    S[table-format=1.2(3)]
	    S[table-format=3.2(1)]
	    S[table-format=3.2(1)]
	    S[table-format=1.2]
	    S[table-format=1.2(3)]
	    S[table-format=1.2(3)]
	    S[table-format=1.2(3)]
	    S[table-format=1.2(3)]}
		\toprule
		method & {$\#c$} & {$sc1$} & {$sc2/sc1$} & {$\%n$} & {$\%p$} & {$\%ao$} & {$\%no$} & {$\%s$} \\
		\toprule
        NWF & 6.30 \pm 0.90 & 370.50 \pm 40.56 & 0.93 \pm 0.06 & 1.00 & 1.00 & 0.00 & 0.00 & 0.00 \\ 
WF$_{EC}$ & 6.30 \pm 1.10 & 378.60 \pm 39.04 & 0.90 \pm 0.07 & 1.00 & 1.00 & 0.00 & 0.00 & 0.00 \\ 
ABACUS$_{31}$ & 69.40 \pm 2.05 & 123.50 \pm 17.02 & 0.92 \pm 0.05 & 0.55 & 0.00 & 0.44 & 0.32 & 0.00 \\
ABACUS$_{42}$ & 31.20 \pm 2.44 & 65.20 \pm 2.22 & 0.92 \pm 0.05 & 0.30 \pm 0.01 & 0.00 & 0.11 \pm 0.01 & 0.07 & 0.00 \\
EMCD & 14.00 & 279.00 & 0.77 & 1.00 & 1.00 & 0.00 & 0.00 & 0.00 \\ 
PMM$_l$ & 9.00 & 992.40 \pm 52.98 & 0.12 \pm 0.02 & 1.00 & 1.00 & 0.00 & 0.00 & 0.07 \pm 0.07 \\
PMM$_h$ & 9.00 & 217.20 \pm 25.84 & 0.94 \pm 0.05 & 1.00 & 1.00 & 0.00 & 0.00 & 0.00 \\
SCML & 9.00 & 256.50 \pm 34.63 & 0.80 \pm 0.05 & 1.00 & 1.00 & 0.00 & 0.00 & 0.00 \\
ML-CPM$_{42}$ & 9.00 & 24.00 & 0.91 & 0.06 & 0.00 & 0.00 & 0.00 & 0.00 \\
Infomap$_{no}$ & 1.00 & 1470.00 & 1.00 & 1.00 & 1.00 & 0.00 & 0.00 & 0.00 \\
Infomap$_o$ & 23.00 \pm 5.77 & 1470.00 & 0.27 \pm 0.11 & 1.00 & 0.65 \pm 0.09 & 0.34 \pm 0.09 & 0.34 \pm 0.09 & 0.45 \pm 0.06 \\
GLouvain$_l$ & 19.50 \pm 0.80 & 347.60 \pm 42.45 & 0.87 \pm 0.05 & 1.00 & 0.65 \pm 0.01 & 0.34 \pm 0.01 & 0.00 & 0.02 \pm 0.02 \\
GLouvain$_h$ & 7.60 \pm 0.66 & 303.89 \pm 11.14 & 0.93 \pm 0.02 & 1.00 & 1.00 & 0.00 & 0.00 & 0.00 \\
MDLPA & 1.00 & 1470.00 & 1.00 & 1.00 & 1.00 & 0.00 & 0.00 & 0.00 \\ 

		\bottomrule
	\end{tabular}}
	\subcaption*{DKPol\rev{. l = 3, a =  493, e =  20226}}
\label{table:global_real_stats:dkpol}
\end{table}

\begin{table}
	\centering
	\caption{\rev{Statistics about the community structures obtained on the \rev{Rattus} network (results averaged over 10 runs). We denote with \textbf{\#c} the number of communities, with \textbf{sc1} the size of the largest community (number of nodes), with \textbf{sc2/sc1} the ratio between the size of the \revm{second} largest community  \revm{and} the  largest, with \textbf{\%n} the percentage of nodes assigned to at least one community, with \textbf{\%p} the percentage of pillars, with \textbf{\%ao} the percentage of actors in more than one community, with \%no the percentage of nodes in more than one community and with \textbf{\%s} the percentage of singleton communities. ML-CPM$_{42}$ is not present because it does not find any community.}}
	\resizebox{\textwidth}{!}{%
	\begin{tabular}{l 
	    S[table-format=1.2(3)]
	    S[table-format=5.2(1)]
	    S[table-format=3.2(1)]
	    S[table-format=1.2]
	    S[table-format=1.2(3)]
	    S[table-format=1.2(3)]
	    S[table-format=1.2(3)]
	    S[table-format=1.2(3)]}
		\toprule
		method & {$\#c$} & {$sc1$} & {$sc2/sc1$} & {$\%n$} & {$\%p$} & {$\%ao$} & {$\%no$} & {$\%s$} \\
		\toprule
        NWF & 136.10 \pm 0.94 & 3264.60 \pm 15.97 & 0.38 \pm 0.01 & 1.00 & 1.00 & 0.00 & 0.00 & 0.04 \\ 
WF$_{EC}$ & 135.40 \pm 1.20 & 3260.40 \pm 18.82 & 0.40 \pm 0.06 & 1.00 & 1.00 & 0.00 & 0.00 & 0.04 \\ 
ABACUS$_{31}$ & 199.70 \pm 3.19 & 544.60 \pm 1.20 & 0.27 \pm 0.05 & 0.18 & 0.00 & 0.15 & 0.02 & 0.00 \\
ABACUS$_{42}$ & 23.70 \pm 1.55 & 29.60 \pm 2.15 & 0.91 \pm 0.06 & 0.01 & 0.00 & 0.00 & 0.00 & 0.00 \\
EMCD & 219.00 & 3306.00 & 0.23 & 1.00 & 1.00 & 0.00 & 0.00 & 0.02 \\ 
PMM$_l$ & 10.00 & 12965.40 \pm 1963.48 & 0.21 \pm 0.18 & 1.00 & 1.00 & 0.00 & 0.00 & 0.14 \pm 0.06 \\
PMM$_h$ & 10.00 & 1864.80 \pm 99.31 & 0.94 \pm 0.03 & 1.00 & 1.00 & 0.00 & 0.00 & 0.00 \\
ML-CPM$_{31}$ & 63.00 & 219.00 & 0.03 & 0.02 & 0.00 & 0.01 & 0.00 & 0.00 \\
Infomap$_o$ & 293.70 \pm 4.02 & 10336.20 \pm 409.14 & 0.12 \pm 0.06 & 1.00 & 0.81 & 0.18 & 0.18 & 0.04 \\
GLouvain$_l$ & 163.80 \pm 2.44 & 3258.50 \pm 30.33 & 0.29 \pm 0.06 & 1.00 & 0.89 & 0.10 & 0.00 & 0.03 \\
GLouvain$_h$ & 137.19 \pm 1.16 & 3261.00 \pm 17.02 & 0.37 \pm 0.05 & 1.00 & 1.00 & 0.00 & 0.00 & 0.04 \\
MDLPA & 287.80 \pm 12.55 & 4411.80 \pm 533.45 & 0.18 \pm 0.02 & 1.00 & 1.00 & 0.00 & 0.00 & 0.03 \\

		\bottomrule
	\end{tabular}}
	\subcaption*{Rattus\rev{. l = 6, a =  2640, e =  3956}}
\label{table:global_real_stats:rattus}
\end{table}

\subsubsection{Accuracy analysis}
\label{sssec:results_glob_acc}

With the aim of answering \rev{\textbf{Q1}} (i.e., ``To what extent are the evaluated methods able to detect ground truth communities?'', cf. Section~\ref{sec:experiments}), we perform here an extensive quantitative analysis about the accuracy obtained by each method with respect to ground truth communities. 
For real-world networks, only two of them have an available ground truth\revm{:} AUCS (i.e., affiliations to research groups) and DKPol (i.e., affiliation to political parties). All synthetic networks  come with controlled ground truth. 

\begin{figure}[!htpb]
	\centering
	\begin{subfigure}[t]{0.45\textwidth}
		\centering
		\captionsetup{justification=centering}
		\includegraphics[width=\textwidth]{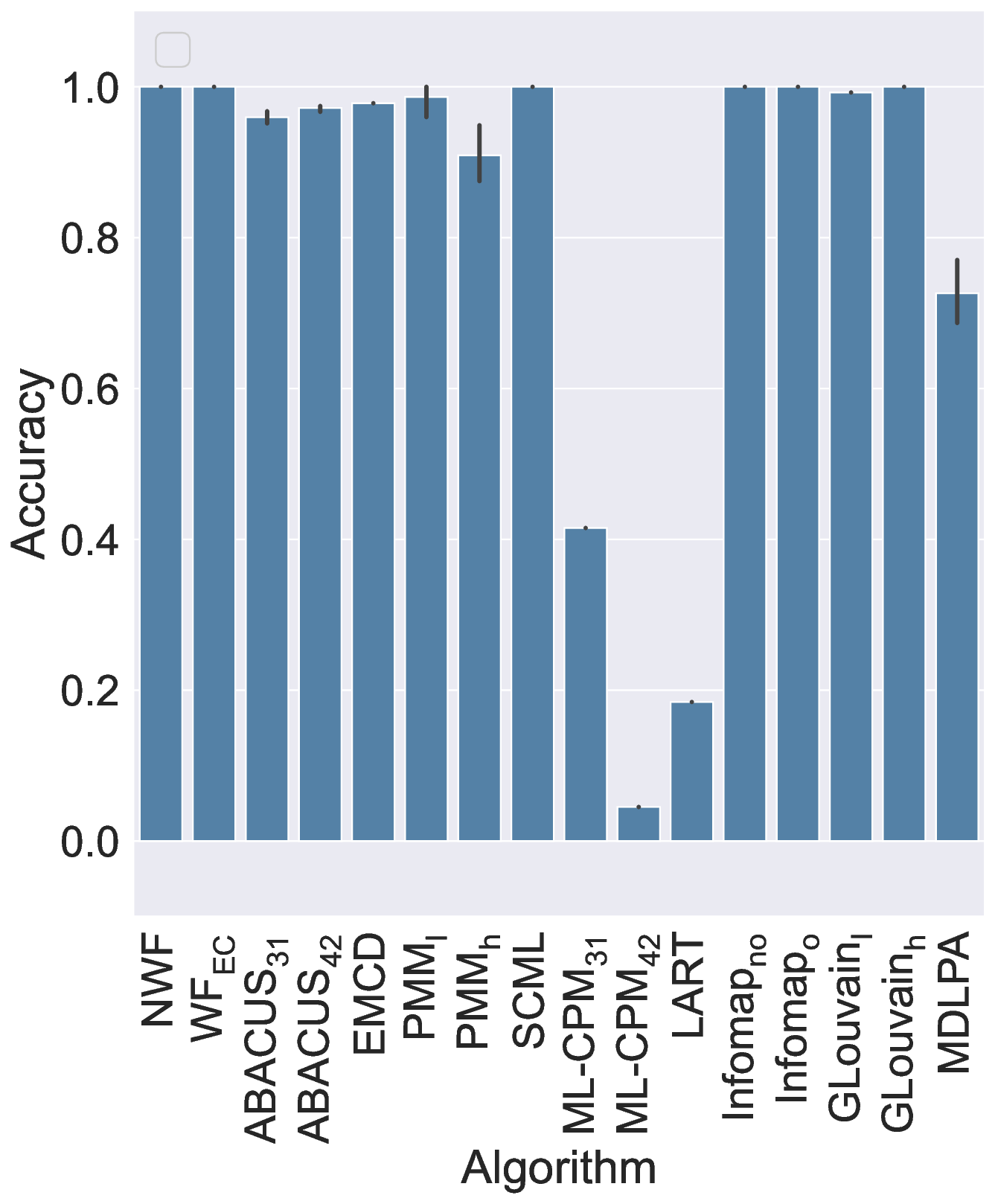}
		\caption{Pillar Equal Partitioning  (PEP)}
	\end{subfigure}
	\begin{subfigure}[t]{0.45\textwidth}
		\centering
		\captionsetup{justification=centering}
		\includegraphics[width=\textwidth]{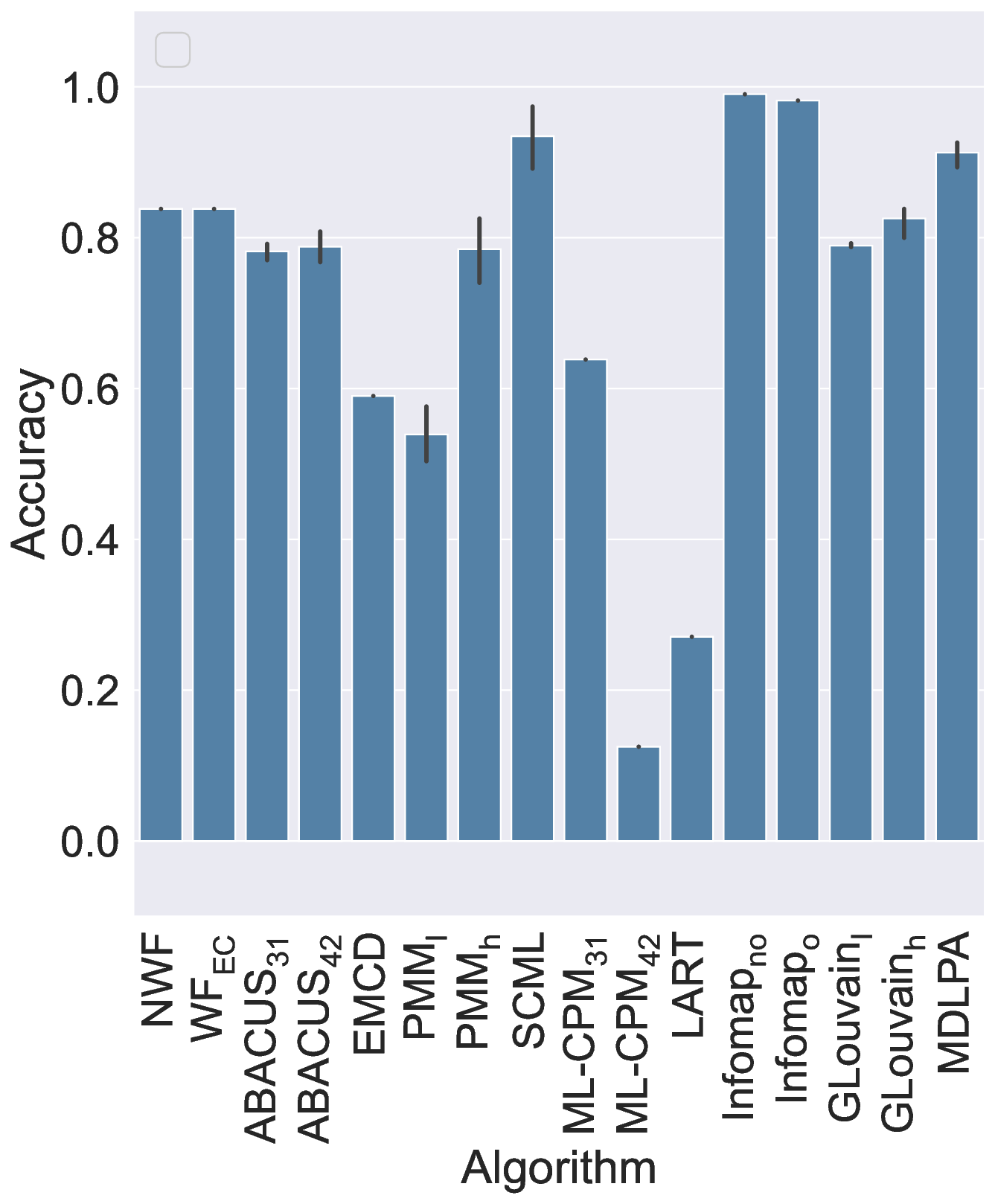}
		\caption{Pillar Non-equal Partitioning (PNP)}
	\end{subfigure}
		\begin{subfigure}[t]{0.45\textwidth}
		\centering
		\captionsetup{justification=centering}
		\includegraphics[width=\textwidth]{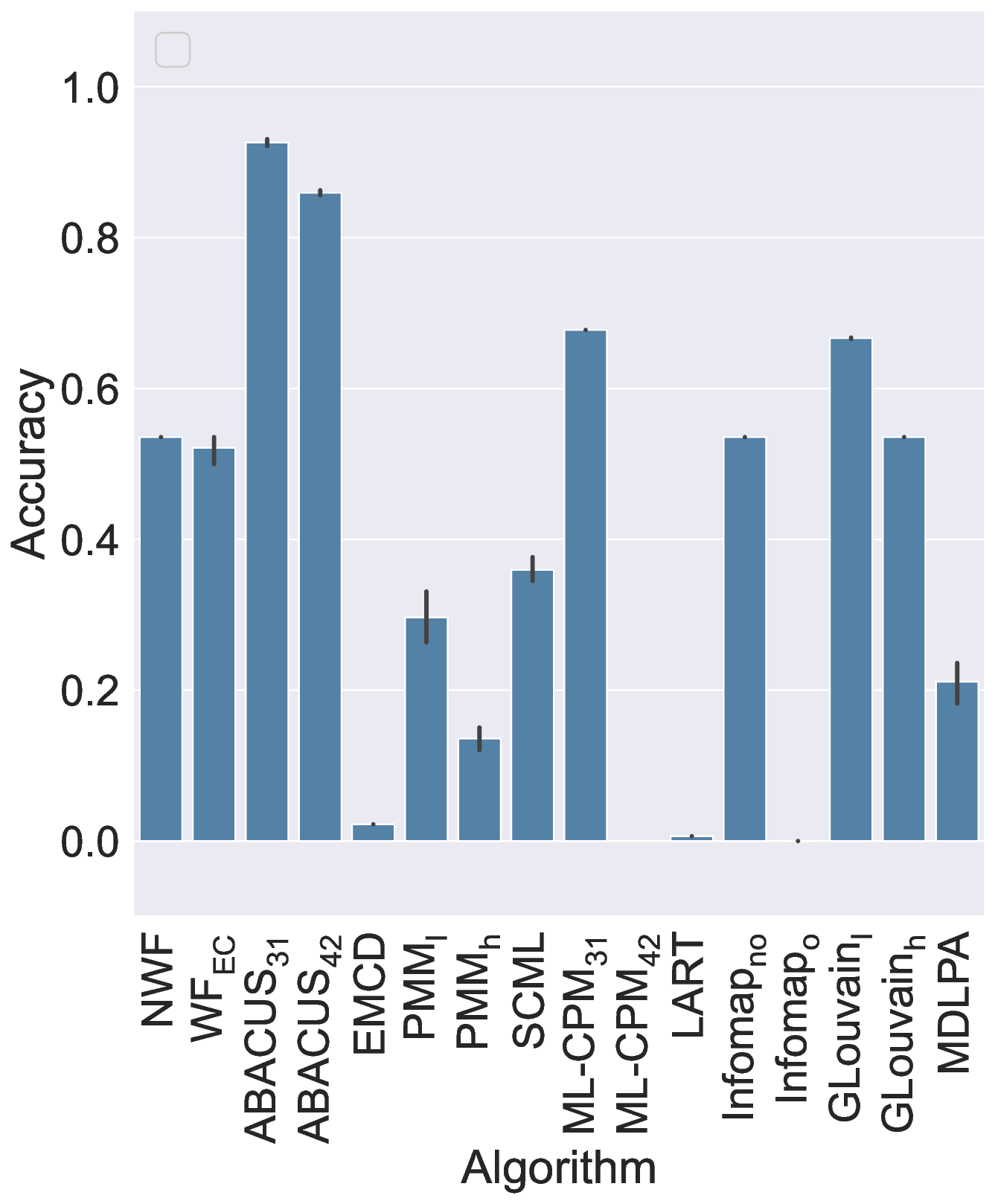}
		\caption{Semi-pillar Equal Partitioning (SEP)}
	\end{subfigure}
 	\begin{subfigure}[t]{0.45\textwidth}
 		\centering
 		\captionsetup{justification=centering}
 		\includegraphics[width=\textwidth]{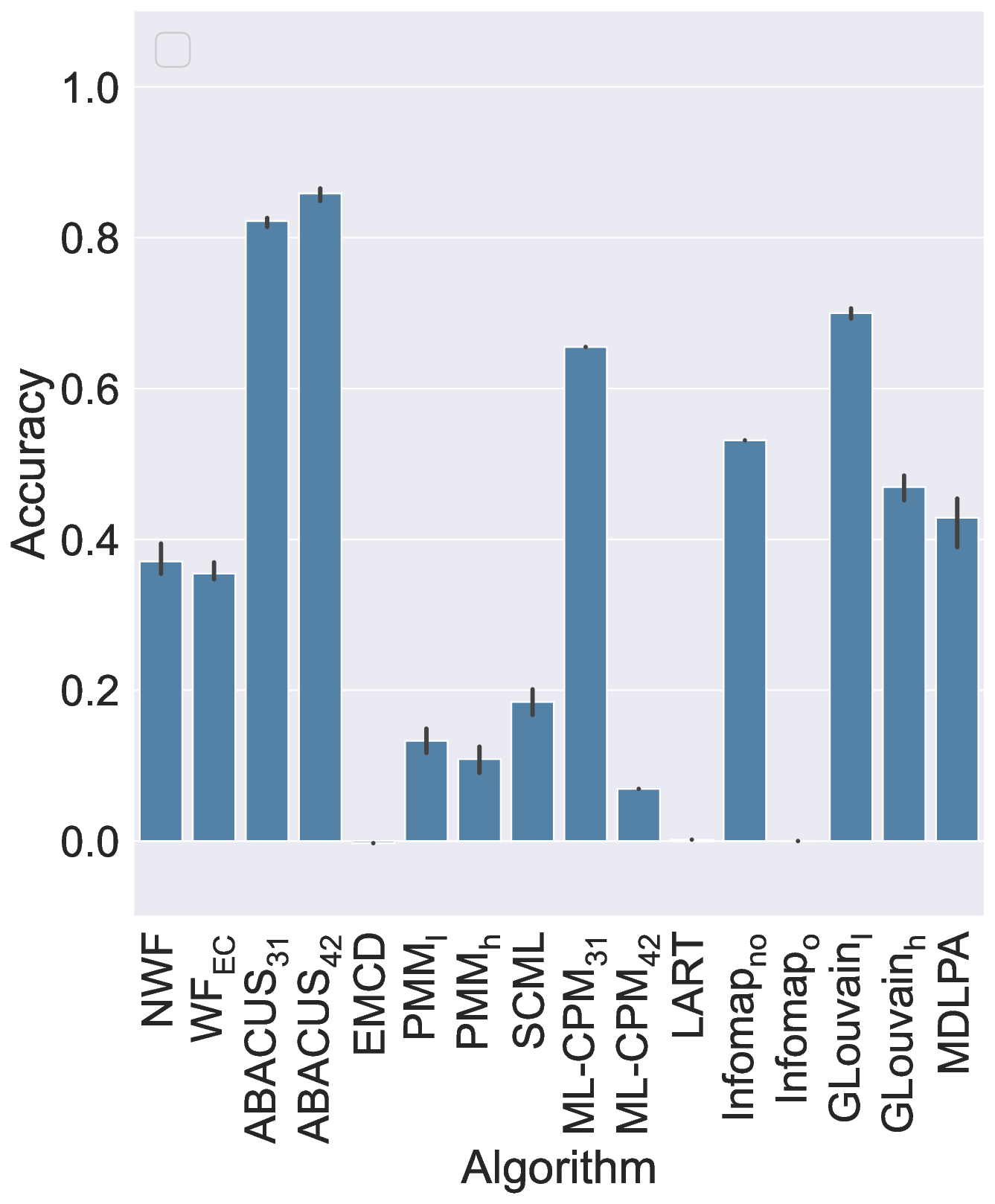}
 		\caption{Semi-pillar Non-equal Partitioning (SNP)}
 	\end{subfigure}
	\caption{Accuracy with respect to a ground truth,  Omega index, pillar communities (100 actors)}
	\label{fig:accuracy_global_syn_1}
\end{figure}

 \begin{figure}[!htpb]
 	\centering
 	\begin{subfigure}[t]{0.45\textwidth}
 		\centering
 		\captionsetup{justification=centering}
 		\includegraphics[width=\textwidth]{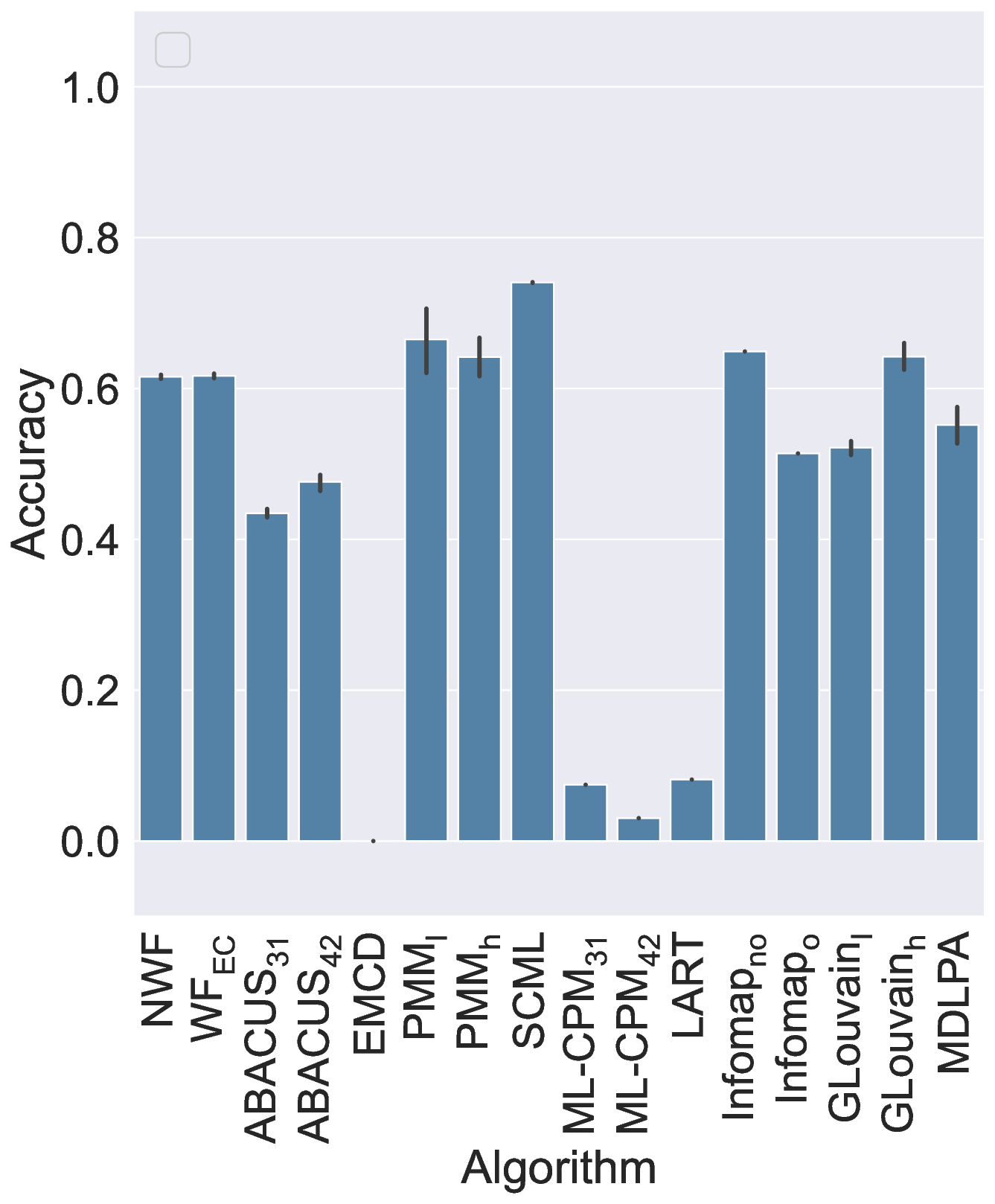}
 		\caption{Pillar Equal Overlapping (PEO)}
 	\end{subfigure}
 	\begin{subfigure}[t]{0.45\textwidth}
 		\centering
 		\captionsetup{justification=centering}
 		\includegraphics[width=\textwidth]{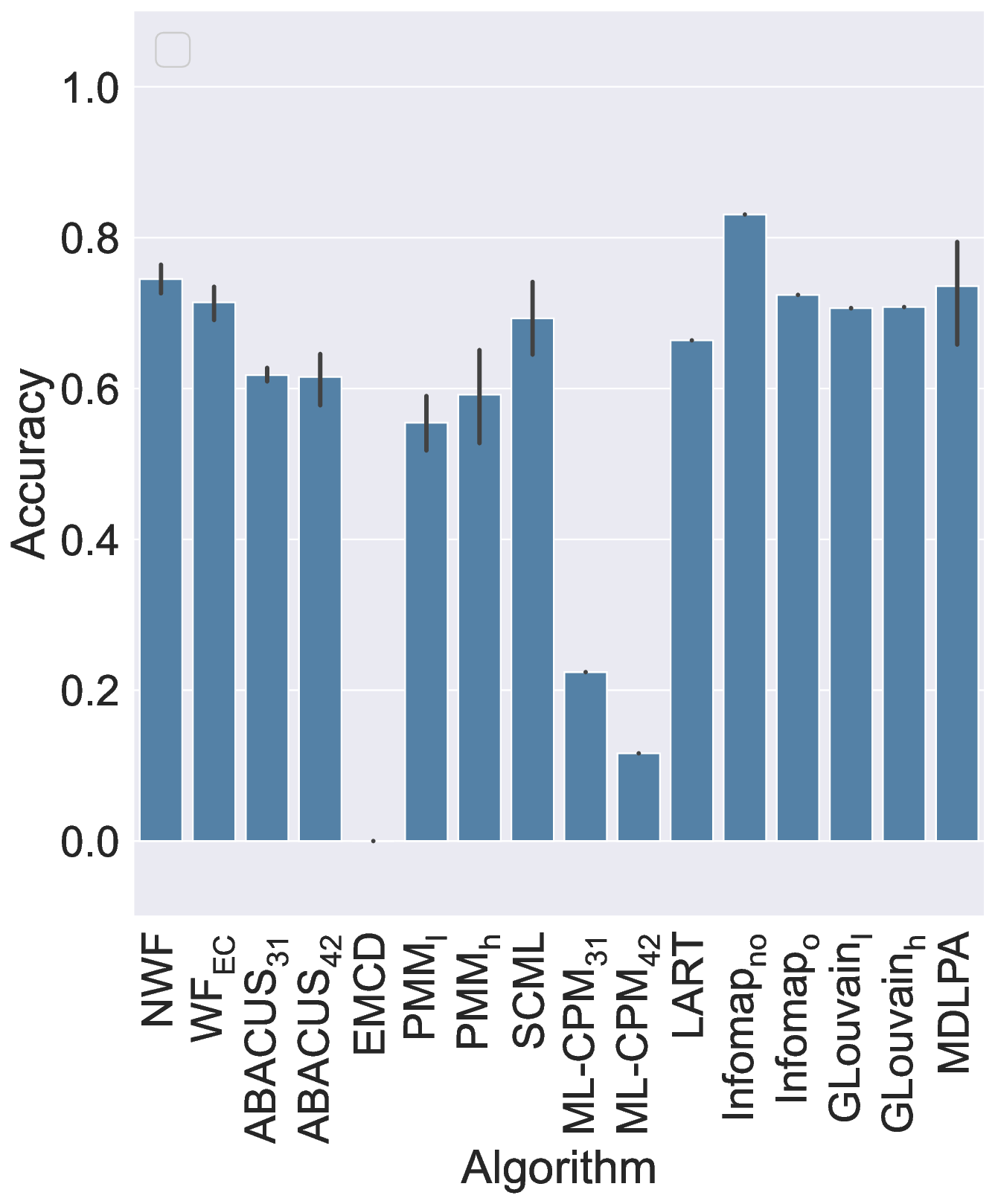}
 		\caption{Pillar Non-equal Overlapping (PNO)}
 	\end{subfigure}
 		\begin{subfigure}[t]{0.45\textwidth}
 		\centering
 		\captionsetup{justification=centering}
 		\includegraphics[width=\textwidth]{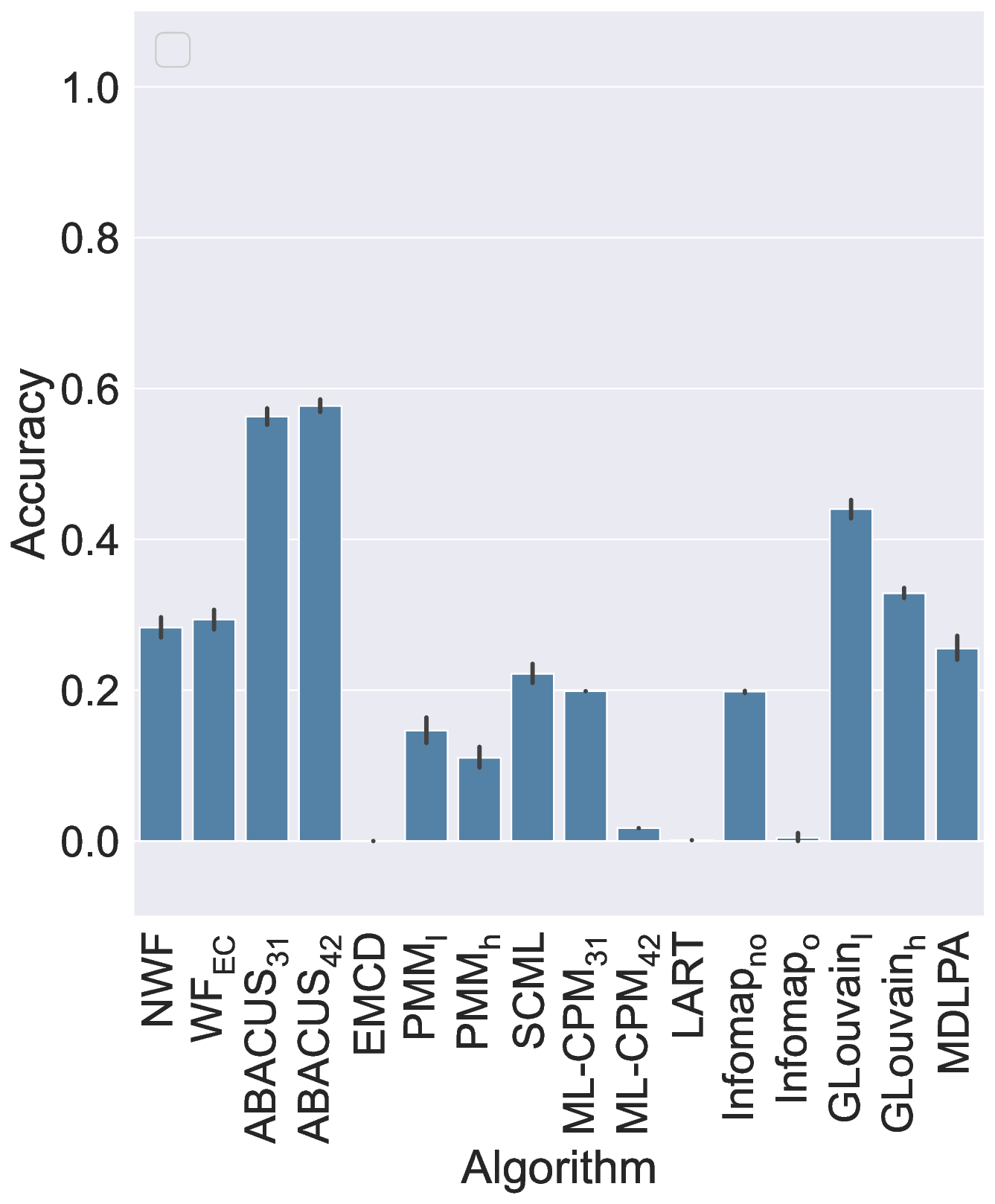}
 		\caption{Semi-pillar Equal Overlapping (SEO)}
 	\end{subfigure}
 	\begin{subfigure}[t]{0.45\textwidth}
 		\centering
 		\captionsetup{justification=centering}
 		\includegraphics[width=\textwidth]{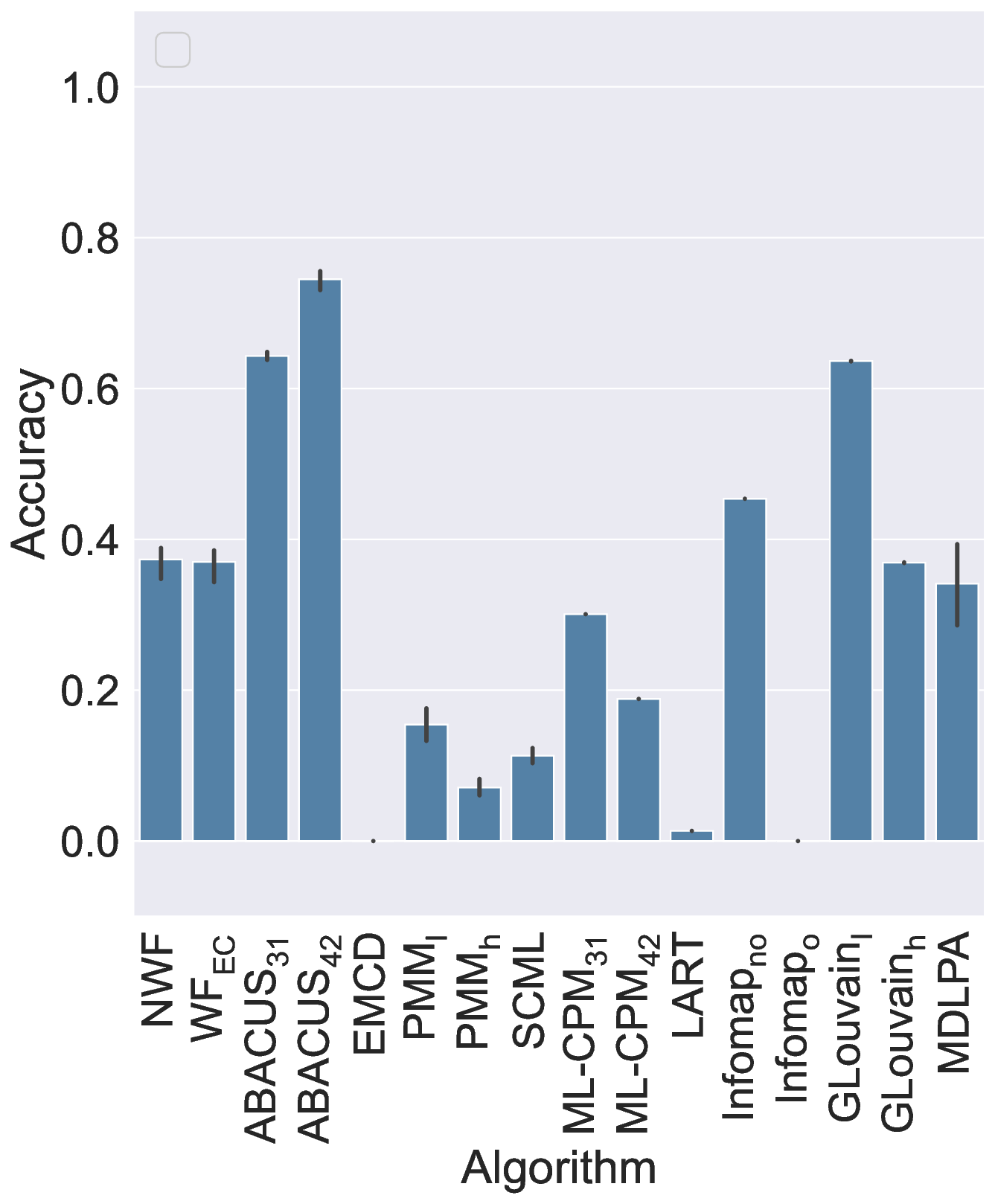}
 		\caption{Semi-pillar Non-equal Overlapping (SEO)}
 	\end{subfigure}
 	\caption{Accuracy with respect to a ground truth,  Omega index, overlapping communities (100 actors)}
 	\label{fig:accuracy_global_syn_2}
 \end{figure}
	
\begin{figure}[!htpb]
	\centering
	\begin{subfigure}[t]{0.45\textwidth}
		\centering
		\captionsetup{justification=centering}
		\includegraphics[width=\textwidth]{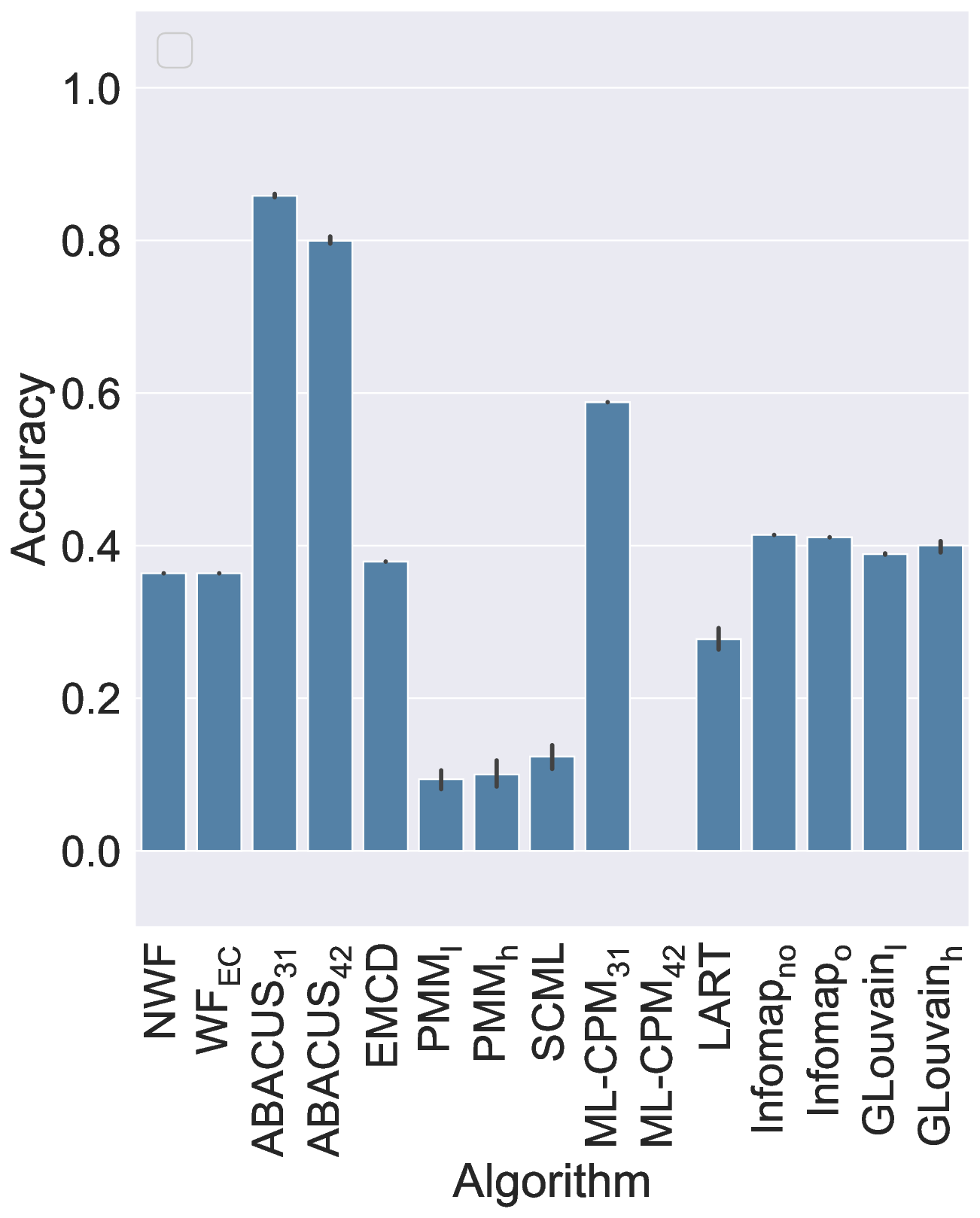}
		\caption{Mixed (MIX)}
	\end{subfigure}
	\begin{subfigure}[t]{0.45\textwidth}
		\centering
		\captionsetup{justification=centering}
		\includegraphics[width=\textwidth]{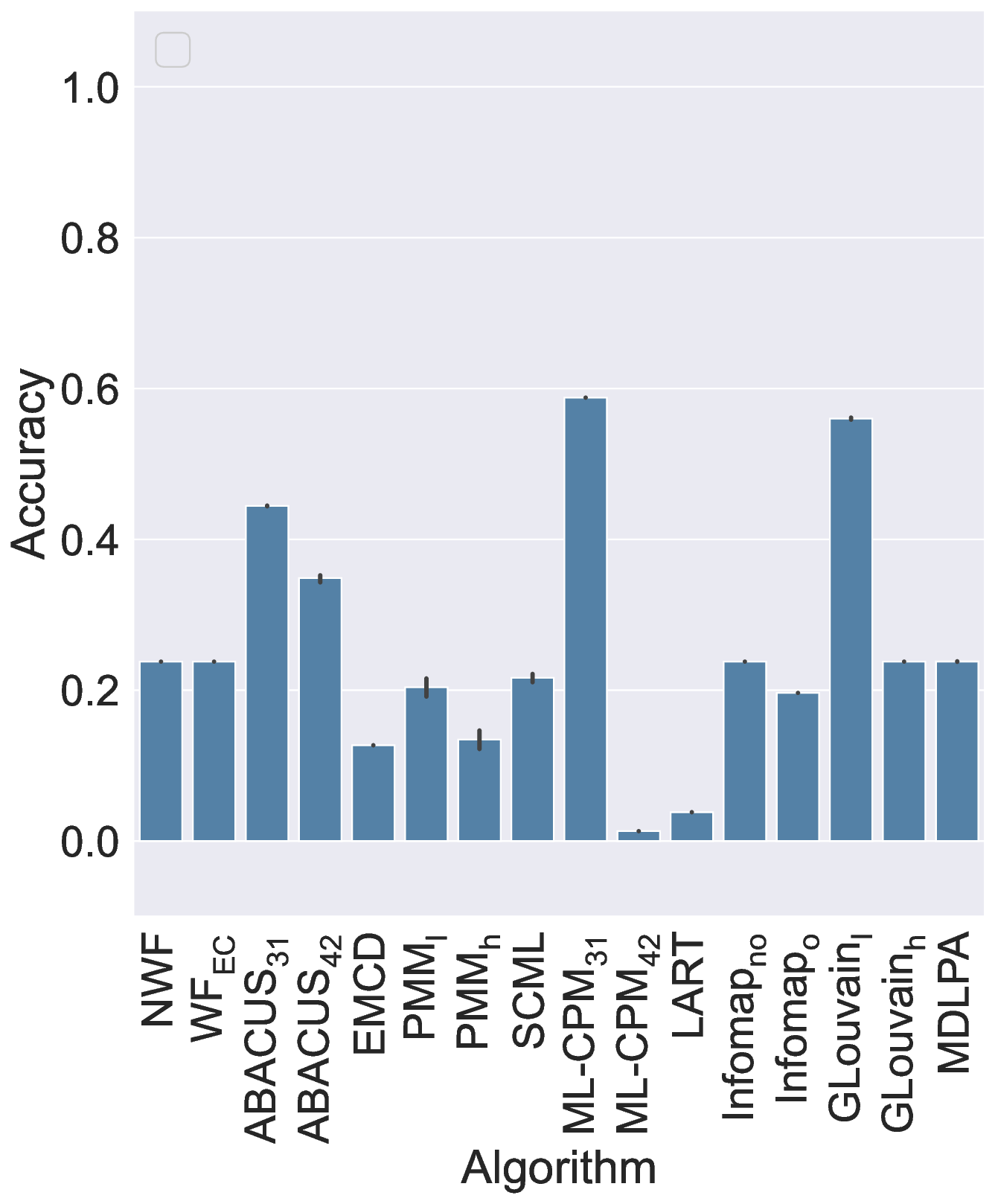}
		\caption{Hierarchical (HIE)}
	\end{subfigure}
	\caption{Accuracy with respect to a ground truth,  Omega index, mixed and hierarchical communities (100 actors)}
	\label{fig:accuracy_global_syn_3}
\end{figure}

\begin{figure}[!htpb]
	\centering
	\begin{subfigure}[t]{0.45\textwidth}
		\centering
		\includegraphics[width=\textwidth]{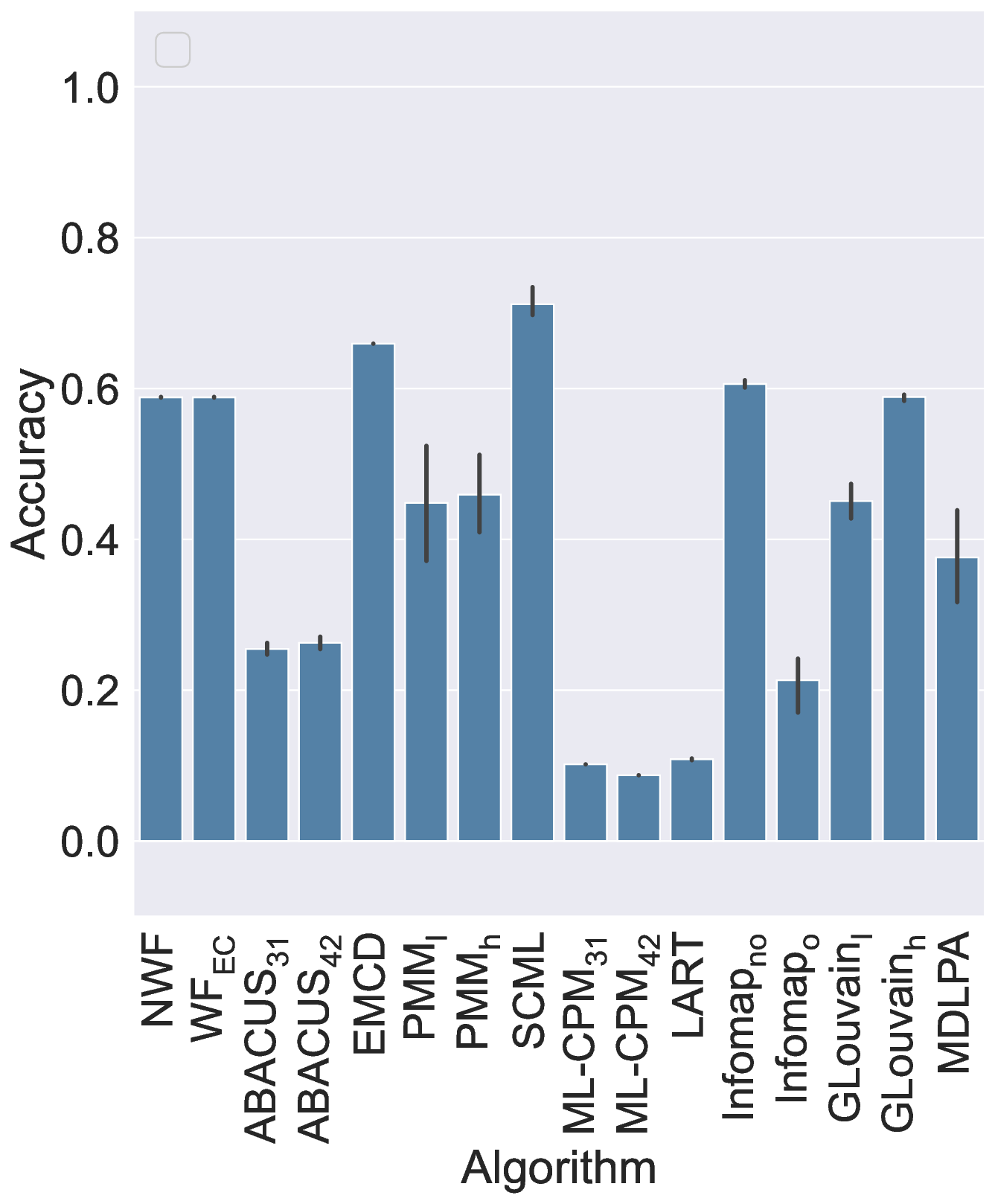}
		\caption{AUCS}
	\end{subfigure}
	\begin{subfigure}[t]{0.45\textwidth}
		\centering
		\includegraphics[width=\textwidth]{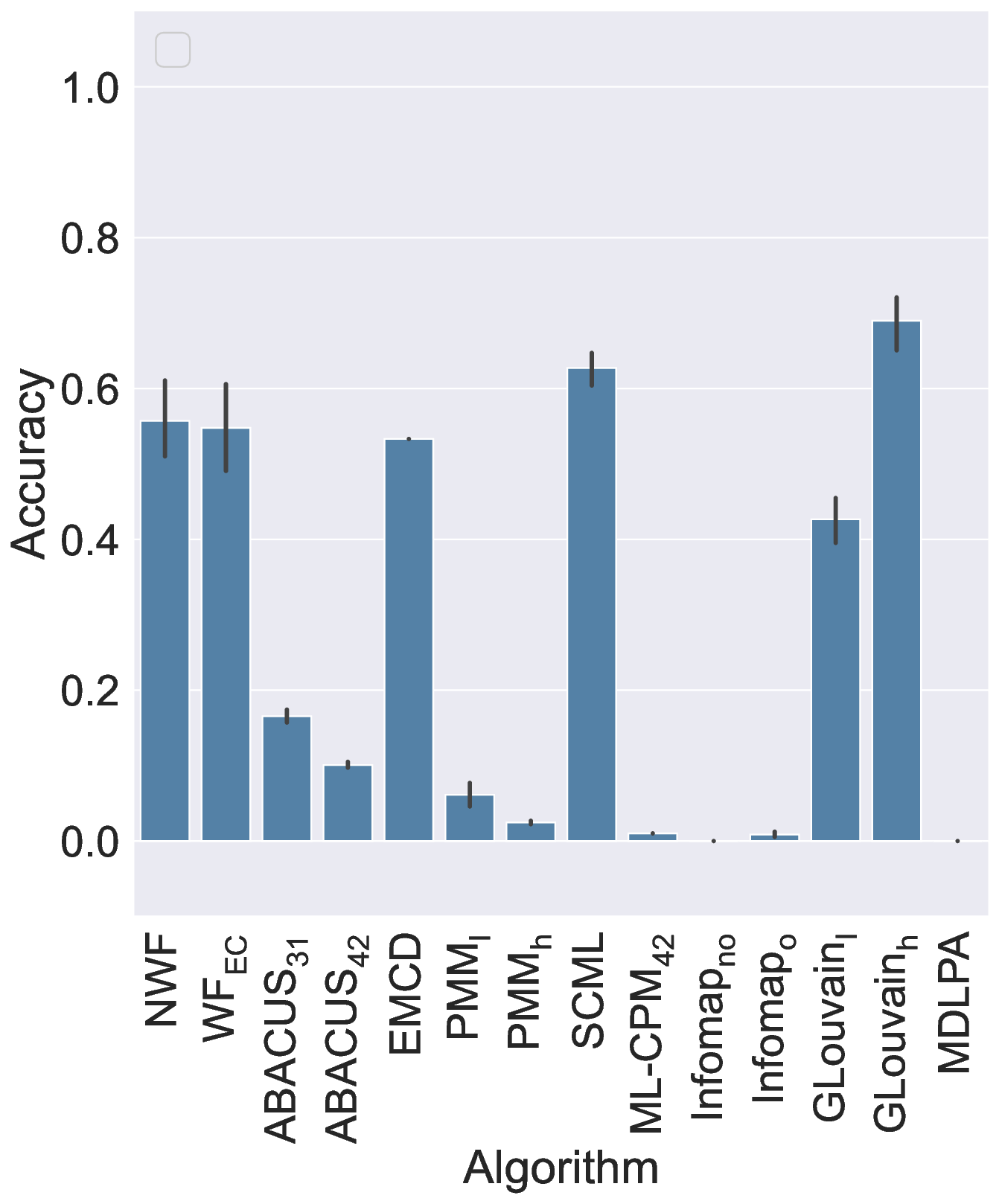}
		\caption{DKPol}
	\end{subfigure}
	\caption{Accuracy with respect to a ground truth for real-world networks, measured using Omega index}
	\label{fig:accuracy_global_r}
\end{figure}

\revm{Our results on networks with 100 actors are} \rev{reported in Figures~\ref{fig:accuracy_global_syn_1}, \ref{fig:accuracy_global_syn_2}, \ref{fig:accuracy_global_syn_3}, while \ref{fig:accuracy_global_r} shows results on real networks. At the end of this section we also show results on larger networks, which generally confirm these results highlighting a few additional behaviors. \revm{From these figures we can see how the main element playing a role in determining the accuracy of the methods is the pillar nature of the community structure.}

In the case of Pillar Equal Partitioning (PEP) structures almost all the methods perform very well, with \textsf{WF$_{EC}$}, \textsf{NWF}, \textsf{Infomap} and \textsf{GLouvain} (both versions) reaching perfect accuracy. Overall, only \textsf{ML-CPM} (both versions) and \textsf{LART} score below 0.5. In the first case, the strict rules imposed by its parameters explain the performance, for the latter, as we saw in Table~\ref{table:global_real_stats:2} \textsf{LART} does not seem to be able to group a considerable number of nodes into communities. Similar patterns, even if with worse levels of accuracy, are visible for all the Pillar structures (PNP, PEO, PNO). Minor notable differences are present in the Pillar Non-equal Partitioning structure where \textsf{Infomap} (both variations) performs better than all the other methods (that also score above $0.8$). Despite the positive results for many methods, one could easily ask if in the general context of pillar community structures proper multilayer methods are necessary since the same (good) results can be achieved with flattening-based methods. }


\rev{The more the network moves away from a pillar structure (with semi-pillar, mixed and hierarchical structures) the worse the results are among most of the methods. A notable exception is \textsf{ABACUS} that, regardless of the variation, keeps performing above the average with Semi-Pillar and Mixed Communities\revm{, with \textsf{\revm{ML-}CPM$_{31}$} also performing better than most other methods on Semi-Pillar structures}. Hierarchical structures are extremely challenging for all the methods with the notable exceptions of \textsf{\revm{ML-}CPM$_{31}$} and \textsf{GLouvain$_{l}$}, although GLouvain is finding communities on individual layers and thus it is not clearly identifying any hierarchy spanning multiple layers.}


\rev{The reason why some methods have an Omega index around 0 is that in these cases these methods only find one or two large communities. This is not surprising if we consider the structures of some synthetic datasets. In the overlapping community structures all the communities are kept together by their overlapping parts, and in the semi-pillar structures the well-separated semi-pillar communities spanning a subset of the layers result connected by the different communities on the remaining layers.}


\rev{These results may indicate that, even though for simple Pillar Equal Partitioning structures multilayer methods do not seem to provide any real advantage over flattening-based methods, more complex structures show how proper multilayer methods can perform better than flattening-based methods.}

\rev{Figure~\ref{fig:accuracy_global_r} reports on the accuracy obtained by the evaluated methods on real-world networks.
It can be observed how accuracy values are relatively low on both networks for all methods, i.e., with Omega index always below $0.8$ and often below $0.5$. More interestingly, the best performing methods do not entirely overlap with the methods that perform the best with the synthetic data.
On AUCS, the best performing method is \textsf{SCML} ($0.70$), followed by \textsf{EMCD}. }

\rev{The results are even more variable on DKPol, where many methods show low results.\footnote{Zero values are a result of identifying a clustering constituted of only one giant component (i.e with \textsf{Infomap$_{no}$)}. The result of \textsf{\revm{ML-}CPM$_{31}$}  is not reported as the execution took more than 24 hours.} An exception to this are the two variants of \textsf{GLouvain}, reaching accuracies of $0.68$ (\textsf{GLouvain$_{h}$})  and $0.43$ (\textsf{GLouvain$_{l}$}) respectively. \textsf{SCML}, \textsf{NWF}, \textsf{WF$_{EC}$} and \textsf{EMCD} also perform relatively well with scores \revm{around} $0.6$. }

As a final remark, the difference in performance between real-world and synthetic networks confirms how the ``ideal'' concept of community, i.e., the one based on topological density that is used to build the synthetic ones and to drive the detection process of the methods,  is often far from the ground truth communities observed in real cases (which are, in turn, often questionable and subjective). This is a well known problem in the community detection field, and poses challenges in both ways, i.e., concerning the need to design both more powerful methods and more reliable ground truths.

\begin{figure}[!htpb]
	\centering
	\begin{subfigure}[t]{0.45\textwidth}
		\centering
		\captionsetup{justification=centering}
		\includegraphics[width=\textwidth]{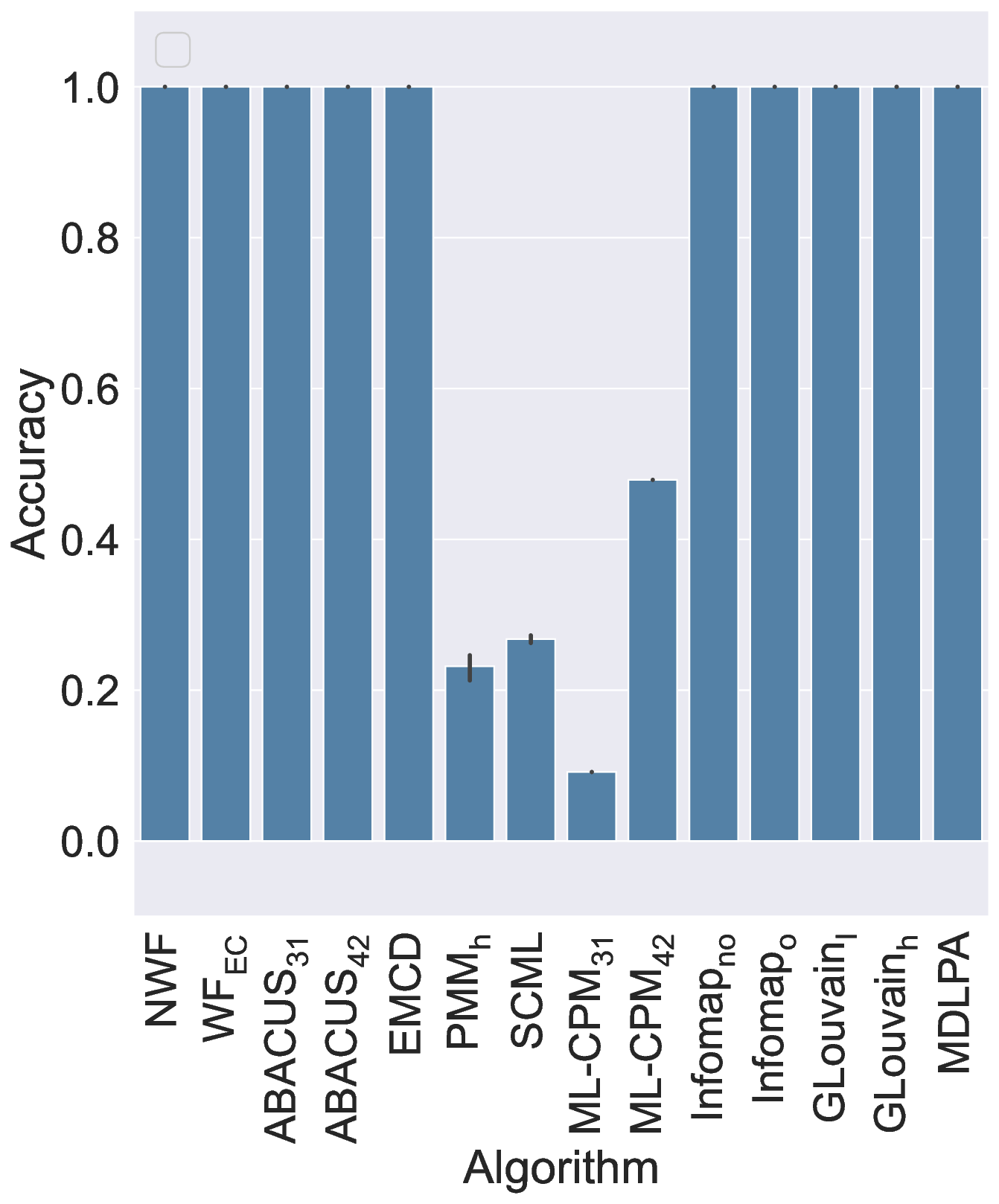}
		\caption{Pillar Equal Partitioning  (PEP)}
	\end{subfigure}
	\begin{subfigure}[t]{0.45\textwidth}
		\centering
		\captionsetup{justification=centering}
		\includegraphics[width=\textwidth]{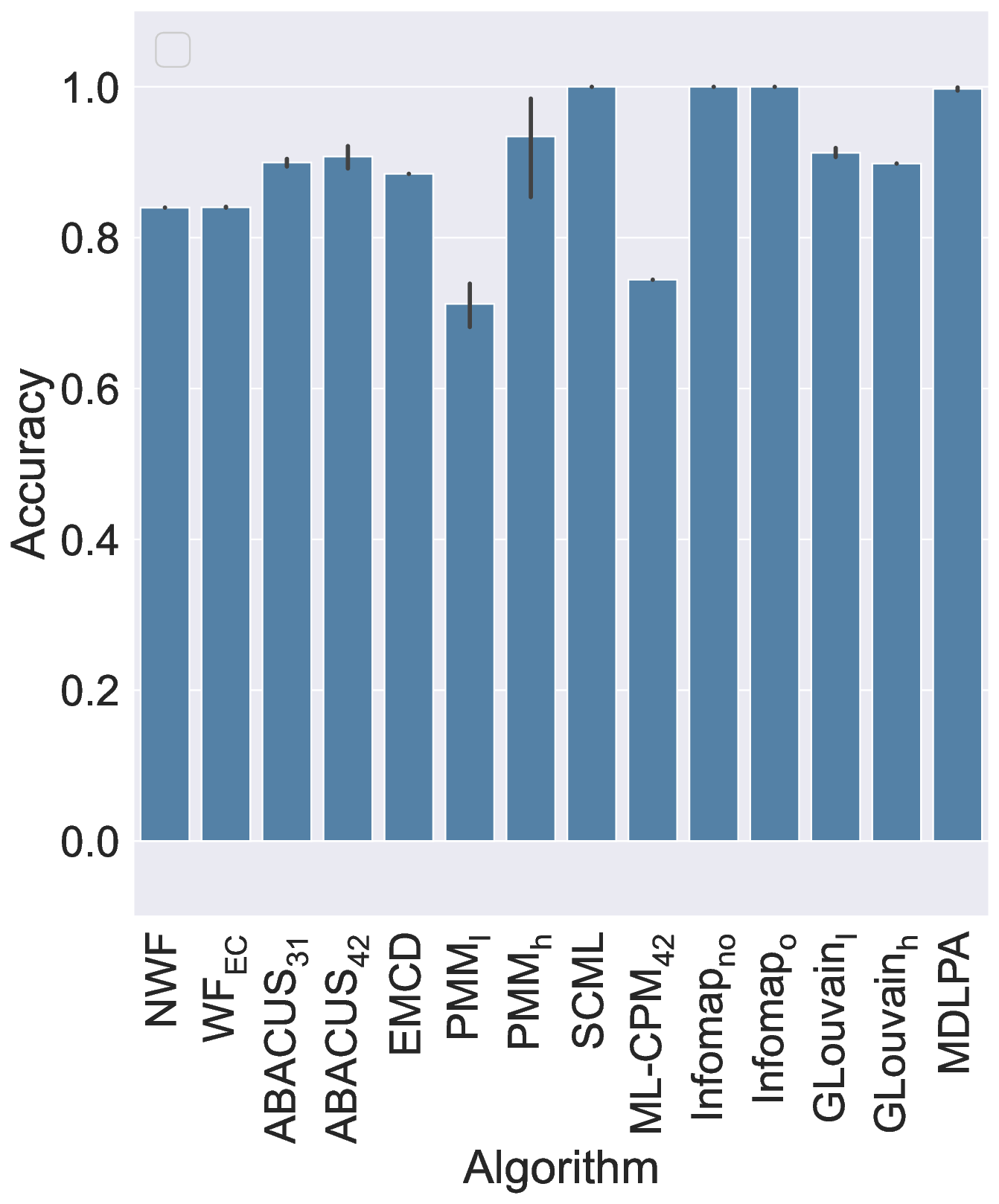}
		\caption{Pillar Non-equal Partitioning (PNP)}
	\end{subfigure}
		\begin{subfigure}[t]{0.45\textwidth}
		\centering
		\captionsetup{justification=centering}
		\includegraphics[width=\textwidth]{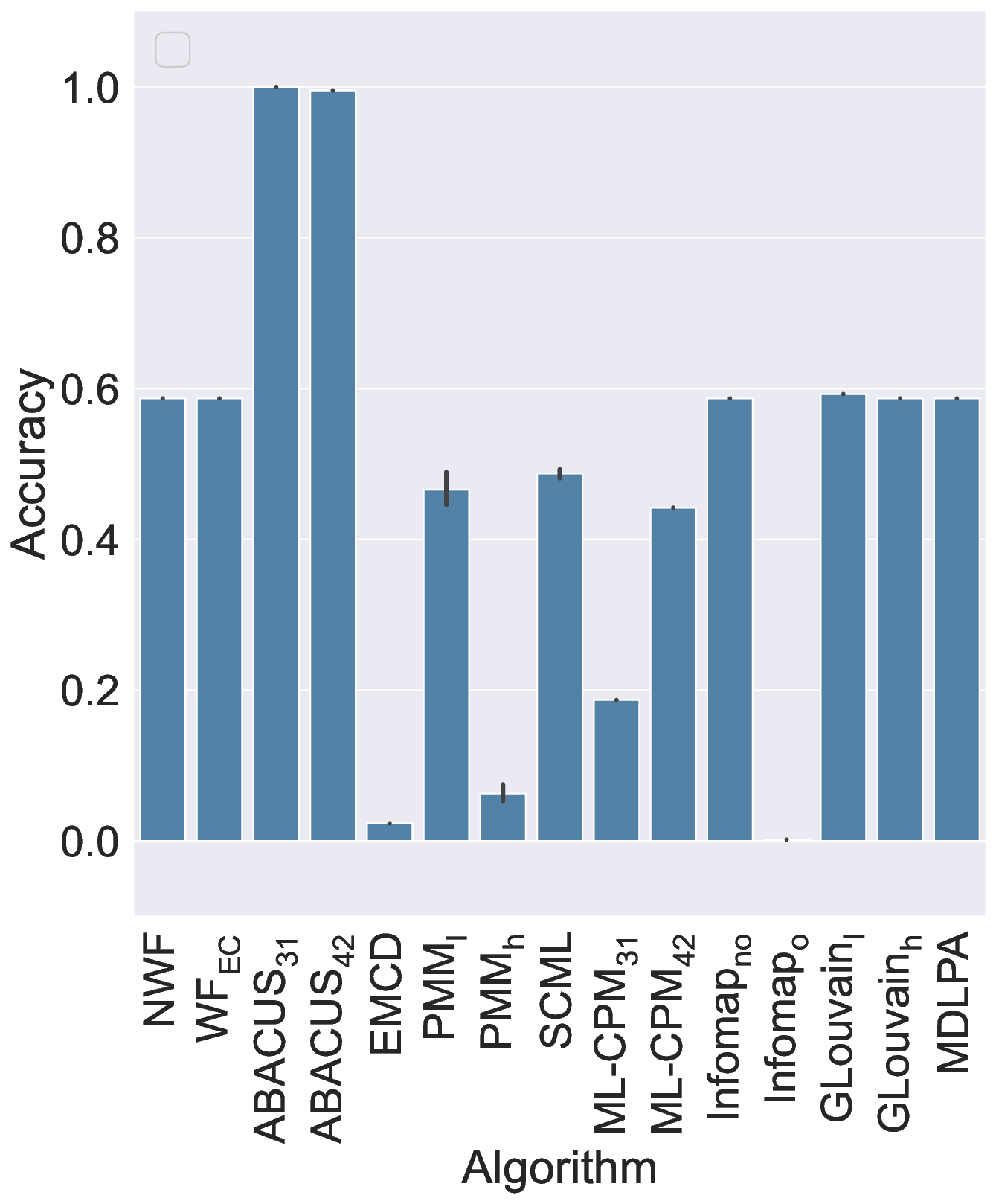}
		\caption{Semi-pillar Equal Partitioning (SEP)}
	\end{subfigure}
 	\begin{subfigure}[t]{0.45\textwidth}
 		\centering
 		\captionsetup{justification=centering}
 		\includegraphics[width=\textwidth]{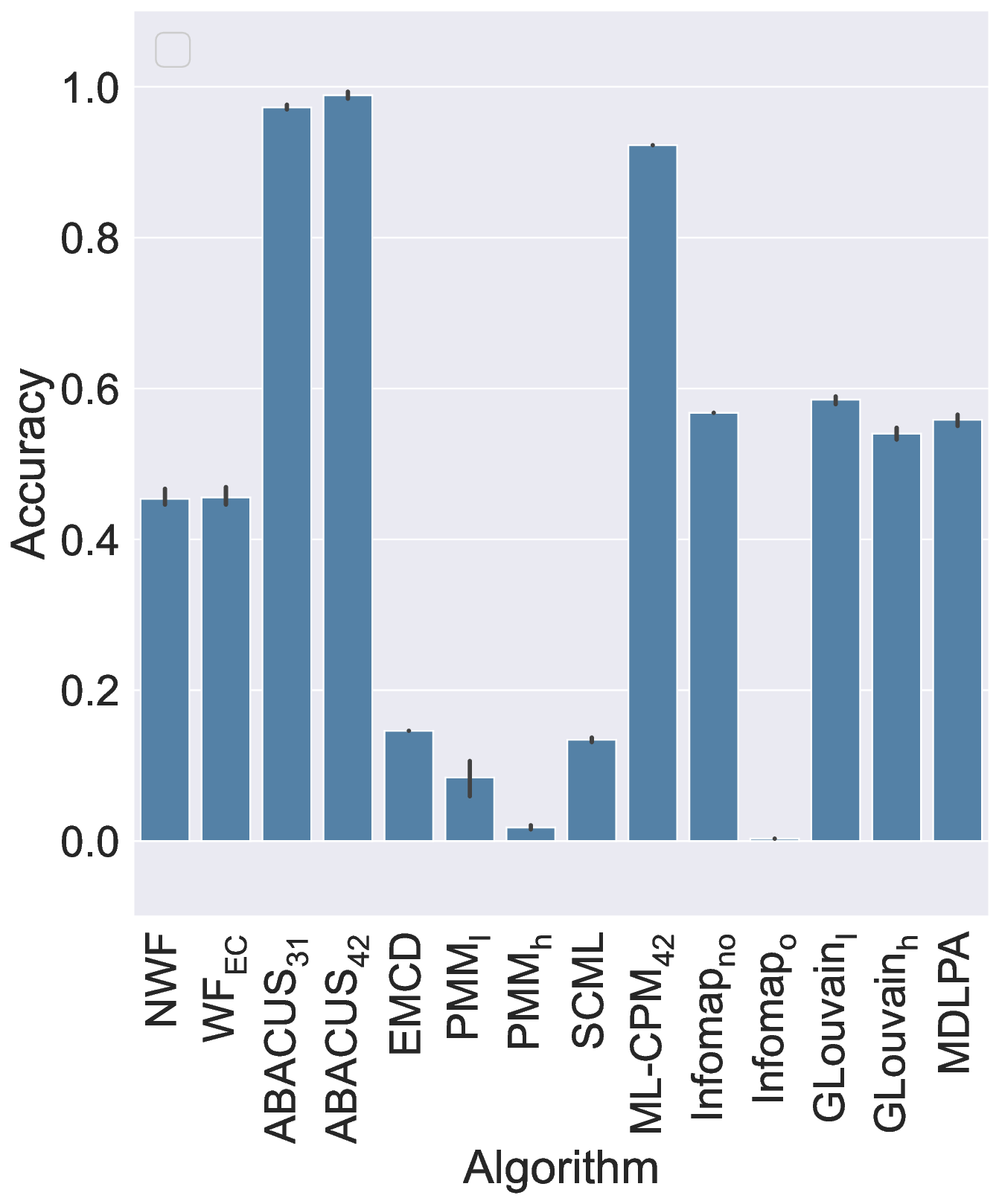}
 		\caption{Semi-pillar Non-equal Partitioning (SNP)}
 	\end{subfigure}
	\caption{Accuracy with respect to a ground truth,  Omega index, pillar communities (500 actors)}
	\label{fig:accuracy_global_syn_1_500}
\end{figure}

 \begin{figure}[!htpb]
 	\centering
 	\begin{subfigure}[t]{0.45\textwidth}
 		\centering
 		\captionsetup{justification=centering}
 		\includegraphics[width=\textwidth]{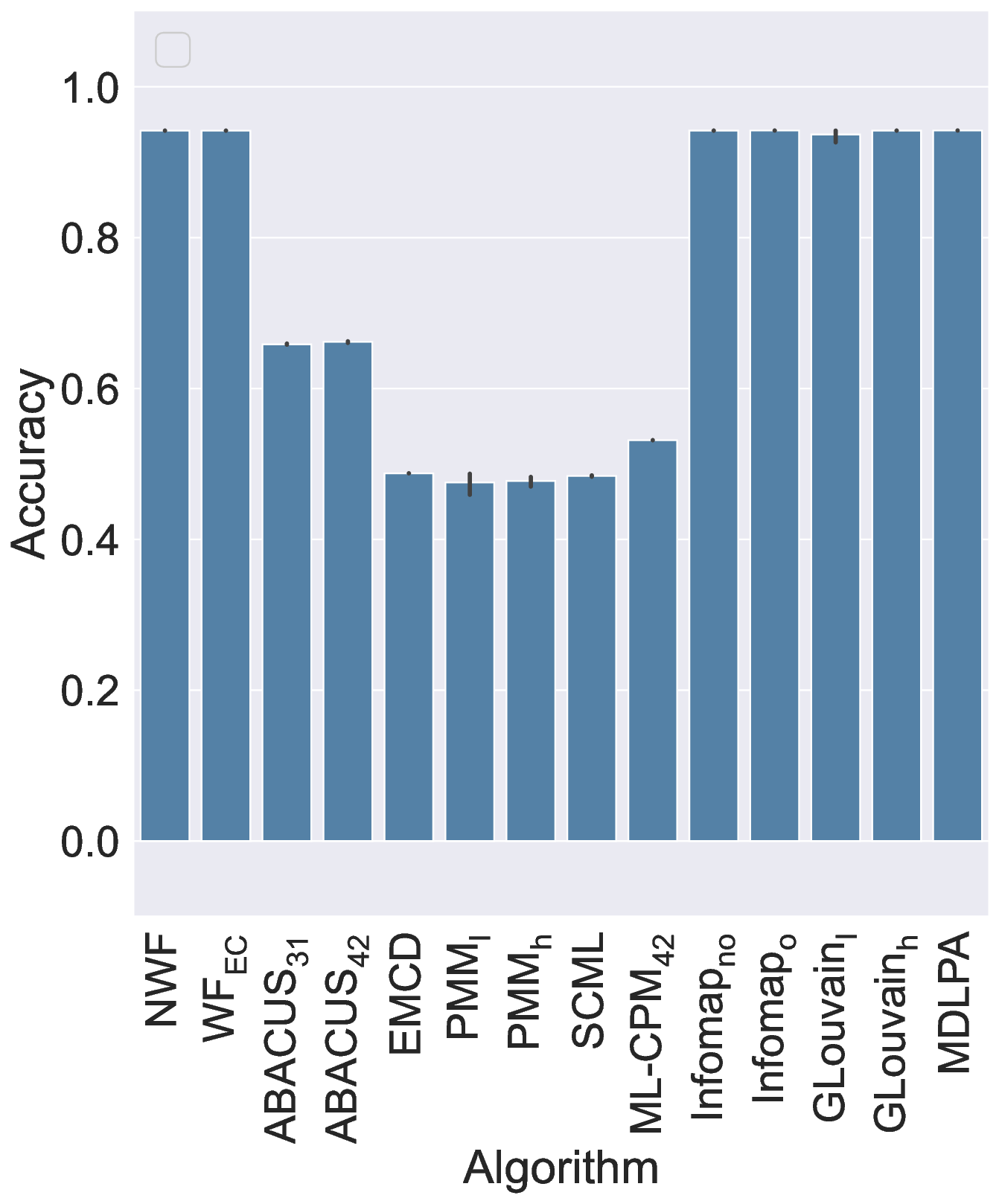}
 		\caption{Pillar Equal Overlapping (PEO)}
 	\end{subfigure}
 	\begin{subfigure}[t]{0.45\textwidth}
 		\centering
 		\captionsetup{justification=centering}
 		\includegraphics[width=\textwidth]{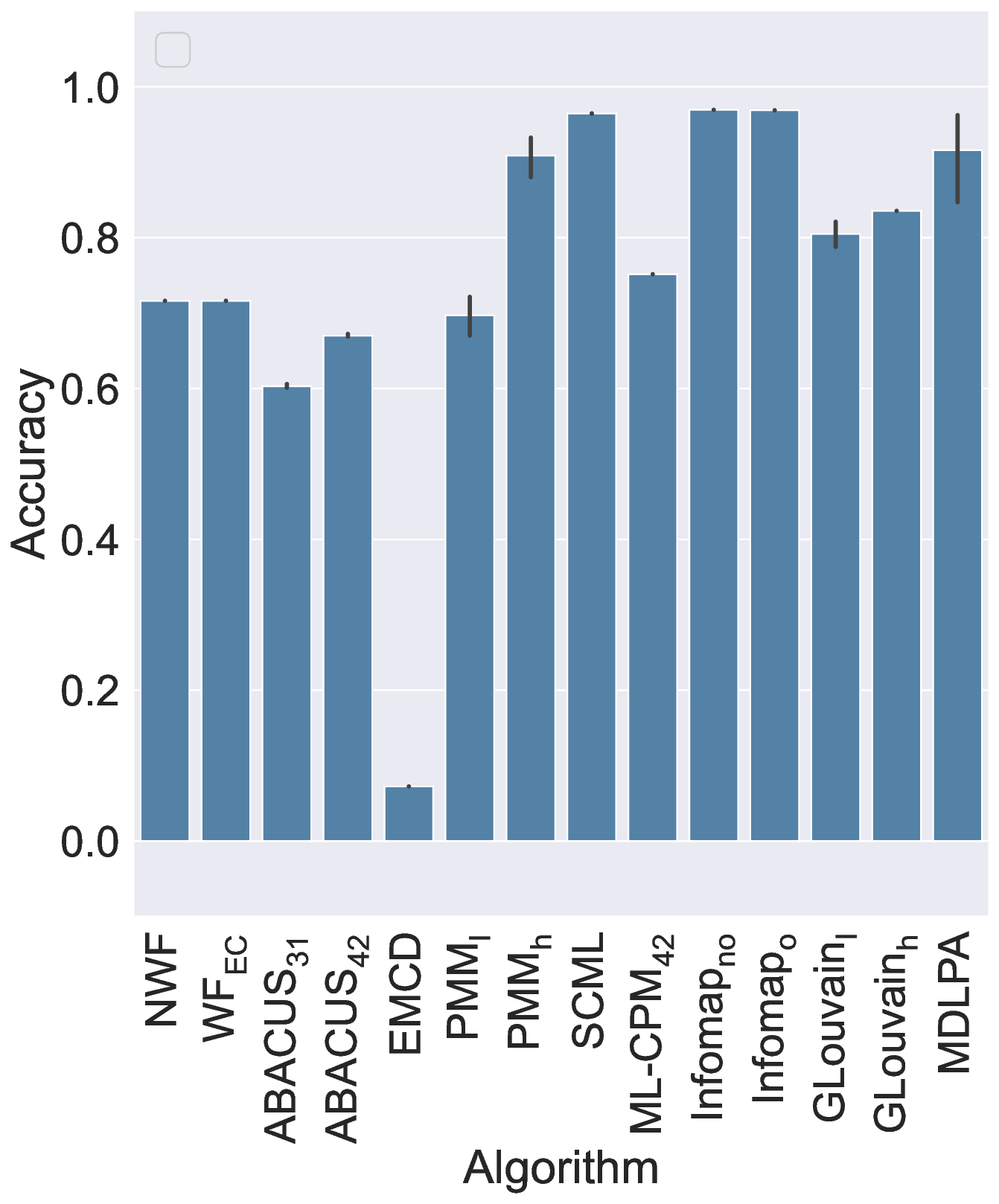}
 		\caption{Pillar Non-equal Overlapping (PNO)}
 	\end{subfigure}
 		\begin{subfigure}[t]{0.45\textwidth}
 		\centering
 		\captionsetup{justification=centering}
 		\includegraphics[width=\textwidth]{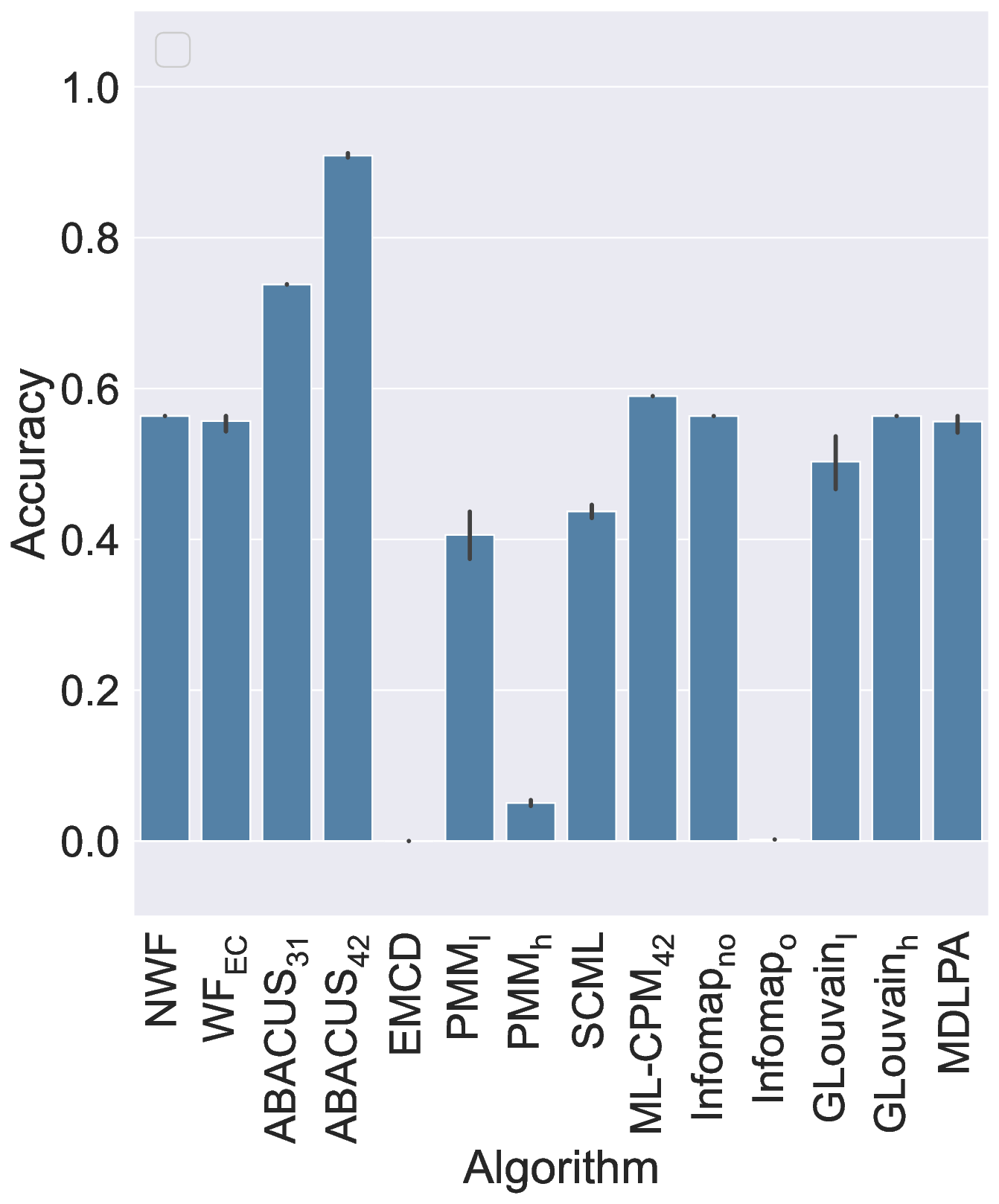}
 		\caption{Semi-pillar Equal Overlapping (SEO)}
 	\end{subfigure}
 	\begin{subfigure}[t]{0.45\textwidth}
 		\centering
 		\captionsetup{justification=centering}
 		\includegraphics[width=\textwidth]{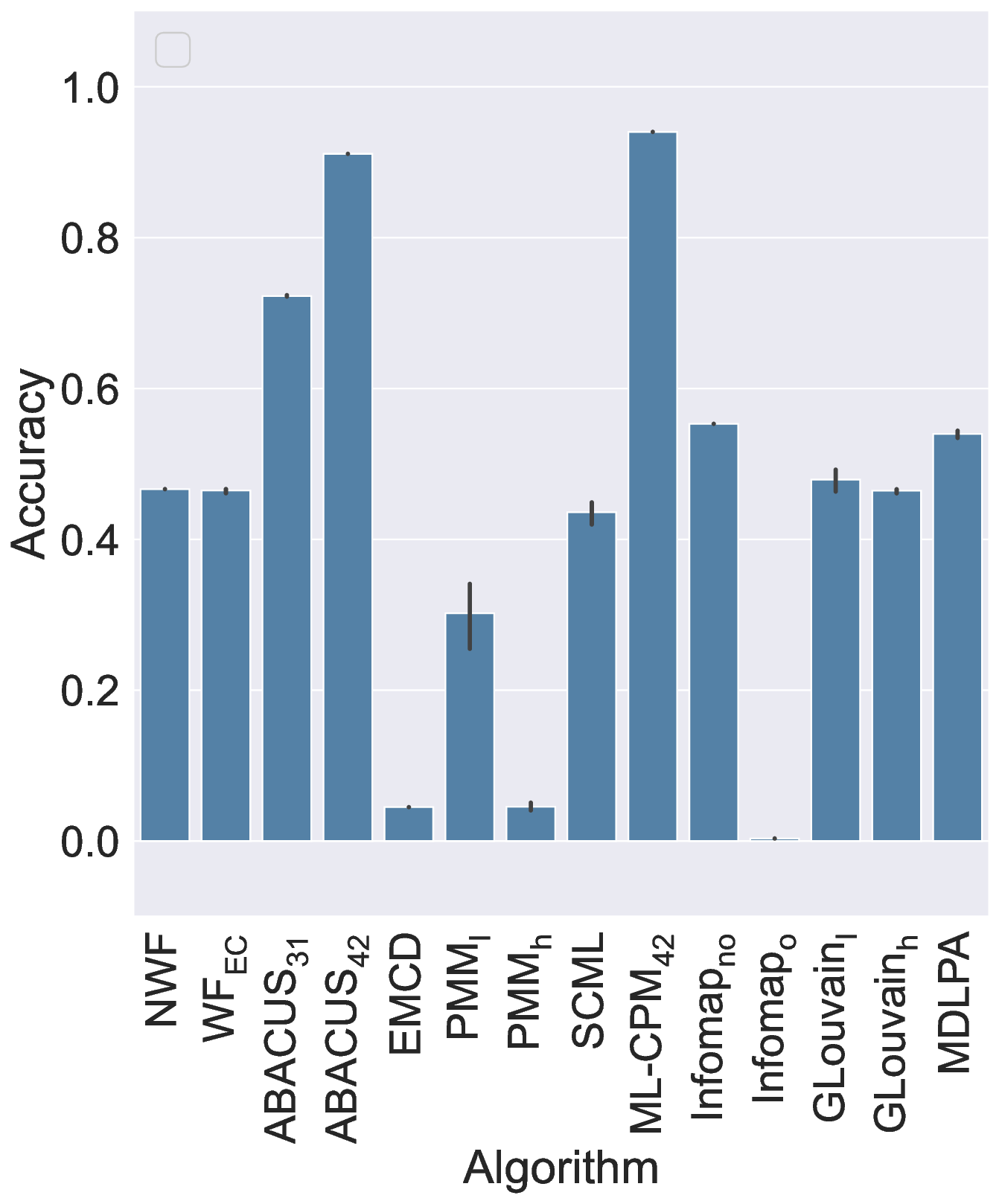}
 		\caption{Semi-pillar Non-equal Overlapping (SEO)}
 	\end{subfigure}
 	\caption{Accuracy with respect to a ground truth,  Omega index, overlapping communities (500 actors)}
 	\label{fig:accuracy_global_syn_2_500}
 \end{figure}
	
\begin{figure}[!htpb]
	\centering
	\begin{subfigure}[t]{0.45\textwidth}
		\centering
		\captionsetup{justification=centering}
		\includegraphics[width=\textwidth]{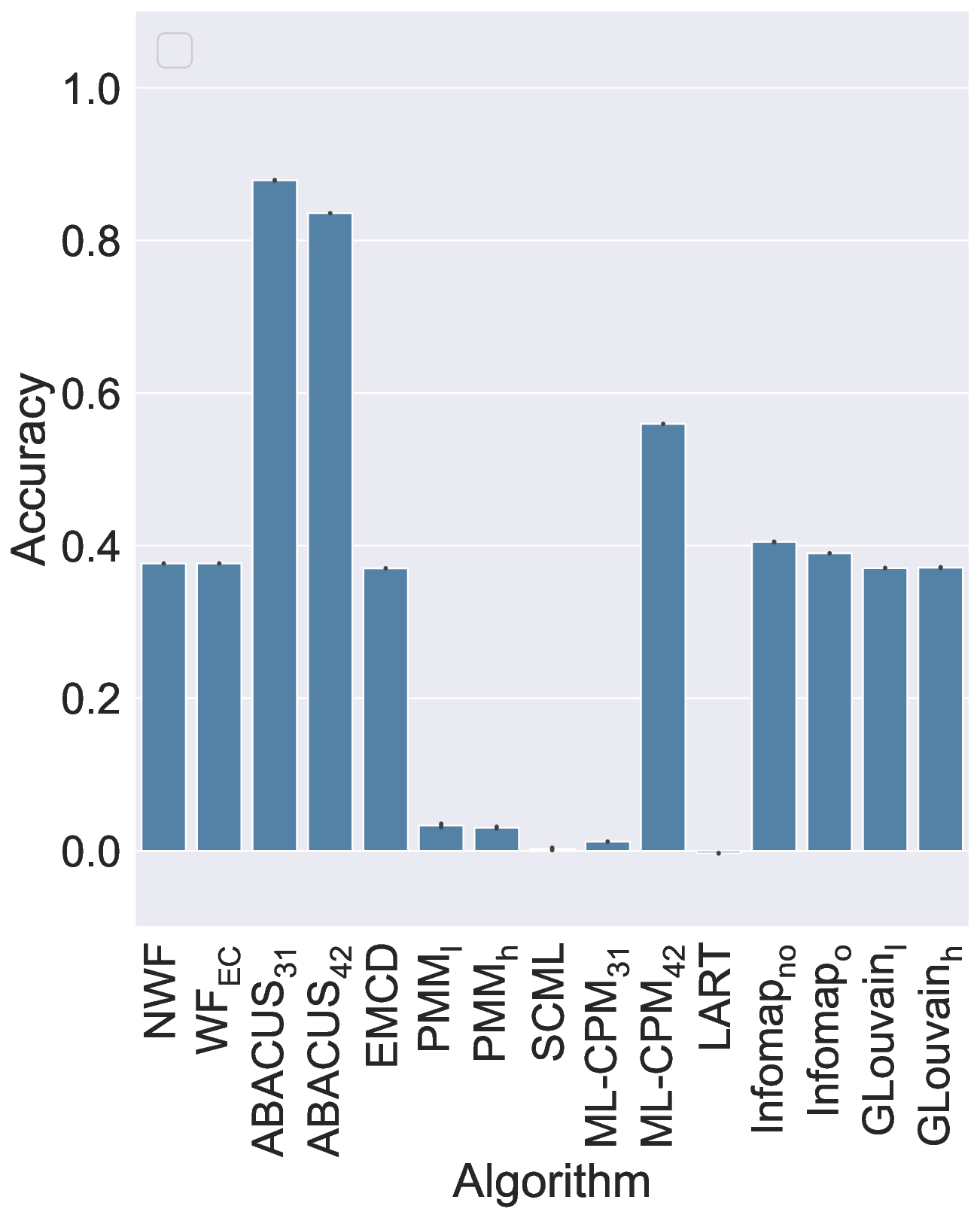}
		\caption{Mixed (MIX)}
	\end{subfigure}
	\begin{subfigure}[t]{0.45\textwidth}
		\centering
		\captionsetup{justification=centering}
		\includegraphics[width=\textwidth]{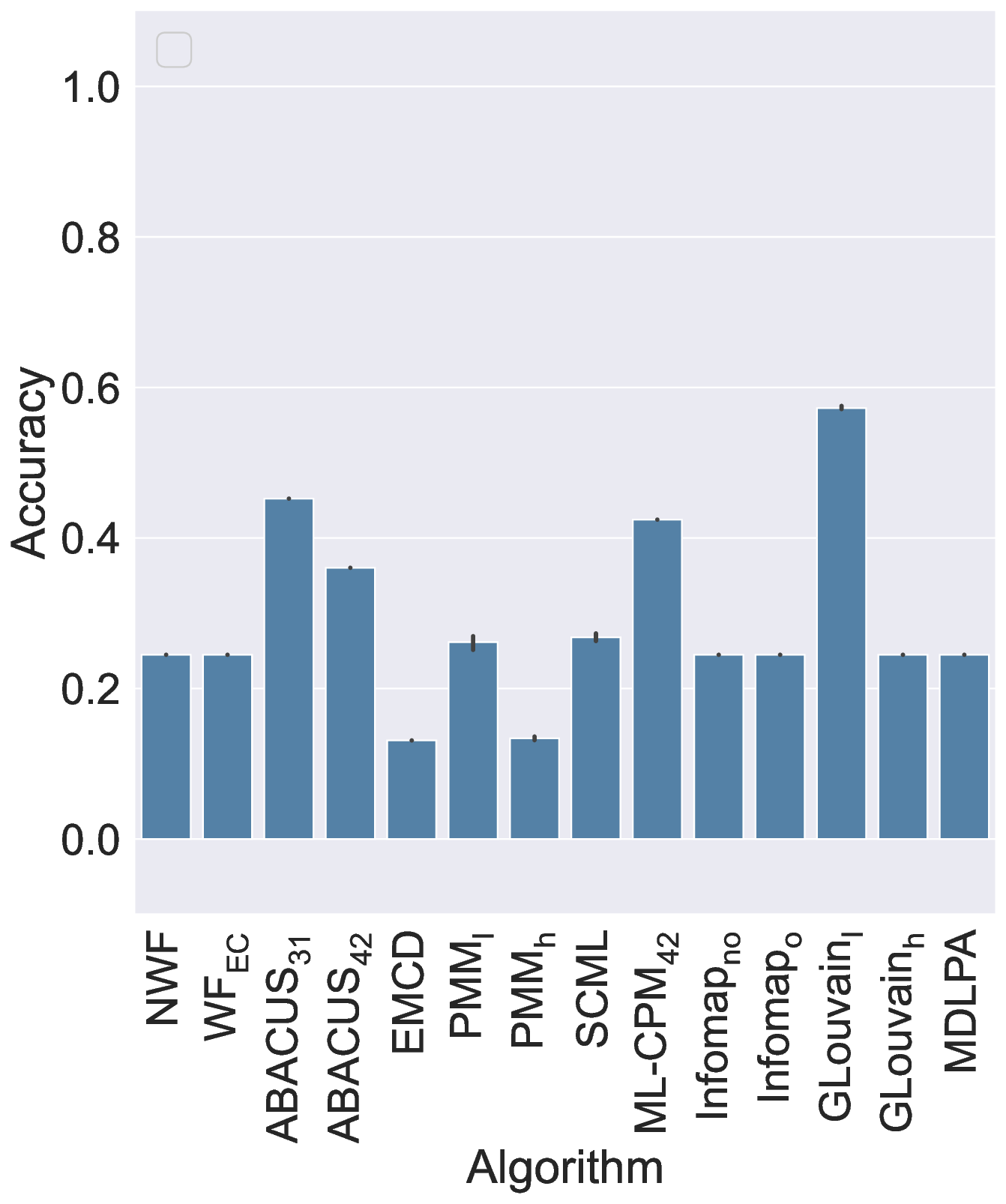}
		\caption{Hierarchical (HIE)}
	\end{subfigure}
	\caption{Accuracy with respect to a ground truth,  Omega index, mixed and hierarchical communities (500 actors)}
	\label{fig:accuracy_global_syn_3_500}
\end{figure}

Figures~\ref{fig:accuracy_global_syn_1_500}, \ref{fig:accuracy_global_syn_2_500}, and \ref{fig:accuracy_global_syn_3_500} show the results of similar experiments on synthetic networks using larger networks. While small networks allowed us to obtain results also with more computationally expensive methods, a potential problem with small networks with communities of 10 actors is that the random generation of edges may increase noise\footnote{By noise we indicate the amount of edges between nodes in differente communities. In our accuracy experiments with synthetic networks the noise is fixed, while we test the effect of variations in the amount of noise in Section~\ref{appendix_noise}} on specific communities. The results are still very similar to the ones with smaller networks, with three main differences. First, some methods (\textsf{PMM} and \textsf{SCML}) show some instability, returning worse results in specific cases. Second, \textsf{ML-CPM} starts working better in the version with a larger minimum clique size. Third, we can see some general improvements in particular in the Pillar Equal Overlapping results, while still confirming the worsening patterns highlighted by the previous experiments when we move to Semi-Pillar and Non-Equal communities.

\subsubsection{Pairwise comparison analysis}
\label{sssec:results_glob_pw}

In order to answer \rev{\textbf{Q2}} (i.e., ``To what extent do the evaluated methods produce similar community structures?'', cf. Section \ref{sec:experiments}), we performed pairwise comparisons between the selected methods, in order to determine the similarity between the community structures produced by each \rev{pair} of methods on each network. 

\begin{figure}[!htbp]
	\centering
	\begin{subfigure}[t]{0.7\textwidth}
		\centering
		\includegraphics[width=\textwidth]{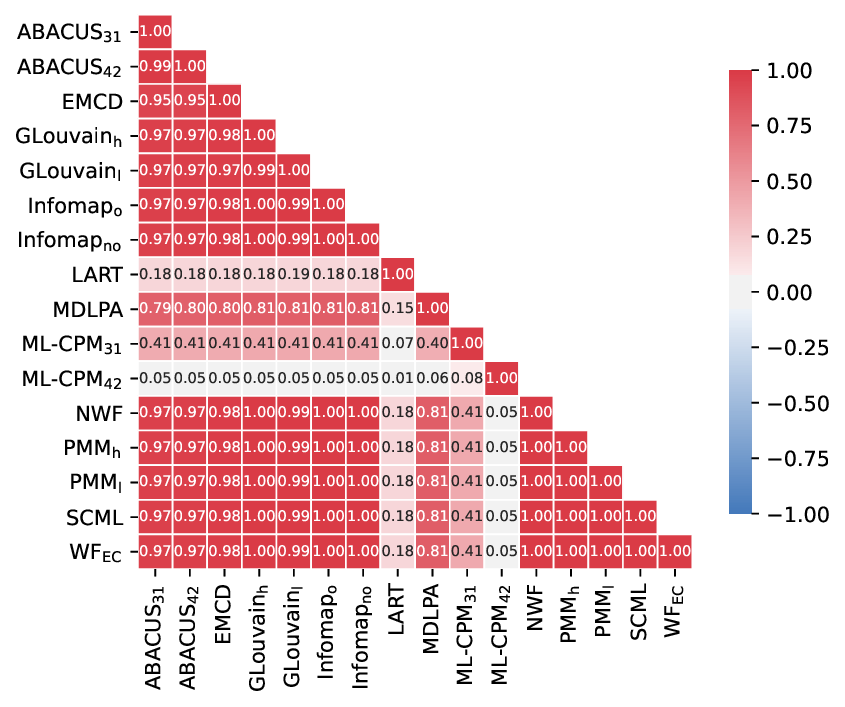}
		\caption{PEP}
	\end{subfigure}
	\begin{subfigure}[t]{0.7\textwidth}
		\centering
		\includegraphics[width=\textwidth]{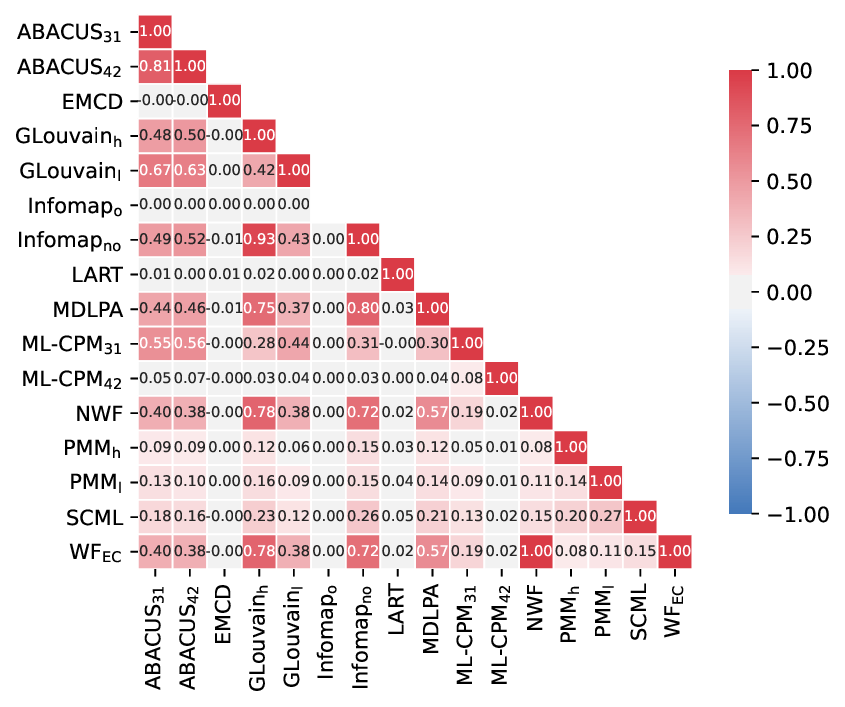}
		\caption{SNP}
	\end{subfigure}
	\caption{Pairwise comparison, Omega index: pillar and semi-pillar partitioning communities}
	\label{fig:pws:omega_n_comparision}
\end{figure}

Figure~\ref{fig:pws:omega_n_comparision} reports on the results of pairwise analysis among Pillar Equal Partitioning and Semi-Pillar Non-equal Partitioning\revm{, with} 
Omega index values for the pairwise similarities. 
We show Omega index values for a matter of homogeneity, since NMI cannot be applied to overlapping solutions. 



The\revm{se} results 
confirm and expand the understanding of the methods we have described so far. In the case of Pillar Equal Partitioning networks, almost all the methods produce very similar structures, with the notable exception of \textsf{ML-CPM} and \textsf{LART}.
In the case of Semi-Pillar \revm{P}artitioning communities the similarities are much smaller with few notable ex\revm{c}eptions: \textsf{Infomap$_{no}$} returns communities extremely similar to those returned by \textsf{GLouvain$_h$} and both also show  a strong similarity ($0.7$) with the communities returned from the flattening-based methods. \revm{Results for other data are not reported here for space reasons, but confirm the same trends highlighted by the analysis of accuracy.} \rev{N}ode-partitioning methods may produce similar community structures on specific cases (i.e., depending on the methods and the target network), suggesting that, when multiple community memberships are not allowed, some communities will often be unambiguously recognized in the network topology. 
Conversely, multiple community memberships allowed by overlapping methods end up in extremely variate solutions, i.e., relatively low similarities are observed regardless of the selected network and pair of methods.

\subsubsection{Scalability Analysis}
\label{sssec:results_glob_scal}

In order to answer \textbf{Q3} (``To what extent are the evaluated methods scalable?'', cf. Section~\ref{sec:experiments}), we tested the scalability of the selected methods with respect to  number of actors and number of layers. \rev{The reported results were obtained on a MacOS Catalina system version 10.15.5 with a 2,4GHz Dual-Core Intel Core i7 processor and 16GB of RAM.} 

\begin{figure}[!htpb]
	\centering
	\begin{subfigure}[t]{0.48\textwidth}
		\centering
		\includegraphics[width=\textwidth]{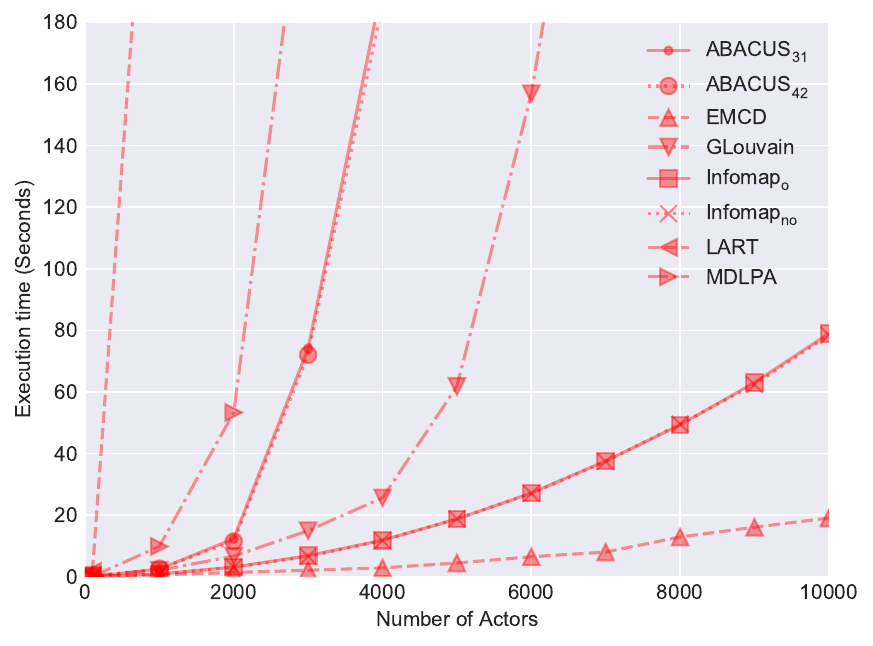}
		\caption{}
	\end{subfigure}
	\begin{subfigure}[t]{0.48\textwidth}
		\centering
		\includegraphics[width=\textwidth]{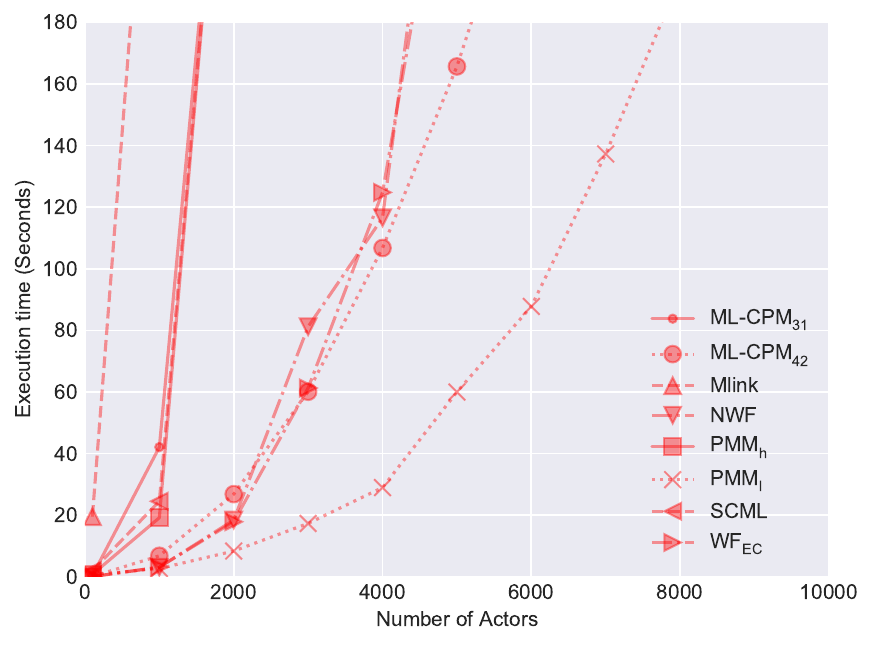}
		\caption{}
	\end{subfigure}
	\caption{Scalability of different community detection methods with respect to the number of actors}
	\label{fig:global_scalability_na}
\end{figure}

\begin{figure}[!htpb]
	\centering
	\begin{subfigure}[t]{0.48\textwidth}
		\centering
		\includegraphics[width=\textwidth]{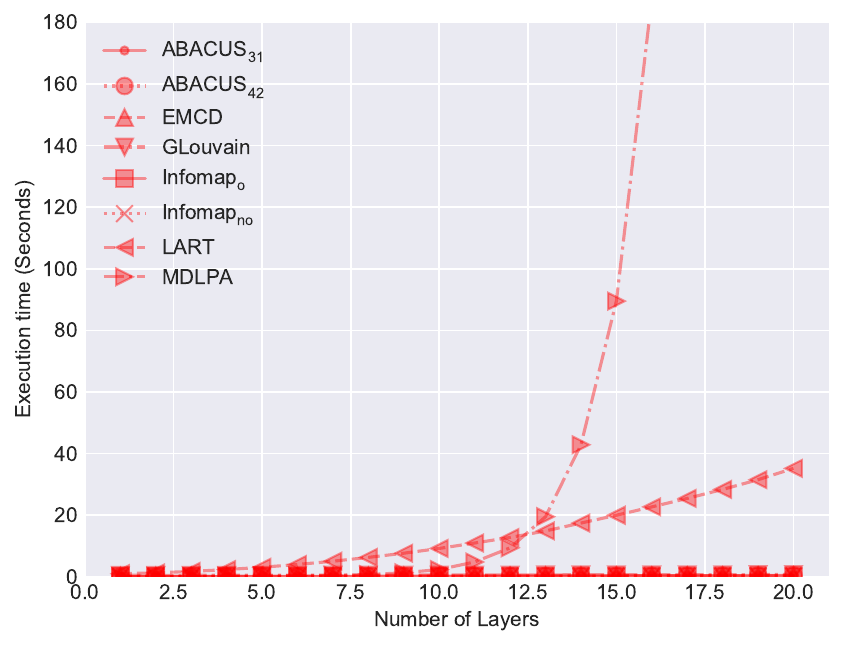}
		\caption{}
	\end{subfigure}
	\begin{subfigure}[t]{0.48\textwidth}
		\centering
		\includegraphics[width=\textwidth]{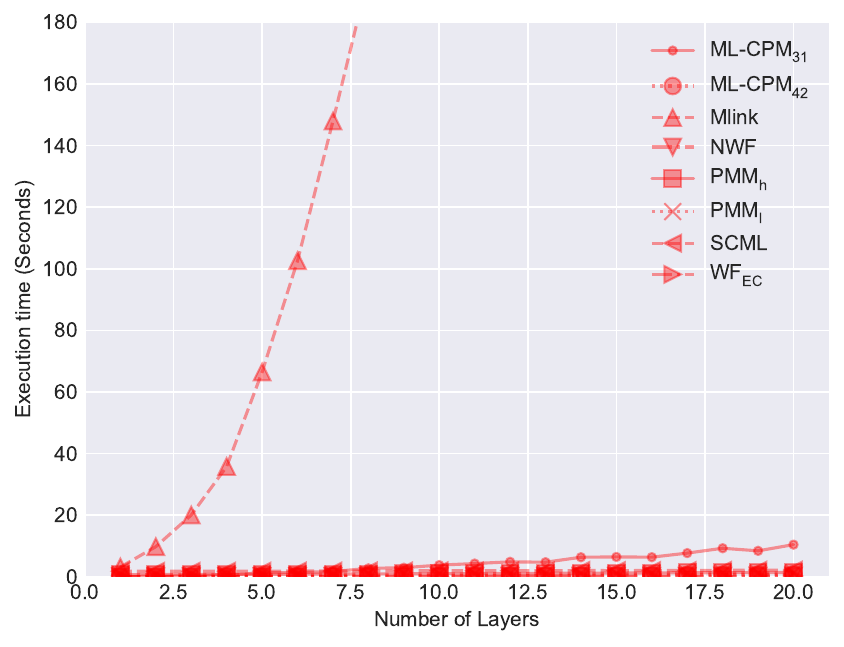}
		\caption{}
	\end{subfigure}
	\caption{Scalability of different community detection methods with respect to the number of layers}
	\label{fig:global_scalability_nl}
\end{figure}

Figures~\ref{fig:global_scalability_na}--\ref{fig:global_scalability_nl} report the scalability of each method with respect to an increment in the number of actors and the number of layers respectively. \rev{Note that in both cases the scalability of the flattening algorithms largely depends on the one of the community detection method used at the final step, since the computational cost of the flattening process is irrelevant. 
Some methods proved to be extremely scalable, more specifically, \textsf{EMCD} and \textsf{Infomap} --- all of which could run in less than a minute on networks containing up to $8000$ actors. However, \textsf{EMCD} takes single-layer community structures as input, therefore the time to find these communities is not counted in the plot. Considering the whole process, we would find \textsf{EMCD} close to the flattening methods. 
ML-C\revm{PM} (both variations), \textsf{MLink} and \textsf{LART} proved to be much less scalable, with a running time quickly increasing with the number of actors.}

\rev{As regards to the scalability in the number of layers (Figure~\ref{fig:global_scalability_nl}), we see that, generally speaking, it affects the results less than the number of actors. Only four methods show some significant increase in execution time: \textsf{ML-CPM} with $m=1$, \textsf{MLink}, \textsf{LART} and \textsf{\revm{MDLPA}}. The behavior of \textsf{\revm{MDLPA}} is in accordance with its theoretical time complexity.}


\subsubsection{Impact of noise}
\label{appendix_noise}

\begin{figure}[!htpb]
	\centering
	\begin{subfigure}[t]{0.48\textwidth}
		\centering
		\includegraphics[width=\textwidth]{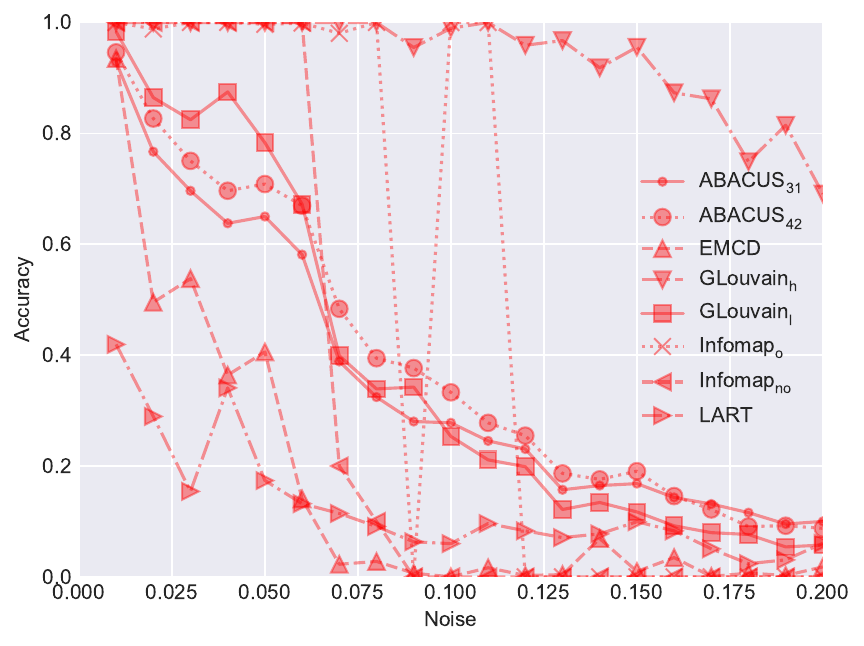}
		\caption{}
	\end{subfigure}
	\begin{subfigure}[t]{0.48\textwidth}
		\centering
		\includegraphics[width=\textwidth]{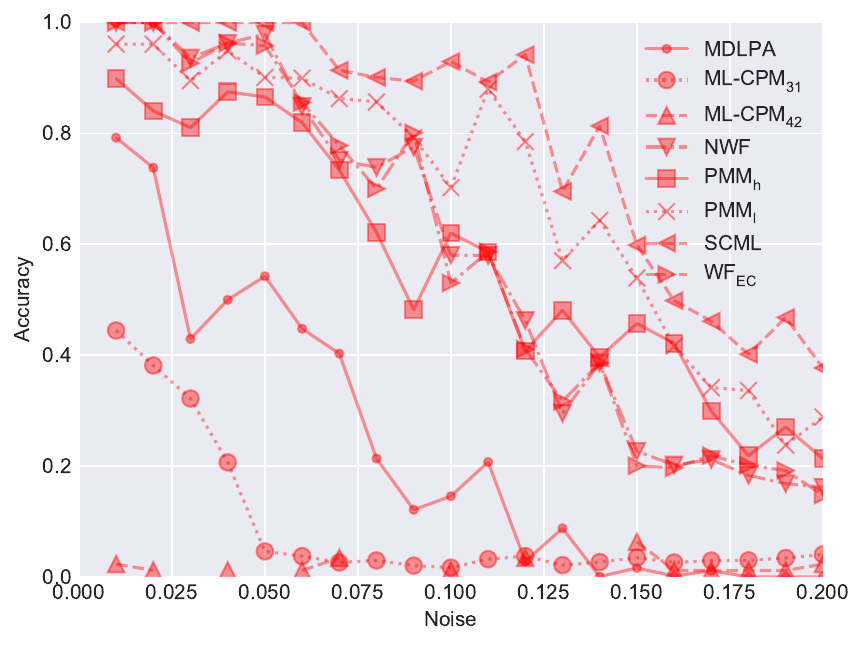}
		\caption{}
	\end{subfigure}
	\caption{Impact of noise on accuracy}
	\label{fig:noise}
\end{figure}

\revm{Figure~\ref{fig:noise} shows how the accuracy of the tested methods changes when we increase the number of edges across ground-truth communities. The basic data has been generated with 100 actors, 3 layers, 10 Pillar Equal Partitioning communities, and the $x$ axis indicates the probability of nodes in different communities to be adjacent. The results show how GLouvain with a high omega is less affected by noise than other methods, although this should be considered a result of the  \revm{presence of} pillar communities and the fact that high values of omega force the generation of pillars. Interestingly, Infomap shows a phase transition: with low noise it can identify the correct communities, then suddenly it starts returning one single community leading to an Omega Index value of 0.}

\subsection{Local Methods}
\label{ssec:results_local}

In this section we report the experimental results of the comparative evaluation of local multiplex community detection methods. The section is structured as follows: Section~\ref{sssec:local_acc_anyalisis} presents the results of the accuracy analysis, Section~\ref{sssec:local_pw_comp} reports on the results of the pairwise comparison between different methods, while Section~\ref{sssec:local_scal_anyalisis} discusses scalability issues.

\subsubsection{Accuracy analysis}
\label{sssec:local_acc_anyalisis}

We performed an accuracy analysis on the local community detection methods, by comparing the local community of each actor to the one that same actor belongs to in the ground truth. \rev{S}imilarity is computed using \rev{the} Jaccard \rev{index}, while the final accuracy value is the average over all actors.

\begin{figure}
		\centering
		\begin{subfigure}{0.33\textwidth}
			\centering
			\includegraphics[width=.95\textwidth]{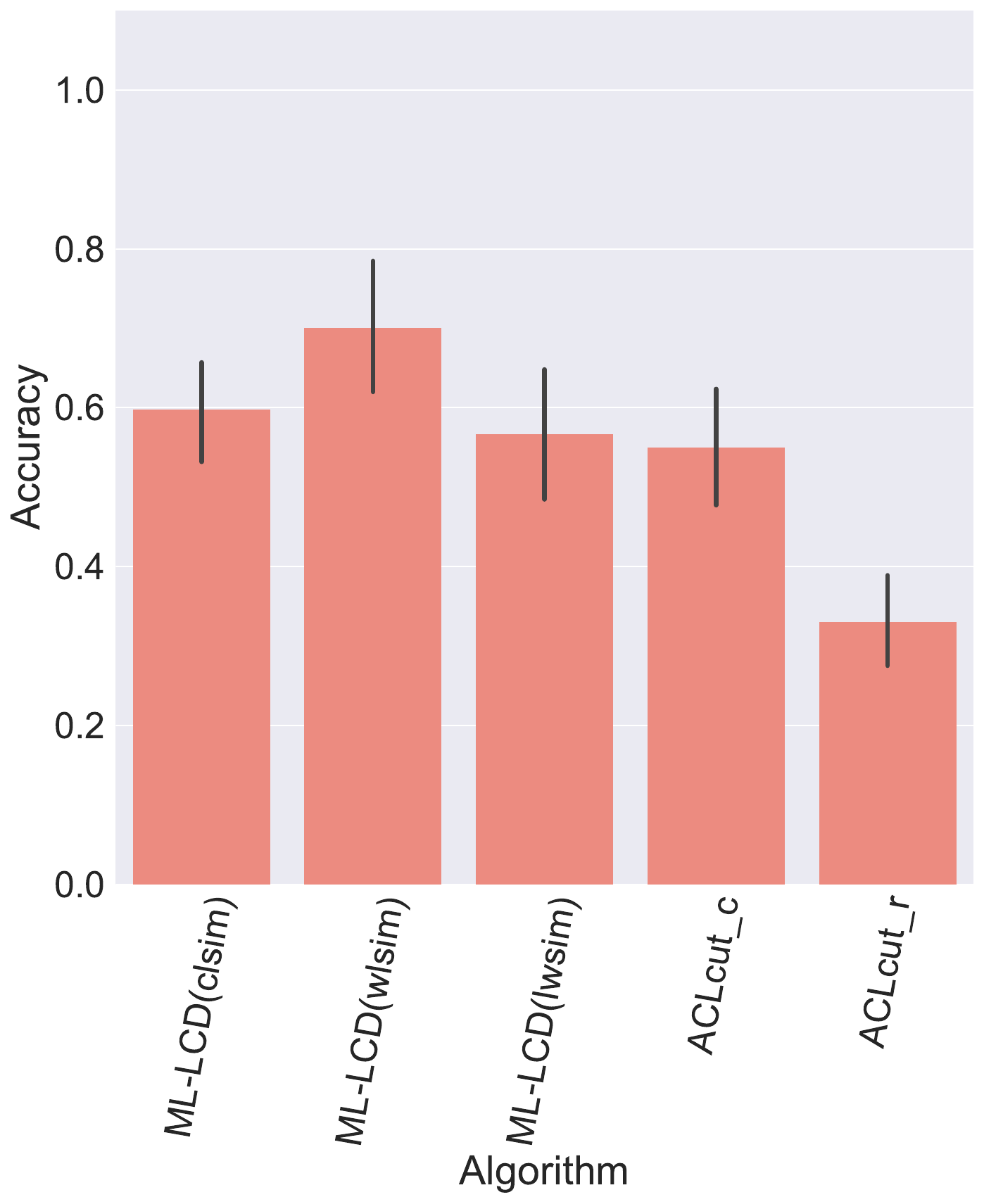}
			\caption{\textbf{AUCS}}
		\end{subfigure}
		\begin{subfigure}{0.33\textwidth}
			\centering
			\includegraphics[width=.95\textwidth]{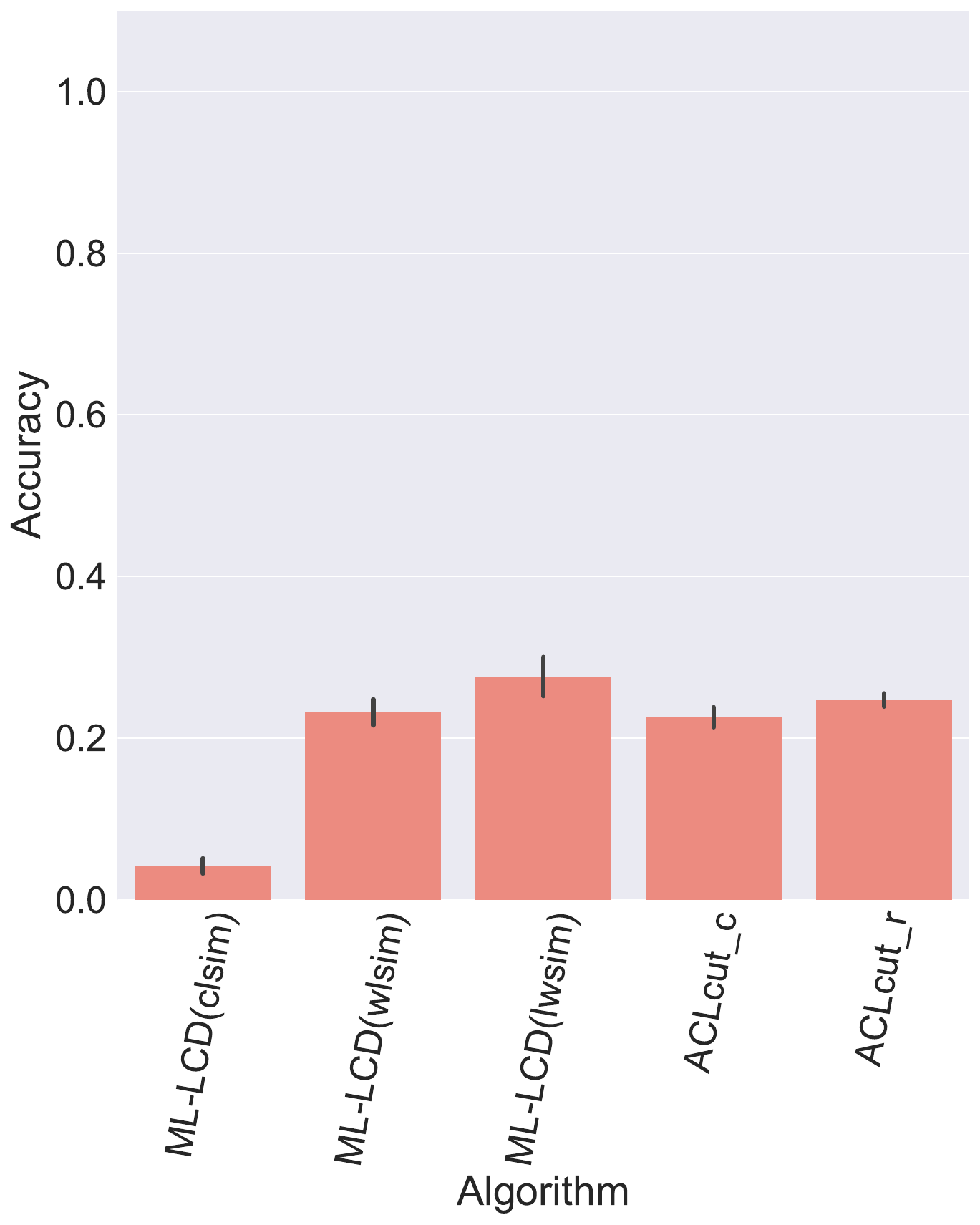}
			\caption{\textbf{DKPol}}
		\end{subfigure}
		\caption{Average accuracy of the local methods with respect to a ground truth, on real-world networks}
		\label{fig:acc_local_real}
\end{figure}

Figure ~\ref{fig:acc_local_real} shows results on real\rev{-}world networks. On AUCS, accuracy is in the range of $0.5$--$0.7$ for $4$ out of $5$ methods, with \textsf{ML-LCD$_{(wlsim)}$} being the best performer ($0.7$). Much lower accuracy values were obtained on \rev{DKP}ol, where the best performing method was \textsf{ML-LC\rev{D}$_{(lwsim)}$} ($0.27$).

Concerning synthetic networks,
\rev{we limited our analysis} to networks with \rev{a} pillar partitioning community structure (PEP and PNP), for compatibility with the methods' output (both return actor communities).
\rev{In these cases, we observed that a}ccuracies are much higher than the ones observed for real-world networks, with all values in the range [$0.8$,$1.0$].  \textsf{ML-LCD$_{(clsim)}$} is the best performing method, since it is able to perfectly identify the ground truth community structure on both networks.
 

Summarizing, while all methods proved to be able to identify synthetic pillar community structures, their performance was much worse on real\rev{-}world networks. These results confirm the behavior observed for global methods (cf. Section~\ref{sssec:results_glob_acc}).
Moreover, it should be pointed out that 
comparing a global community structure (i.e., the ground truth) to a set of local ones (i.e., the results obtained by local methods on all actors) may not be completely fair. 
\rev{The ground truth in this case represents a global partitioning of the network, while local communities are actor-centered, query dependent and, in general, they overlap with each other. 
Moreover, they may be discovered without having  a complete knowledge of the network graph, which is the case for   \textsf{ML-LCD}.  Although  based on the comparison of  conceptually different objects (i.e., global and local communities), our   accuracy analysis is     still significant as it quantifies to what extent the local community formed around  a certain actor falls inside the community found in the global structure that contains the actor.}  Unfortunately, no networks with an associated ground truth of multiplex local communities are available at the time of writing.

\subsubsection{Pairwise comparison}
\label{sssec:local_pw_comp} 

As seen in Section \ref{sssec:results_glob_pw} for global methods, we set up an equivalent evaluation stage based on pairwise comparison between the local methods.
In this case, we resorted to \rev{the} Jaccard index to measure the similarity of the community solutions produced by two local methods. Since these methods are query-dependent (i.e., they return the local community of a given query/seed node), we computed the Jaccard similarity between each pair of communities obtained using the same actor as seed, and then averaged the results over all actors. 

\begin{figure}[!htpb]
 		\centering
 		\begin{subfigure}{0.4\textwidth}
 			\centering
 			\includegraphics[width=.95\textwidth]{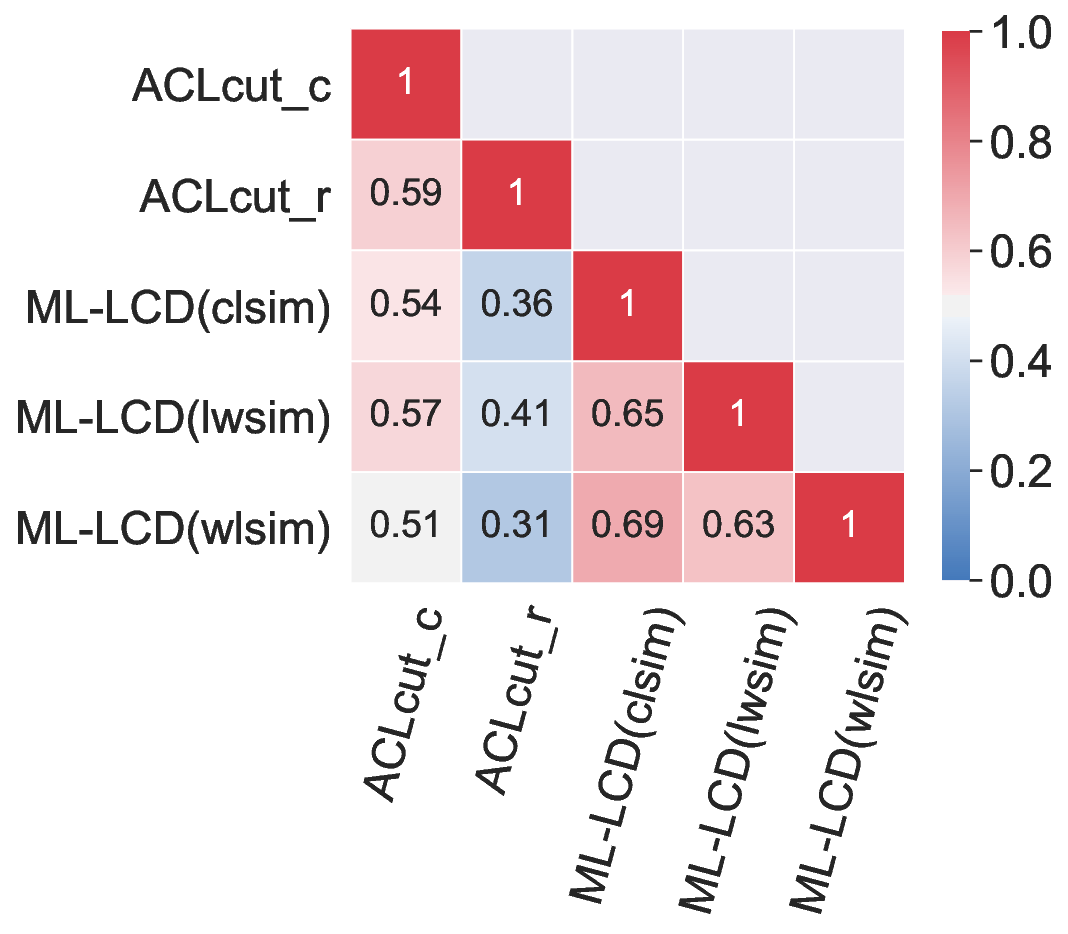}
 			\caption{\textbf{AUCS}}

 		\end{subfigure}
 		\begin{subfigure}{0.4\textwidth}
 			\centering
 			\includegraphics[width=.95\textwidth]{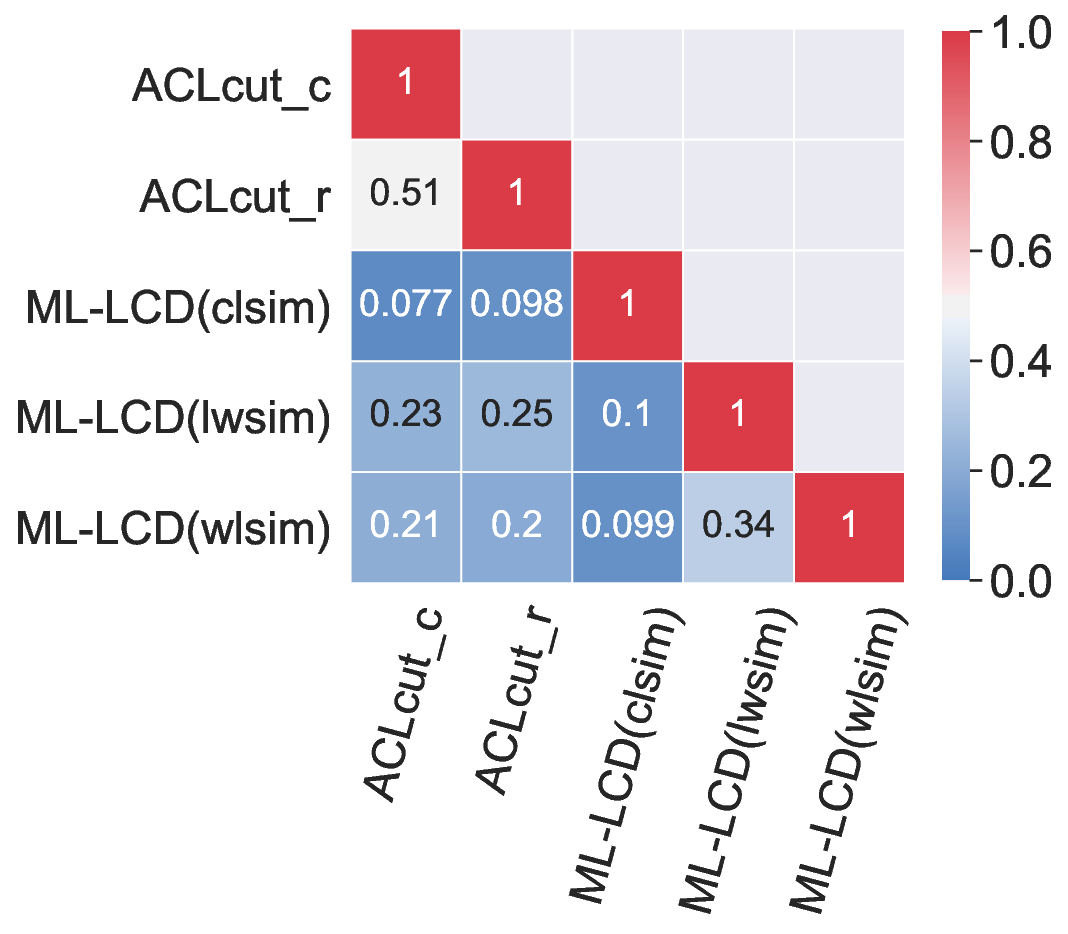}
 			\caption{\textbf{DKPol}}

 		\end{subfigure}


		\caption{Average pairwise similarity among the different local methods on real-world networks}
		\label{fig:local_pw_sim_real}
 \end{figure}

Figure~\ref{fig:local_pw_sim_real} reports on the results obtained on real-world networks. On most of these networks (\rev{DKP}ol, Airports, and Rattus), we can note that communities identified by different variants of \textsf{ML-LCD} and \textsf{ACLcut} tend to be very different. \rev{L}ooking at AUCS, the communities identified by all variants of both \textsf{ML-LCD} and \textsf{ACLcut} tend to be less different and a higher similarity can be observed among the three variants of \textsf{ML-LCD}.

\begin{figure}[!htpb]
	\centering
	\begin{subfigure}{0.4\textwidth}
		\centering
		\includegraphics[width=.95\textwidth]{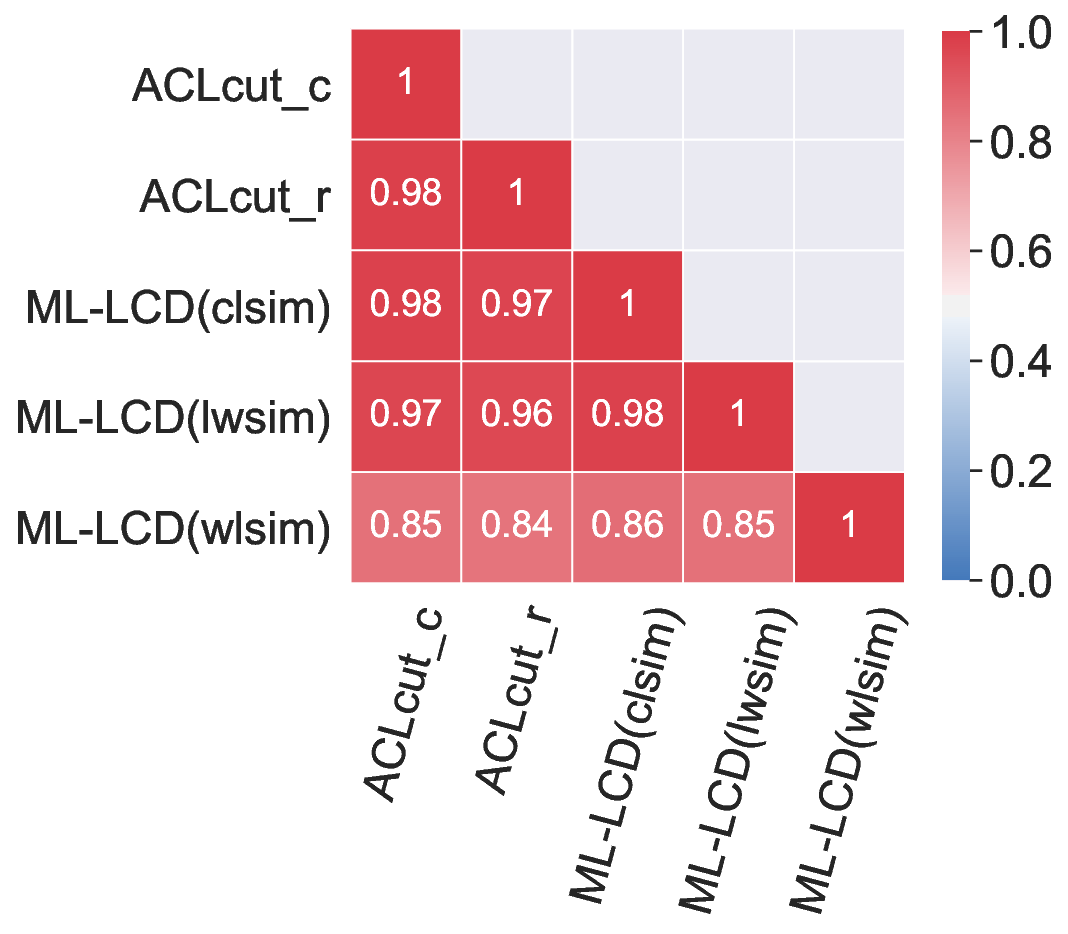}
		\caption{\textbf{PEP}}

	\end{subfigure}
	\begin{subfigure}{0.4\textwidth}
		\centering
		\includegraphics[width=.95\textwidth]{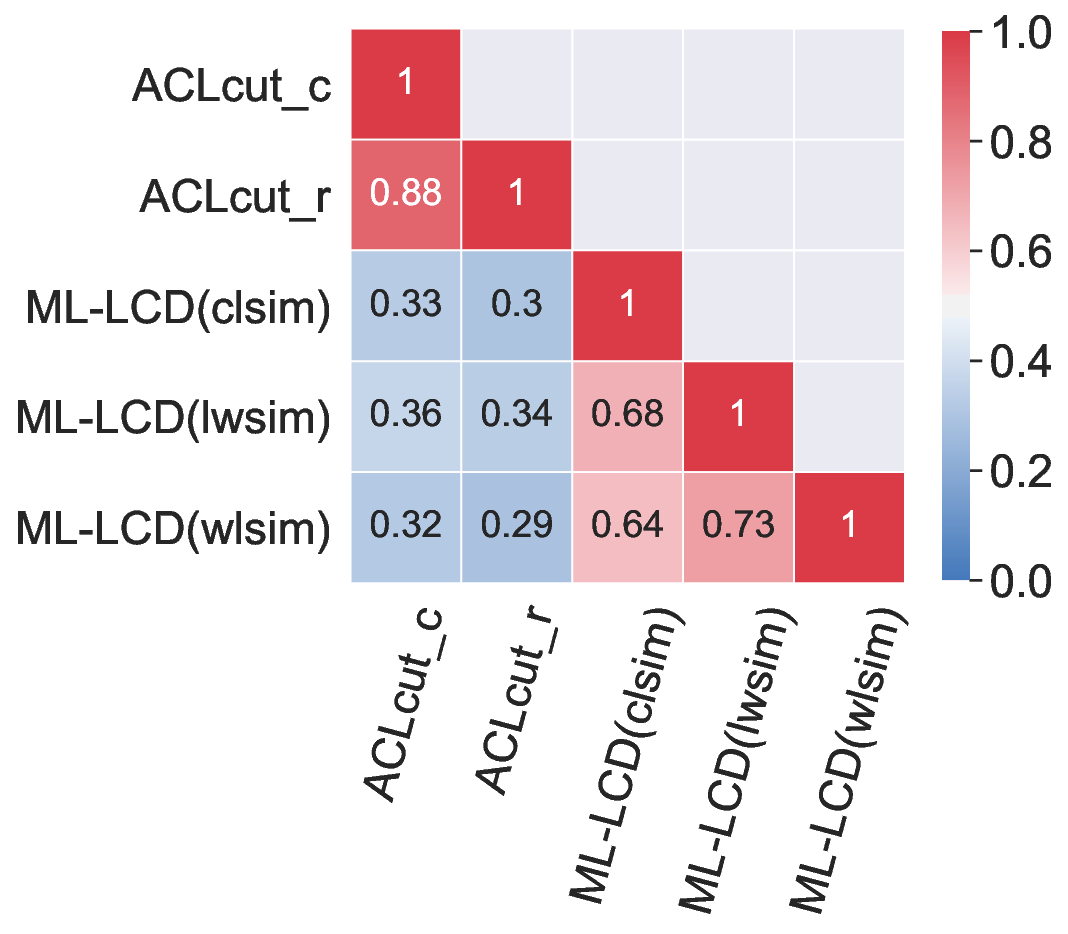}
		\caption{\textbf{PEO}}
	\end{subfigure}
	\begin{subfigure}{0.4\textwidth}
		\centering
		\includegraphics[width=.95\textwidth]{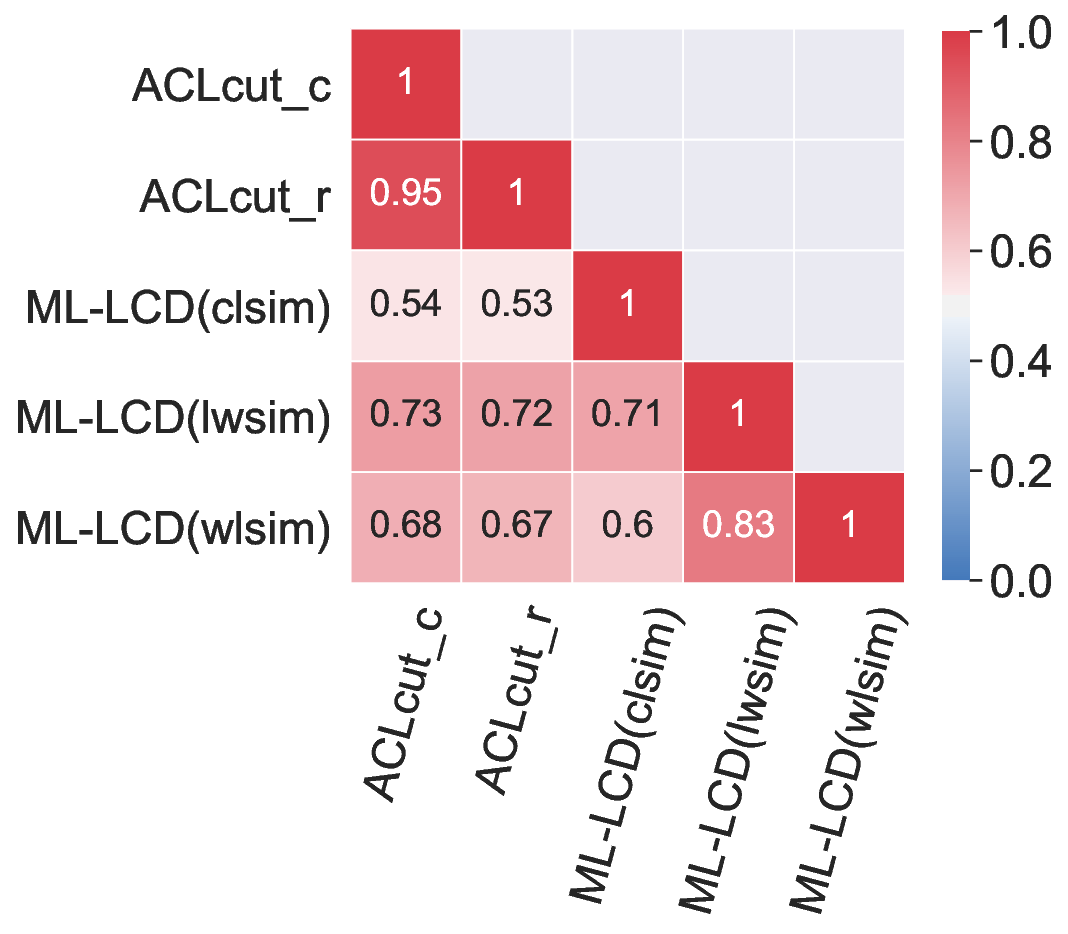}
		\caption{\textbf{HIE}}
	\end{subfigure}
	\begin{subfigure}{0.4\textwidth}
		\centering
		\includegraphics[width=.95\textwidth]{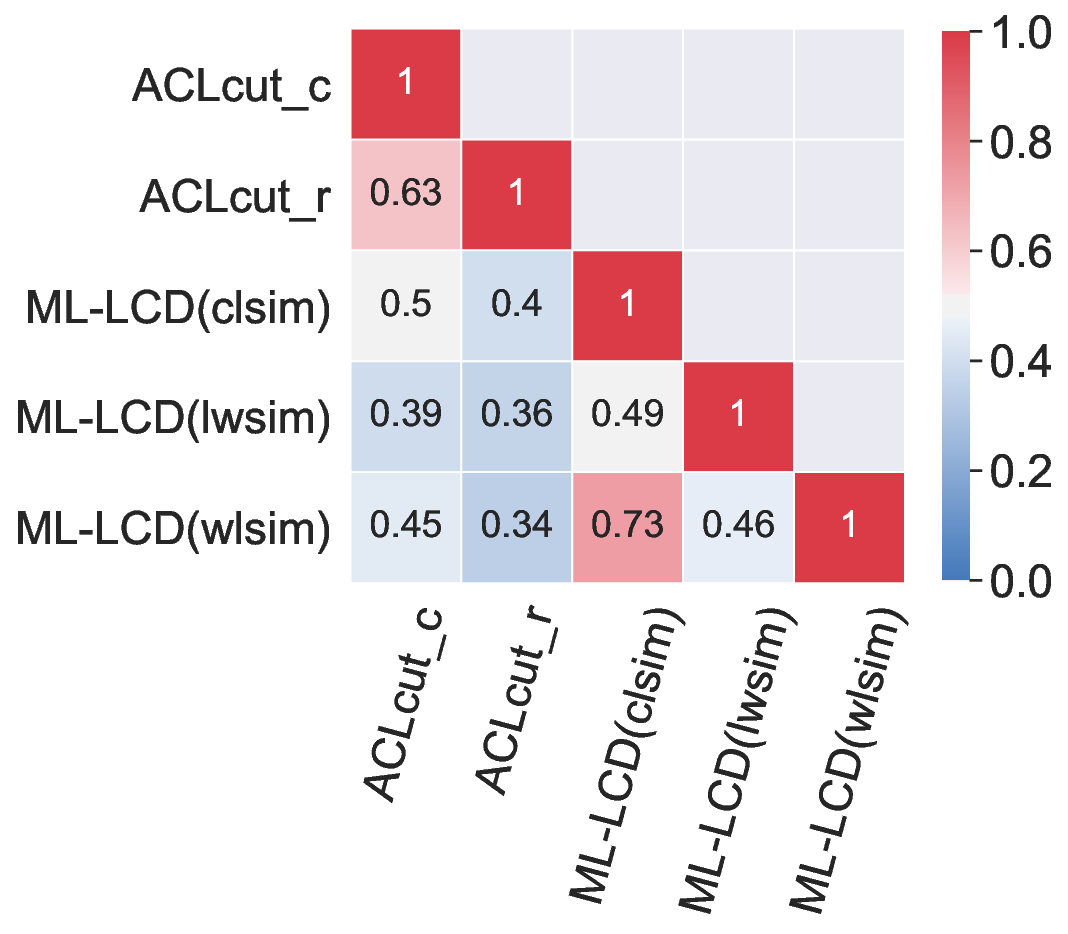}
		\caption{\textbf{\rev{MIX}}}
	\end{subfigure}
	\caption{Average pairwise similarity among the different local methods when the same seed is used as an input, on selected synthetic networks}
	\label{fig:local_pw_sim_syn}
\end{figure}

For synthetic networks (Figure~\ref{fig:local_pw_sim_syn}), it can be noted how similarities are higher for networks based on pillar community structures. In some cases (i.e., PEP and PNP) all methods are practically interchangeable, with all similarities equal or near to $1.0$.
 In other networks with pillar (i.e., PEO and PNO), semi-pillar (i.e., SEP and SEO) or both (MIX and HIE) community structures, similarities are stronger between the different variants of each method. 
Summing up, we observed some similarities in the behavior of all local methods on some real-world and synthetic networks, with an expected tendency of the variants of a same method to identify similar local communities. Nevertheless, this cannot be taken as a general rule, since we also observed specific cases where all methods behaved differently from each other, both on real-world and synthetic networks.

\begin{figure}
	\centering
	\begin{subfigure}{0.3\textwidth}
		\centering
		\includegraphics[width=.95\textwidth]{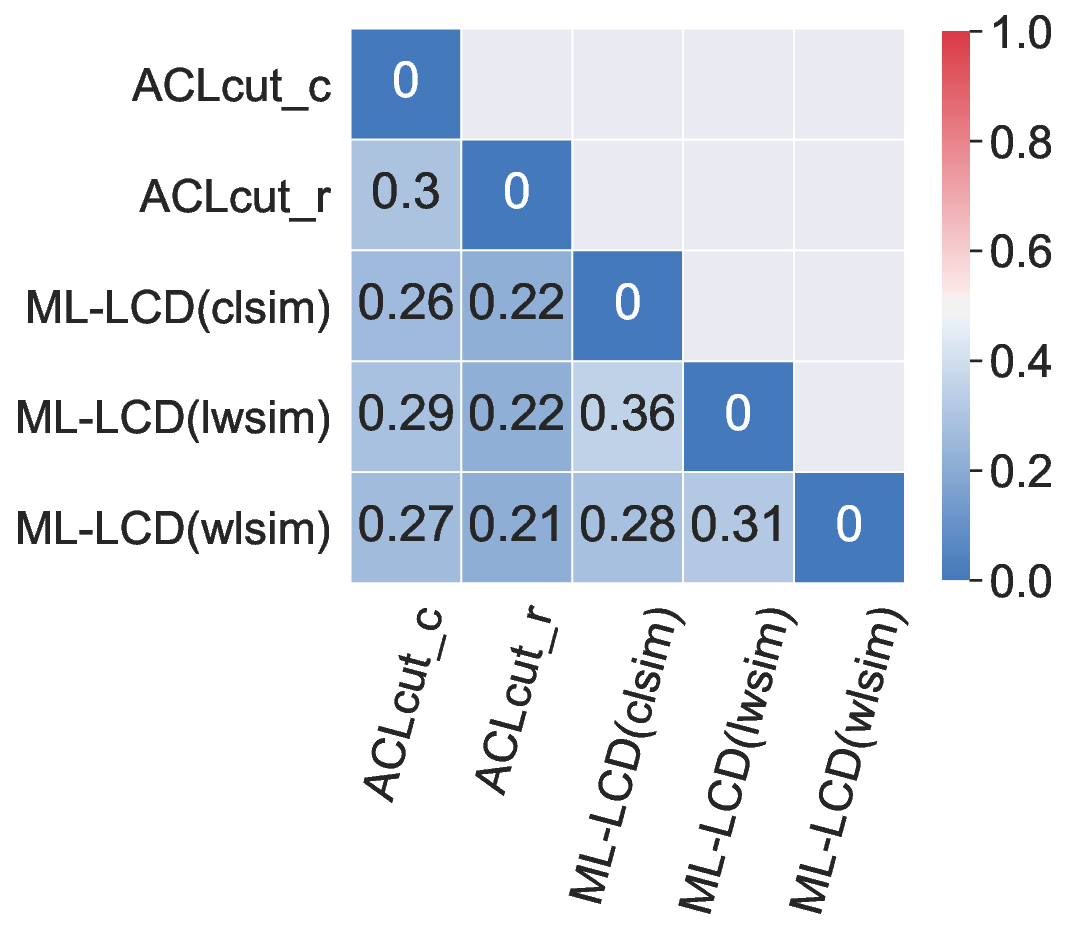}
		\caption{\textbf{AUCS}}
		\label{fig:aucs_local_pw}
	\end{subfigure}
	\begin{subfigure}{0.3\textwidth}
		\centering
		\includegraphics[width=.95\textwidth]{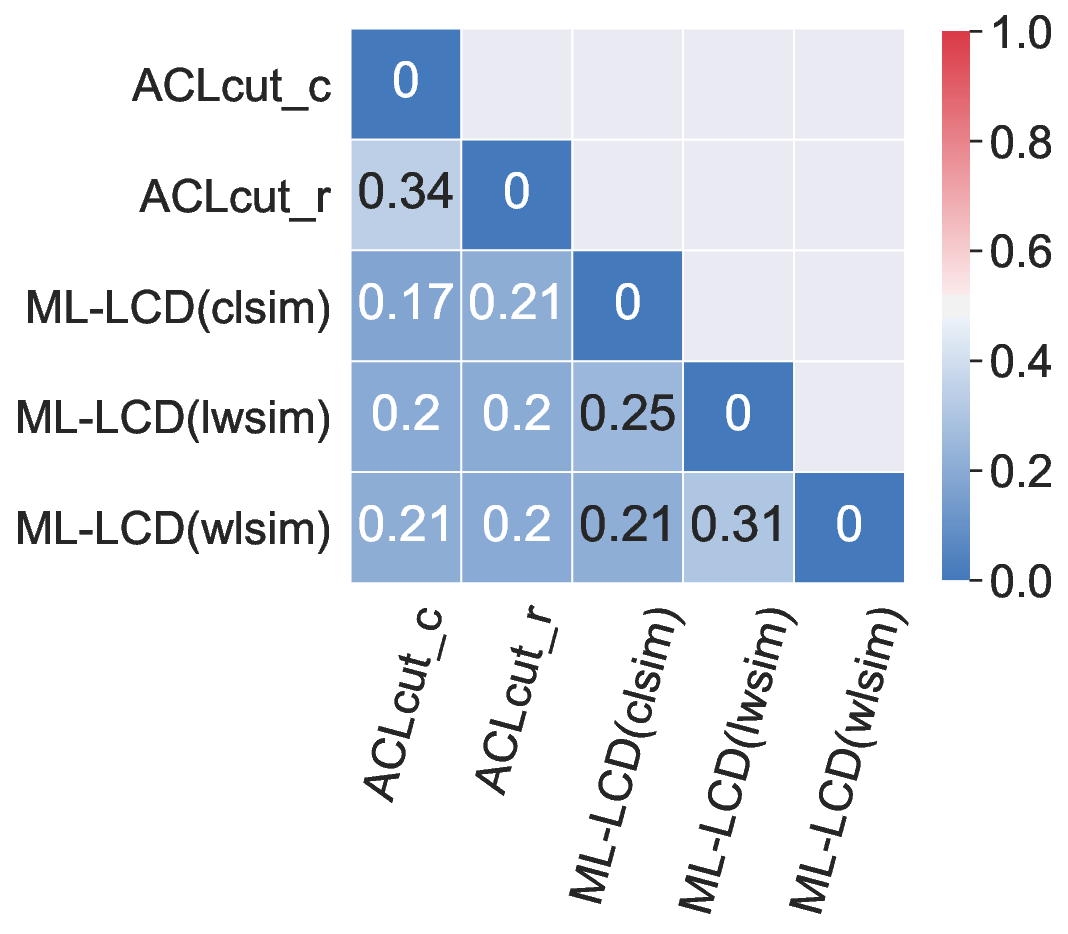}
		\caption{\textbf{DKPol}}
		\label{fig:dkpol_local_pw}
	\end{subfigure}
	\begin{subfigure}{0.3\textwidth}
		\centering
		\includegraphics[width=.95\textwidth]{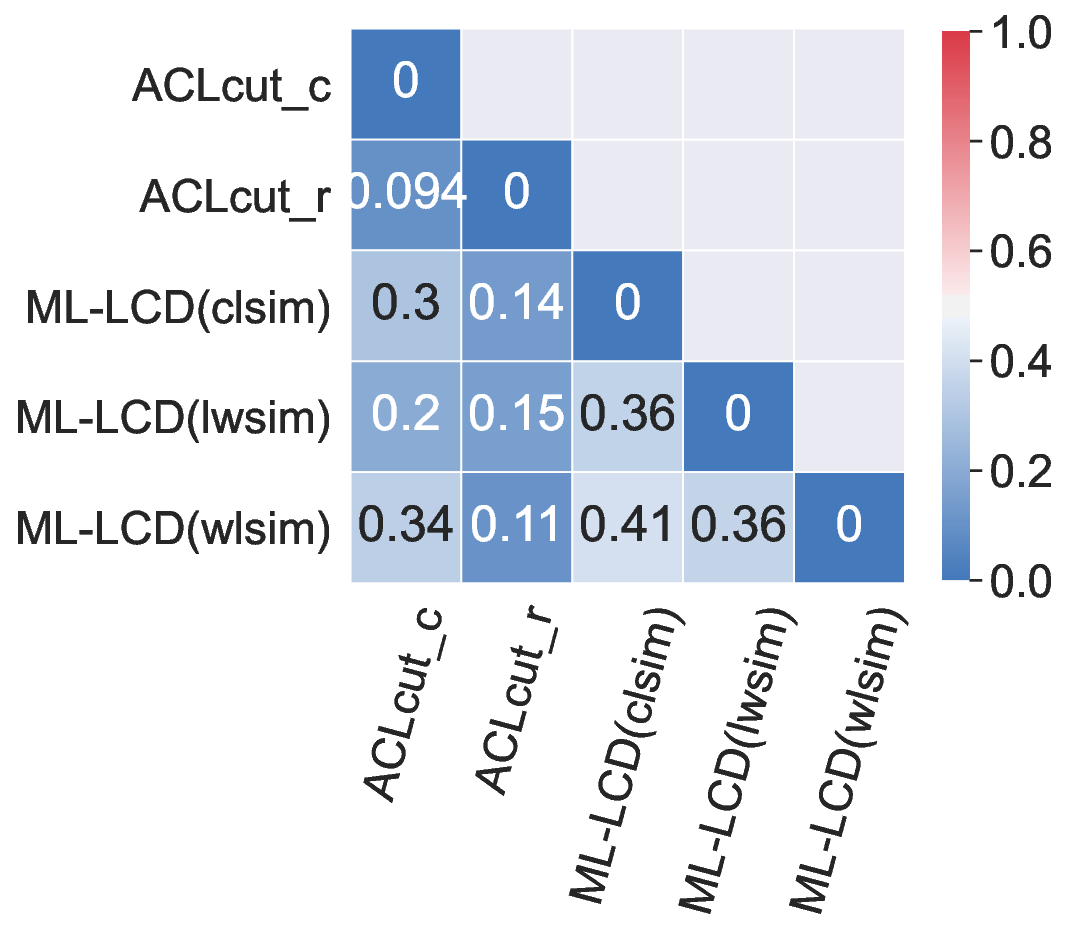}
		\caption{\textbf{Airports}}
		\label{fig:airports_local_pw}
	\end{subfigure}
	\begin{subfigure}{0.3\textwidth}
		\centering
		\includegraphics[width=.95\textwidth]{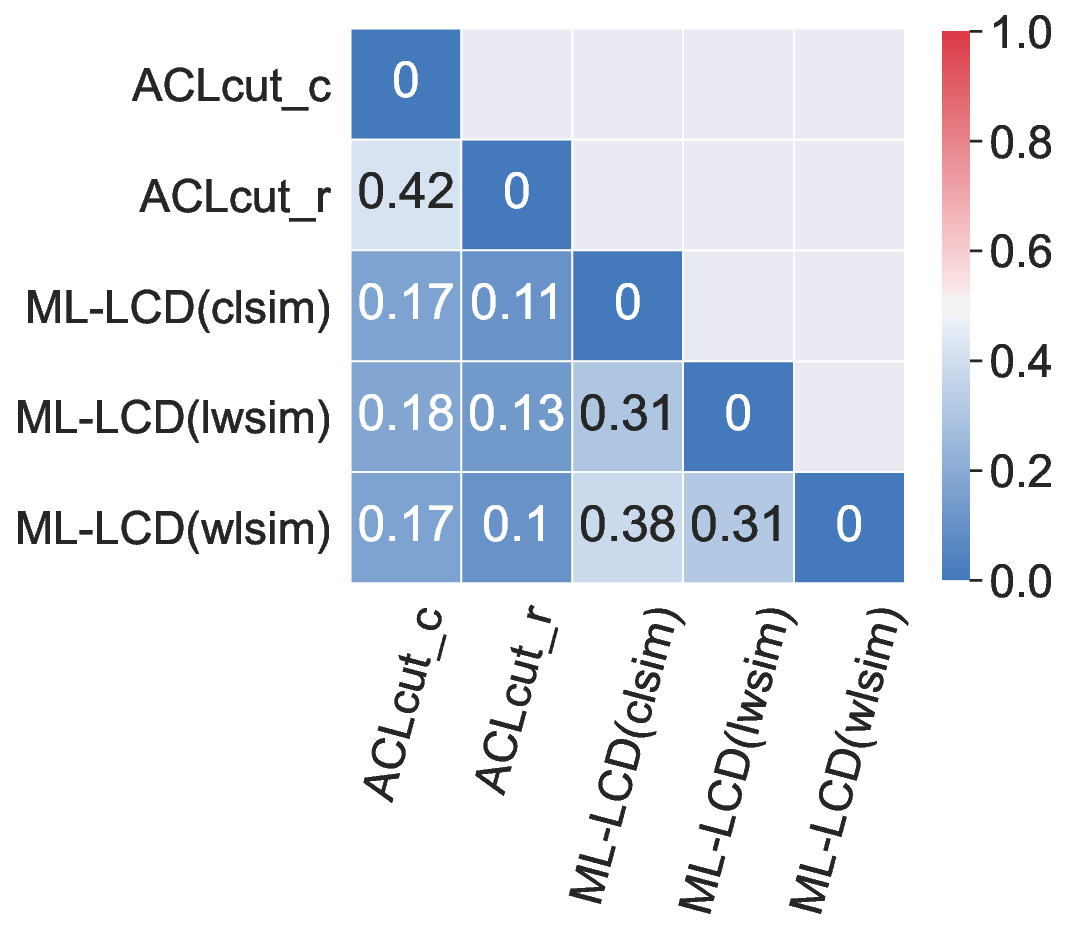}
		\caption{\textbf{Rattus}}
		\label{fig:rattus_local_pw}
	\end{subfigure}
	\caption{Standard deviation for average pairwise similarity among the different local method\revm{s} (\revm{r}eal networks)}
	\label{fig:local_pw_sim_real_sd}
\end{figure}

\begin{figure}[!htpb]
	\centering
	\begin{subfigure}{0.3\textwidth}
		\centering
		\includegraphics[width=.95\textwidth]{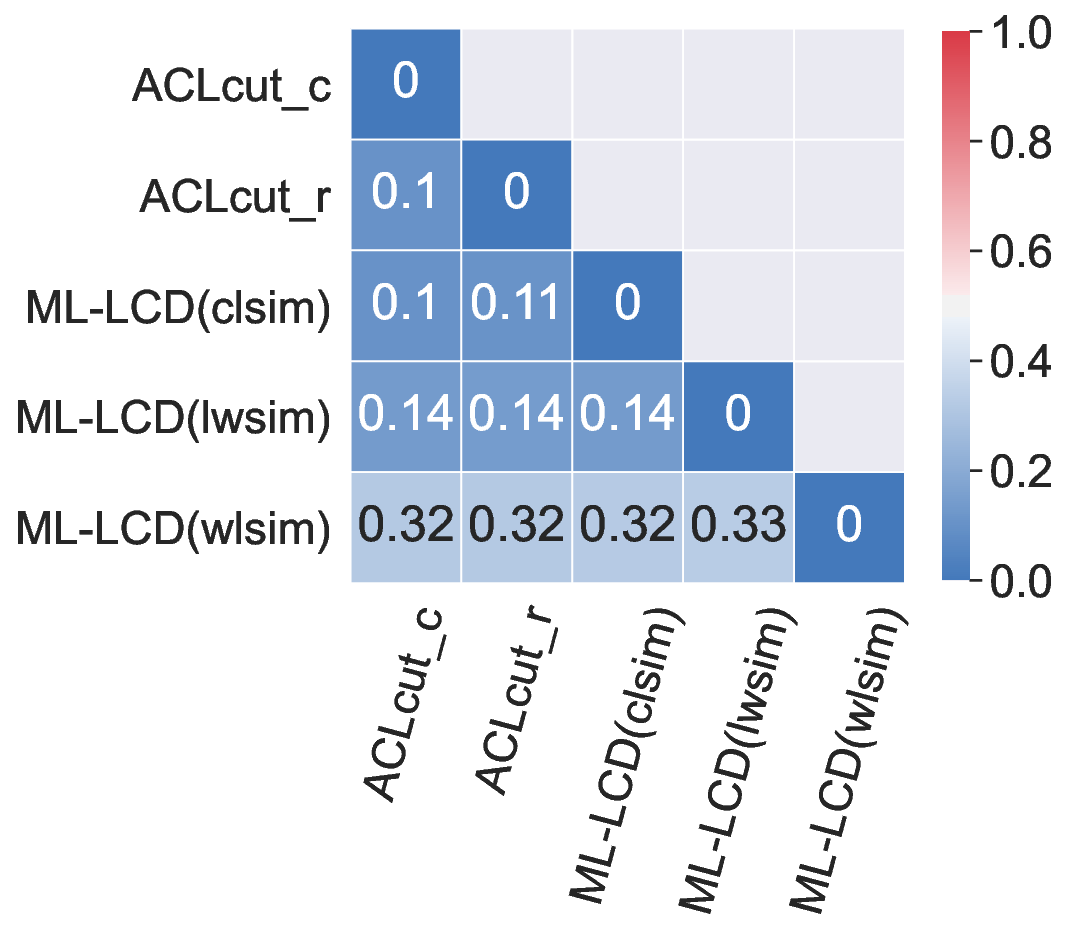}
		\caption{\textbf{PEP}}
		\label{fig:pep_local_pw}
	\end{subfigure}
	\begin{subfigure}{0.3\textwidth}
		\centering
		\includegraphics[width=.95\textwidth]{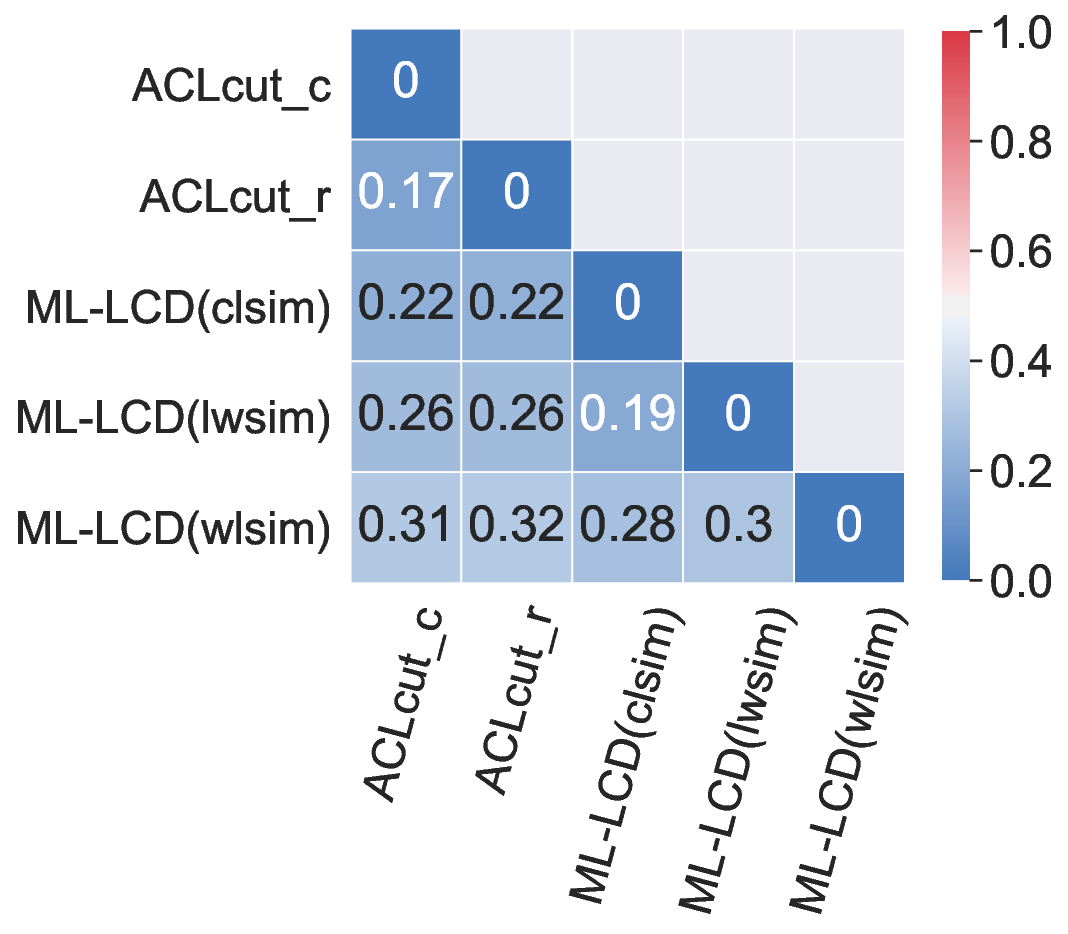}
		\caption{\textbf{PNP}}
		\label{fig:pnp_local_pw}
	\end{subfigure}
	\begin{subfigure}{0.3\textwidth}
		\centering
		\includegraphics[width=.95\textwidth]{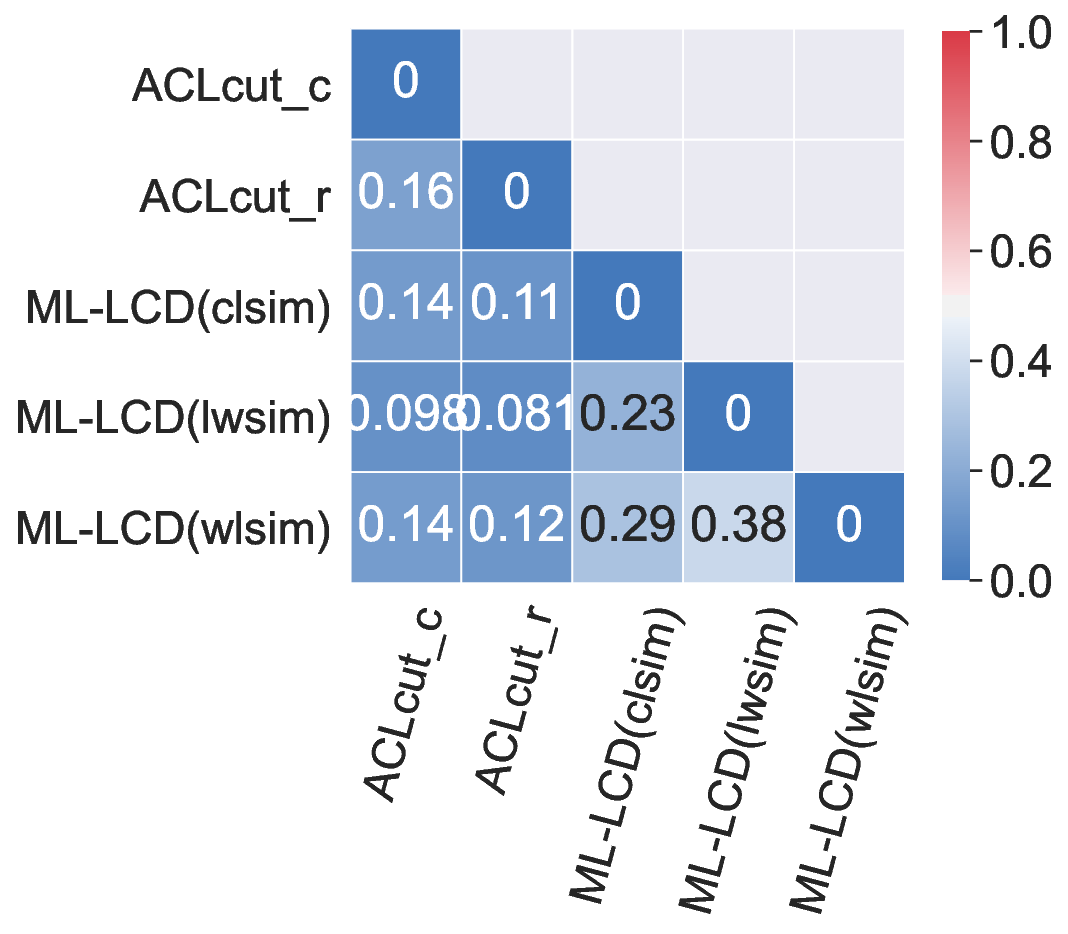}
		\caption{\textbf{PEO}}
		\label{fig:peo_local_pw}
	\end{subfigure}
	\begin{subfigure}{0.3\textwidth}
		\centering
		\includegraphics[width=.95\textwidth]{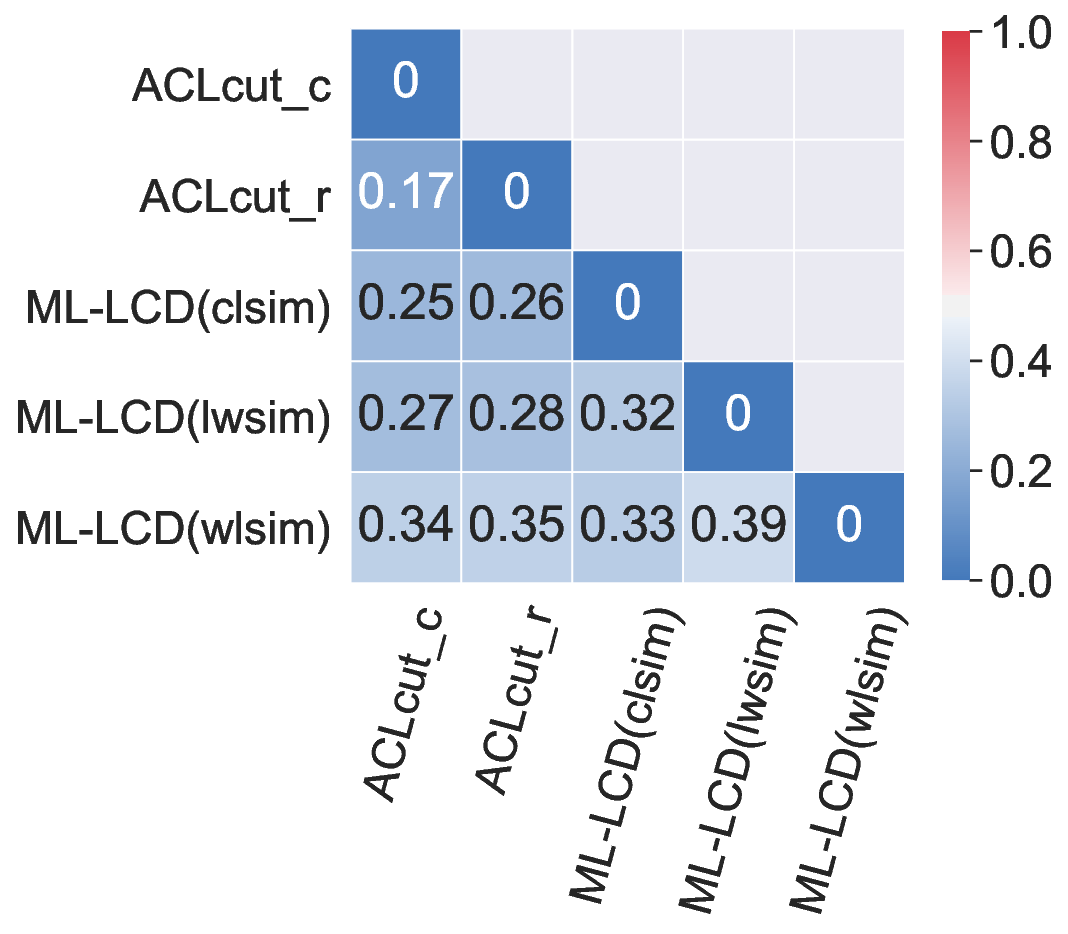}
		\caption{\textbf{PNO}}
		\label{fig:pno_local_pw}
	\end{subfigure}
	\begin{subfigure}{0.3\textwidth}
		\centering
		\includegraphics[width=.95\textwidth]{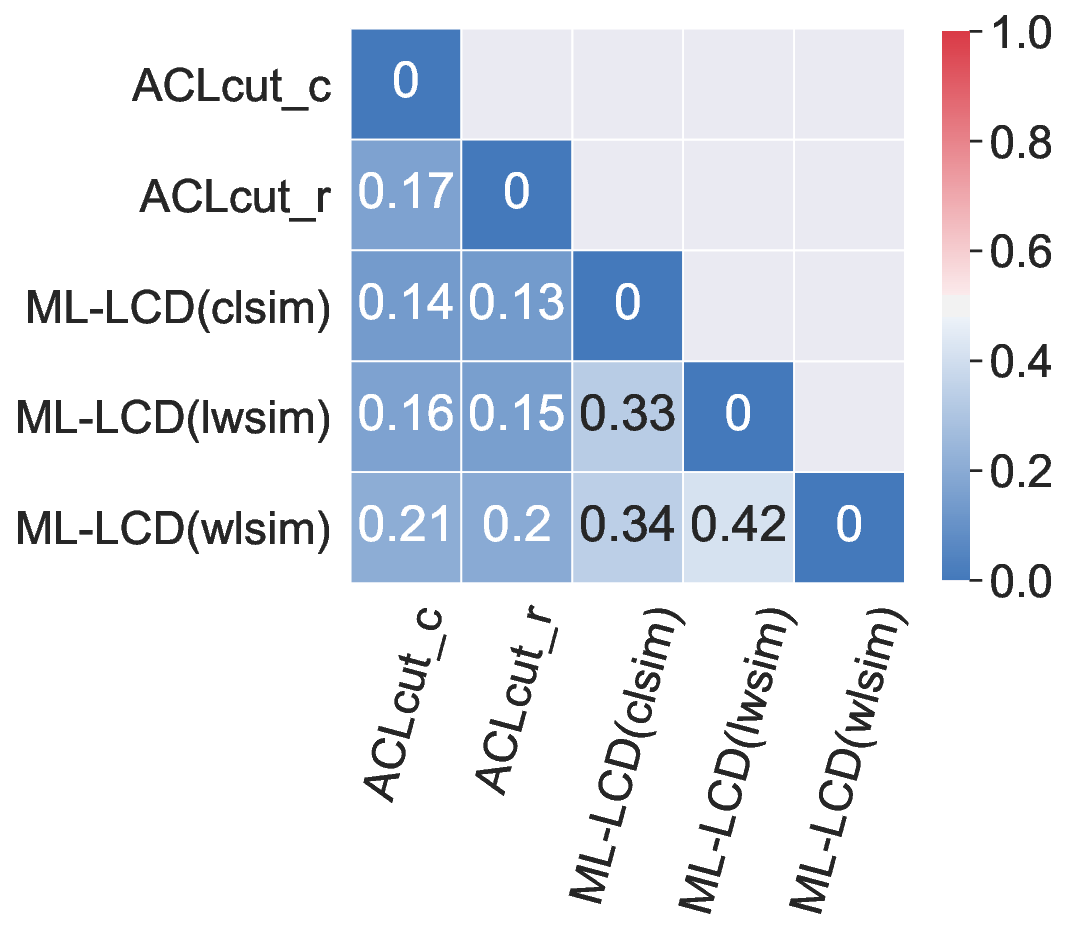}
		\caption{\textbf{SEP}}
		\label{fig:sep_local_pw}
	\end{subfigure}
	\begin{subfigure}{0.3\textwidth}
		\centering
		\includegraphics[width=.95\textwidth]{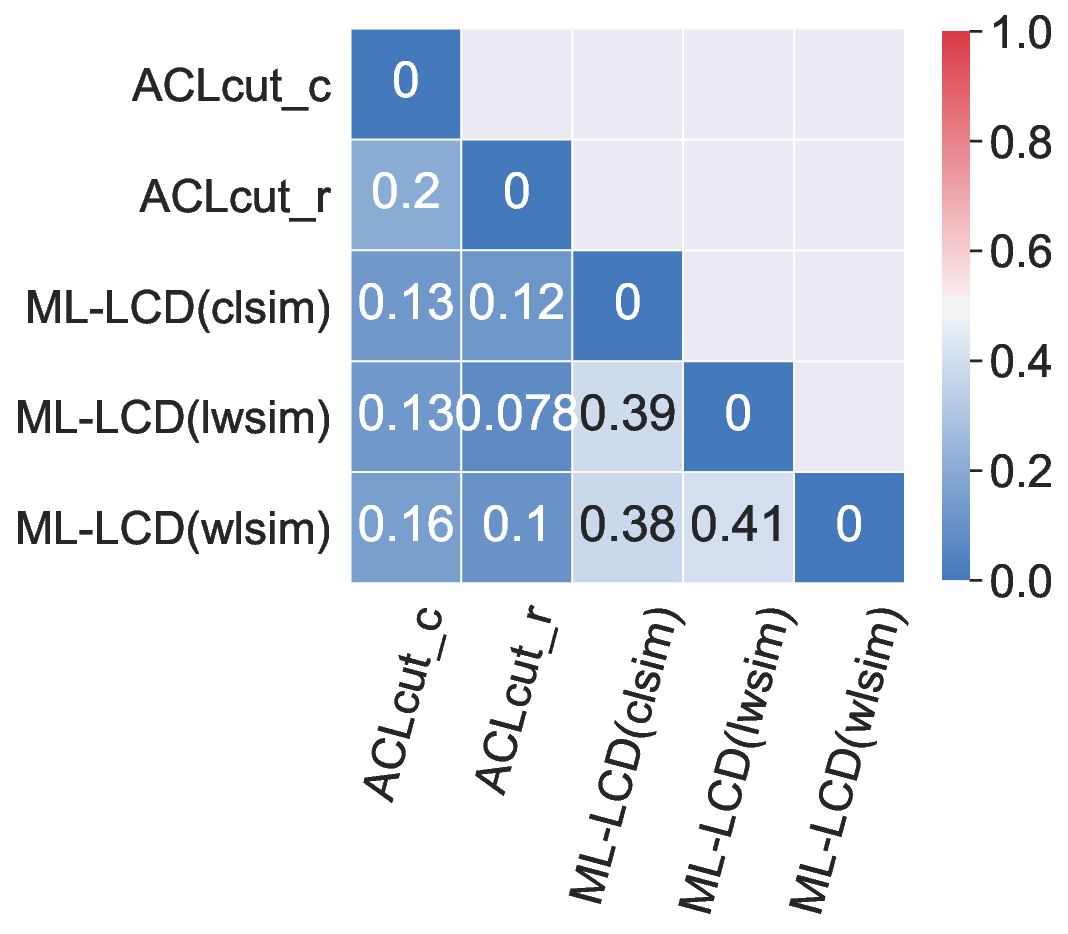}
		\caption{\textbf{SEO}}
		\label{fig:seo_local_pw}
	\end{subfigure}
	\begin{subfigure}{0.3\textwidth}
		\centering
		\includegraphics[width=.95\textwidth]{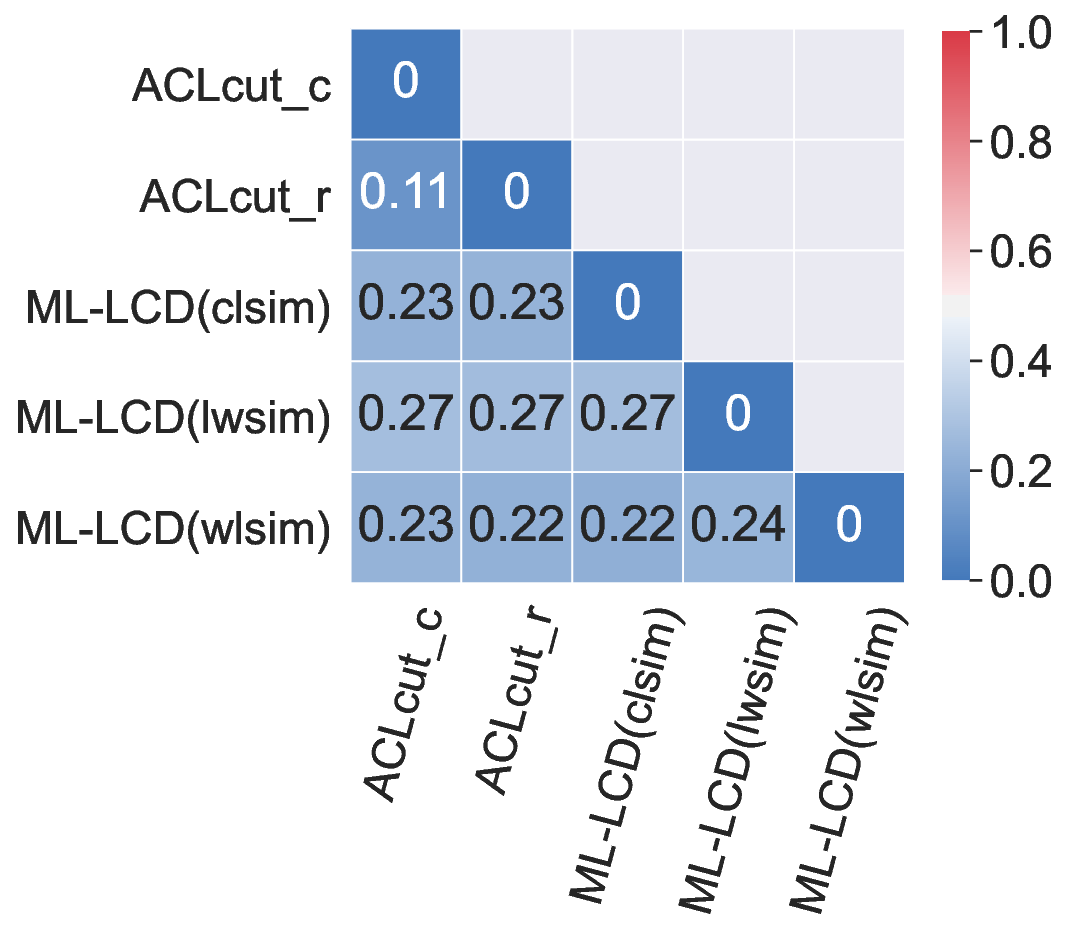}
		\caption{\textbf{NHN}}
		\label{fig:nhn_local_pw}
	\end{subfigure}
	\begin{subfigure}{0.3\textwidth}
		\centering
		\includegraphics[width=.95\textwidth]{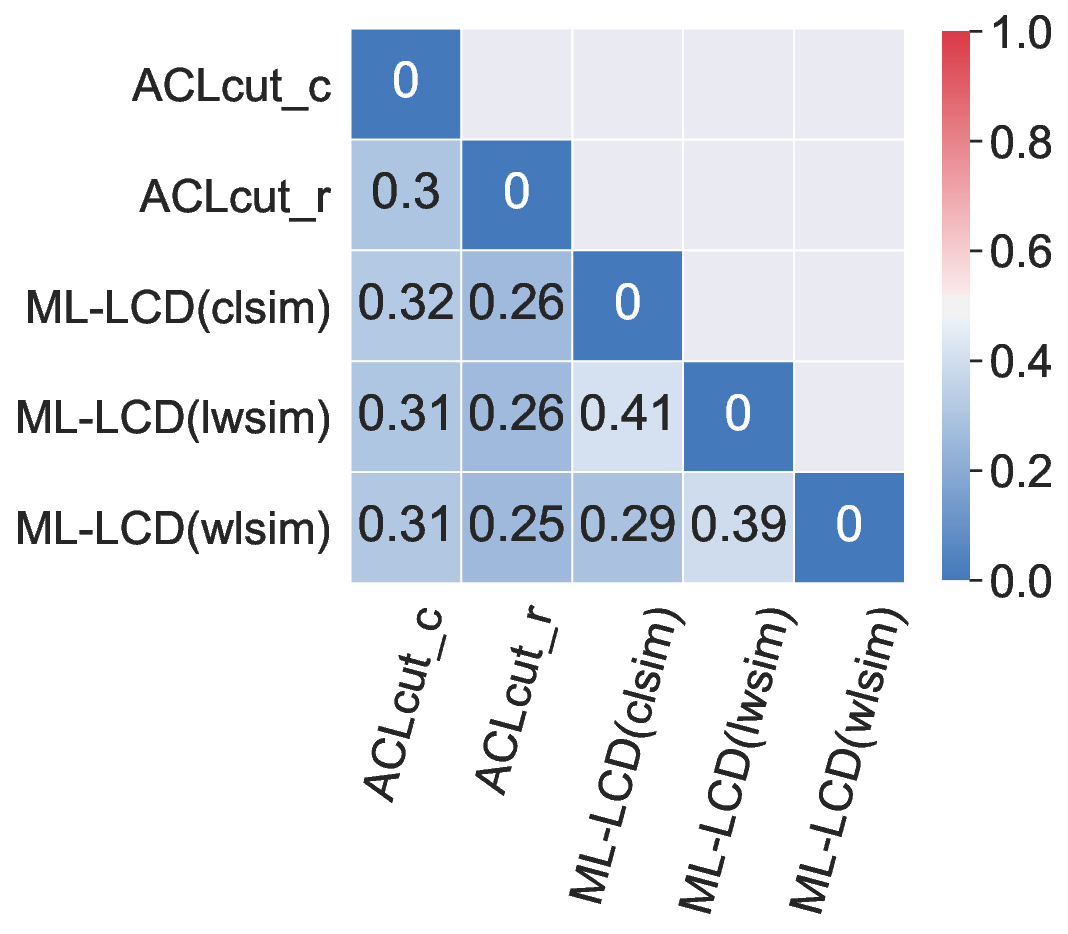}
		\caption{\textbf{NNM}}
		\label{fig:nnm_local_pw}
	\end{subfigure}
	\caption{Standard deviation for average pairwise similarity among the different local method\revm{s} (\revm{s}ynthetic networks)}
	\label{fig:local_pw_sim_syn_sd}
\end{figure}

Figures~\ref{fig:local_pw_sim_real_sd}, \ref{fig:local_pw_sim_syn_sd} report the standard deviation of the average pairwise similarity among local methods on real-world datasets and synthetic datasets reported in Section~\ref{sssec:local_pw_comp}\rev{.}

\subsubsection{Scalability Analysis}
\label{sssec:local_scal_anyalisis}

We tested the scalability of local community detection methods in terms of number of actors and number of layers. To carry out the experiment we used the synthetic networks already used for the global case (Section~\ref{sssec:results_glob_scal}). For each network, we present \rev{median} execution times obtained on 100 random seeds. For each method, we choose the least scalable variant as a representative of that method's scalability.  

\begin{figure}[!htpb]
	\centering
	\begin{subfigure}[t]{0.4\textwidth}
		\centering
		\includegraphics[width=\textwidth]{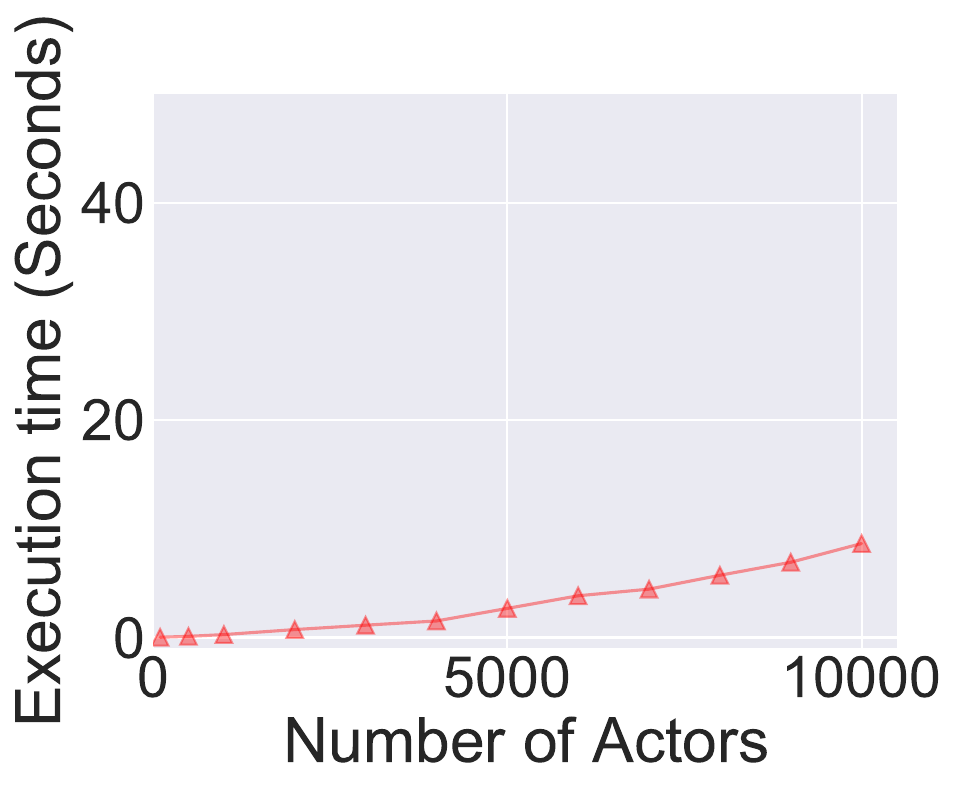}
		\caption{\textbf{ACLcut\rev{$_c$}}}
	\end{subfigure}
	\begin{subfigure}[t]{0.4\textwidth}
		\centering
		\includegraphics[width=\textwidth]{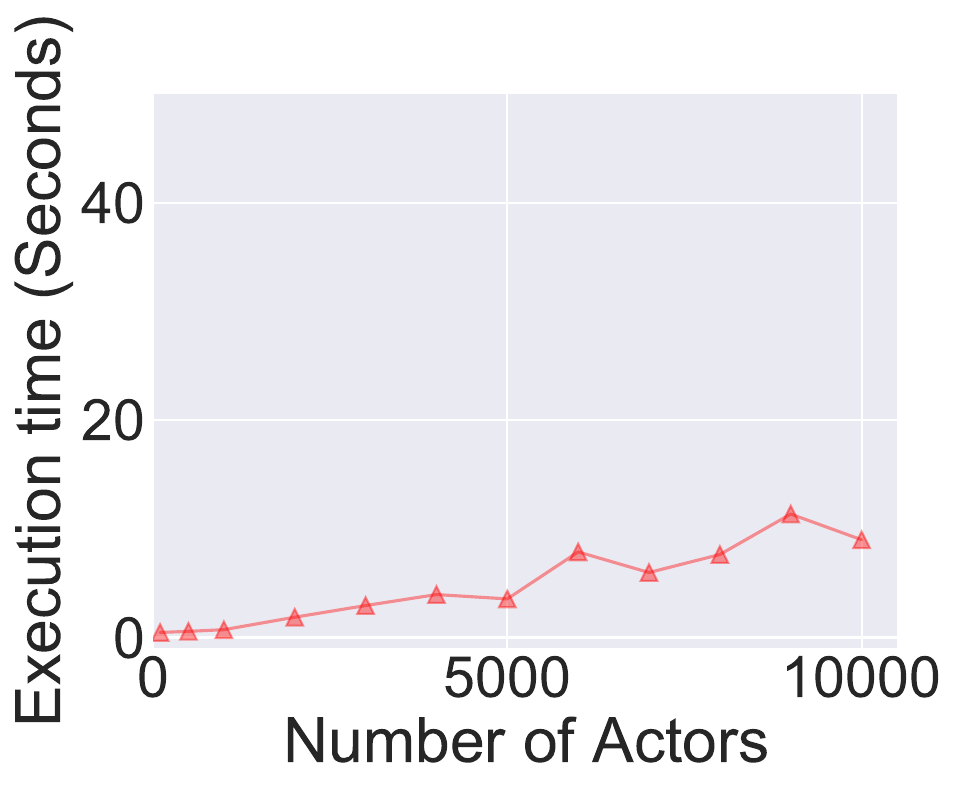}
		\caption{\textbf{ML-LCD\rev{$_{(clsim)}$}}}
	\end{subfigure}
	\caption{\rev{Median} scalability of local methods with respect to the number of actors in the multiplex network}
	\label{fig:actors_scalability_local}
\end{figure}
\begin{figure}[!htpb]
	\centering
	\begin{subfigure}[t]{0.4\textwidth}
		\centering
		\includegraphics[width=\textwidth]{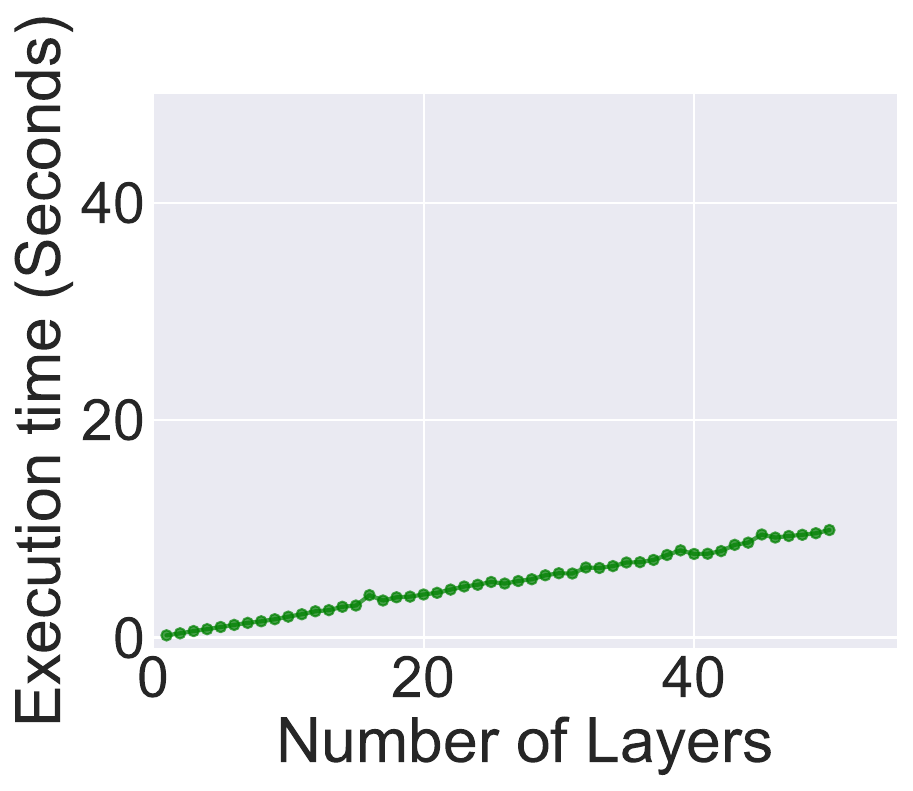}
		\caption{\textbf{ACLcut\rev{$_c$}}}
	\end{subfigure}
	\begin{subfigure}[t]{0.4\textwidth}
		\centering
		\includegraphics[width=\textwidth]{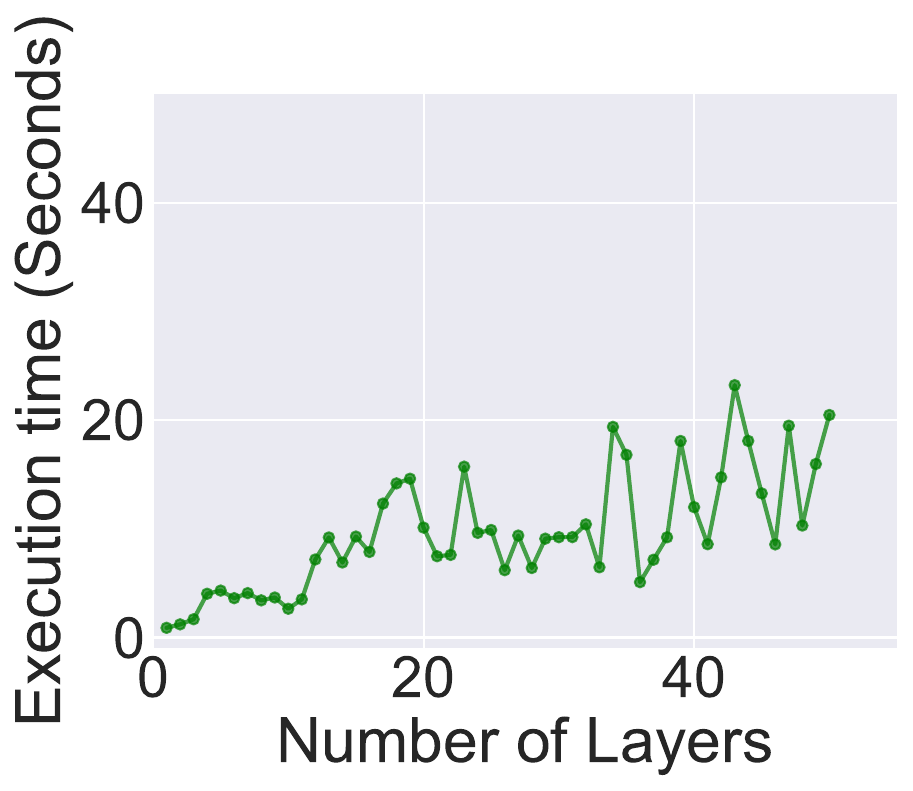}
		\caption{\textbf{ML-LCD\rev{$_{(clsim)}$}}}
	\end{subfigure}
	\caption{\rev{Median} scalability  of local methods with respect to the number of layers in the multiplex network}
	\label{fig:layers_scalability_local}
\end{figure}

Figures~\ref{fig:actors_scalability_local}--\ref{fig:layers_scalability_local} show results related to scalability in terms of number of actors and of layers respectively. \rev{Both methods showed a similar good scalability, with \textsf{ML-LCD} showing a higher dispersion depending on the chosen seeds}.

\section{Discussion}
\label{sec:discussion}



\rev{Our  experimental study had two main outcomes. First, it} allowed us to \rev{identify} guidelines \rev{about which methods can be the most appropriate} for the data and the task at hand. \rev{Second,} observing in which cases the \rev{reviewed} methods  consistently failed in identifying the \rev{expected} communities 
allowed us to identify the multiplex community structures that are challenging with the currently available community detection algorithms. \revm{While the comparative evaluation of community detection methods is a complex process, and additional types of analysis should also be considered in the future, such as the approach used by \cite{Ghasemian2019} for single-layer methods, our work highlights} a set of open problems for community detection methods in multiplex networks.

Accuracy analysis on synthetic networks
has revealed that \rev{most of the} methods perform \rev{very well} when the community structure 
\rev{is made of disjoint} pillar\rev{s}. \rev{Among the many well-performing methods}, Infomap \rev{and SCML are consistently} discovering community structures that are close \rev{or equal} to the ground truth, \rev{GLouvain and the methods using Louvain are also performing well but have some issues with communities of varying size.} whereas  ML-LCD$_{(clsim)}$ appears to be the best choice among the local methods. \rev{It is worth noticing that simpler flattening methods are also among the best methods.}

With regard to non-pillar community structures, we have observed a considerable reduction in the achieved accuracy scores for almost all methods. This observation raises the following question: what kind of assumptions are considered by different methods when multiplex communities are identified? It is clear that there is a tendency, even if not \rev{always} explicitly declared, to assume that multiplex communities are pillar communities expanding over all the layers of the multiplex network.  For instance, multi-slice modularity~\citep{Mucha2010} rewards pillar communities when calculating the modularity score, \rev{and spectral methods assume the existence of a latent community structure at actor level}. While pillar community structures are perfectly reasonable and can be \rev{assumed to exist} in many  scenarios, they \rev{are} also \rev{the simplest possible cases we tested in this article}. As multiplex approaches have been developed 
to overcome the oversimplification of monoplex networks, relying on a single type of ideal community structure seems, at least, a missed opportunity. Thus, more work has to be done on improving the accuracy of community detection methods for non-pillar community structures. 

A second set of considerations can be drawn by looking at the results obtained by the evaluated methods when applied to real-world datasets. Our experiments have shown that, on real-world datasets, the \rev{detected} community structures largely differ from the ground truth. This raises two interesting questions\rev{. First,} to which extent is the assumed ground truth itself a valid assumption? In other words, does the ground truth given for a real dataset always describe the community structures identified by a community detection method, or does it capture only one part of the whole picture?  The answer to this question  is never trivial even in monoplex networks. Nevertheless it is easy to see how adding more layers makes it further complicated. For example, both \rev{DKPol and AUCS} ground truths group together individuals belonging to the same organization (political parties in one case and research groups in the other). The question then becomes whether it is reasonable to assume that the selected relations, observed in the multiplex networks, will produce a community structure corresponding to this formal grouping, and to some extent, how different relations (thus different layers) can be more or less aligned with the hypothesis described above. Will members of the same research group work together, or publish together? Have lunch and fun together? Will members of the same political party retweet each other on Twitter, and reply to each other? Indeed, looking at the accuracy of the community structures identified for the real world dataset\rev{s}, especially in the case of \rev{DKPol}, one might ask whether we are observing a generalized failure of the community detection methods, or conversely, whether the community detection methods were actually able to observe relevant structures that \rev{were} just  different from the community structures assumed in the ground truth. 
 
The second question, which is strongly related to the first one, is whether all the layers included in these datasets positively contribute to an accurate identification of the community structure in these datasets, or whether some of them add more noise that heavily affect the identification process. Indeed, the fact that \rev{most of the} community detection method\rev{s} always give an output\revm{,} no matter what layers are included in the input multiplex network\revm{,} \rev{makes the inclusion of more input layers potentially problematic}. Layers, besides being defined by a specific internal topology, are also defined by internal logics that might or might not be coherent with those of the other layers. The \rev{DKPol} dataset represents a good example of this problem. \revm{S}ome \rev{detailed} analysis of the three layers composing the multiplex network has shown that retweets and following/follower interactions follow relatively \rev{assortative} dynamics for political parties. The replies, however, \rev{are more frequent between members of different} political parties. Here we think that more efforts have to be made in the modelling phase of the multiplex network and some layer-specific measures should be developed to lead the choice of the layers that contribute to the identification of the communities. Several such \emph{multilayer network simplification} methods exist, and more can be developed, as reviewed in
\citep{DBLP:journals/csr/InterdonatoMPTV20}.

A \rev{separate} consideration should be made about the similarities of the obtained results. Focusing, for the above-mentioned reason, mainly on the results obtained from the synthetic networks, it is possible to observe some general patterns. Global partitioning methods show a remarkable level of similarity in detecting community structures based on a pillar-like model. Semi-pillar and hierarchical community structures show a lower degree of similarity between the retrieved community structures. 
\rev{We should also consider that differences in the results of different algorithms may be partially due to the fact that some algorithms use heuristics to optimize an objective function (e.g., generalized Louvain), therefore they might not achieve the optimal value.}

Local methods show a behavior that is, to some extent, similar to the global partitioning methods. When tested on pillar communities they show a remarkable similarity between the produced communities, which can easily lead to calling them interchangeable. Nevertheless, the less pillar-like the community structure in the data is, the higher the differences seem to be at first between \textsf{ACLcut} and \textsf{ML-LCD} and then also between \rev{different settings of} the same algorithm.

\rev{Scalability analysis has also provided useful information about specific methods with scalability issues, which can be used to select feasible approaches depending on the data.}

\rev{
We would also like to draw additional remarks that might be considered mainly by  practitioners. Community detection remains a challenging task, and further complicated in multilayer networks, which is testified by the plethora of available approaches and methods, most of which have been studied in our extensive survey, while new others are currently under development at the time of this writing. From a practical viewpoint, the core problems are, on the one hand, \textit{i}) to select the most suited algorithm and parameterization for a target application domain, and on the other hand, \textit{ii}) to have it clear in mind what kind of community we are interested in or we expect to detect. 
Problem \textit{i}) should be addressed by taking into account that community detection methods, especially if belonging to different methodological approaches, will easily discover different patterns in a multilayer network, mainly because every method has its own bias resulting from the optimization of different criteria. 
We believe this variety of choice should not be seen as a negative point, but rather as an opportunity to find out communities with different structures and related meanings. Also, if the need for having a unified solution from different available ones still remains as a priority, the ensemble-based consensus approach could be considered as the way to go. 
Understanding problem \textit{ii}) will nonetheless be crucial in most cases, as it may  pose a requirement for the  structure of the communities to be discovered, thus possibly impacting on the choice of the method to be used. In any case, this will also depend on the actual presence of communities of a desired form in the input network; for instance, any method based on the identification of cliques of a given size will likely fail if such cliques  are rare or missing at all in the input network. Therefore, one suggestion in this regard  would be to deepen as much as possible the study of structural micro/mesoscopic characteristics of the input network, both in its entirety as a complex system and at \revm{the} level of its constituent layers, to better prepare the subsequent analysis for the community detection task. 
} 

\rev{Despite the complexity of the multiplex community detection task emerging from our study, we would like to conclude our discussion on a positive note. There are many cases where we have a good expectation of what type of community structures could be found in the data. One example is the simple case of actor communities that expand over multiple layers, as in the AUCS network where people inside the same research group work together, publish papers together and go to lunch together -- although the multilayer data allows us to appreciate how administrative people are part of the community only on some layers, and not for example of the co-authorship one. Another example are hierarchical communities where the layers represent different organizational levels, e.g., University-level interactions, Department-level interactions, research-group-level interactions, etc. Overlapping can also be expected inside data describing flexible organizations with people having multiple roles. These examples share the same features of some of our synthetic networks (Pillar, Hierarchical, Overlapping). 
Therefore, domain knowledge about what type of communities to expect can be used together with our accuracy (and scalability, in case of larger networks) plots to determine which algorithms to prioritize.}

\section{Concluding remarks}
\label{sec:conclusion}

\revm{This work has highlighted some facts. When the community structure is simple (pillar, non overlapping communities of similar size) we can expect most of the reviewed methods to work well. Therefore, scalability considerations may be used to choose the best algorithm. When we depart from this simple type of communities it becomes more difficult to identify them, and the concept of type of community structure itself requires more research. We can however see how some more sophisticated approaches not relying on flattening can be more successful in specific cases. Given the different types of communities identified by different methods, it can be valuable to try multiple approaches while exploring multiplex network data. The difficulty to handle some of the data suggests that more research in network preprocessing would be valuable \citep{DBLP:journals/csr/InterdonatoMPTV20} in addition to the development of new community detection methods.}
\revm{Community detection in multiplex networks is an active area, and it will be interesting to see how new algorithms address the challenges highlighted in this work. The code used for the experiments is available at https://bitbucket.org/uuinfolab/20csur and can be extended to include additional methods and data.}






\end{document}